\documentclass[fleqn,usenatbib]{mnras}

\usepackage{newtxtext,newtxmath}

\usepackage[T1]{fontenc}

\DeclareRobustCommand{\VAN}[3]{#2}
\let\VANthebibliography\thebibliography
\def\thebibliography{\DeclareRobustCommand{\VAN}[3]{##3}\VANthebibliography}

\usepackage{gensymb}
\usepackage{graphicx}
\usepackage{wrapfig}
\usepackage{multirow}
\usepackage{graphicx}	
\usepackage{amsmath}	

\newcommand{\ipole}{{\fontfamily{qcr}\selectfont IPOLE}}
\newcommand{\ebhlight}{{\fontfamily{qcr}\selectfont EBHLIGHT}}

\newcommand{\rhigh}{R_{\rm high}}
\newcommand{\rlow}{R_{\rm low}}
\newcommand{\munit}{M_{\rm unit}}
\newcommand{\unit}[1]{~{\rm #1}}

\defcitealias{M872018_PaperI:2024}{EHTC~2018M87~Paper~I}
\defcitealias{M872018_PaperII:2025}{EHTC~2018M87~Paper~II}
\defcitealias{M87_PaperI:2019}{EHTC~M87~Paper~I}
\defcitealias{SgrA_PaperI:2022}{EHTC~Sgr~A*~Paper~I}
\defcitealias{SgrA_PaperV:2022}{EHTC~Sgr~A*~Paper~V}
\defcitealias{M87_2017_2021:2025}{EHTC~M87 EHT Multi-epoch}
\defcitealias{M87_movie:2026}{EHTC~M87 Movie Campaign PR}
\defcitealias{M87_PaperV:2019}{EHTC~M87~Paper~V}
\defcitealias{M87_PaperVII:2021}{EHTC~M87~Paper~VII}
\defcitealias{M87_PaperIX:2023}{EHTC~M87~Paper~IX}

\title[Relativistic electrons around M87 black hole]{Probing the details of relativistic electrons with multifrequency observations of M87 black hole}

\author[J. Keuper et al.]{
J.G.J. Keuper$^{1}$\thanks{E-mail: joep.keuper@ru.nl},
M. Moscibrodzka$^{1}$\thanks{E-mail: m.moscibrodzka@astro.ru.nl}
\\
$^{1}$Department of Astrophysics/IMAPP, Radboud University, P.O. Box 9010, 6500 GL Nijmegen, The Netherlands}

\date{Accepted XXX. Received YYY; in original form ZZZ}

\pubyear{2015}

\begin{document}
\label{firstpage}
\pagerange{\pageref{firstpage}--\pageref{lastpage}}
\maketitle

\begin{abstract}
Beginning in March 2026, the Event Horizon Telescope (EHT) conducted a two-month observing campaign aimed at obtaining the first time-resolved sequence of images of the black hole M87* at the center of the Virgo A galaxy. The primary scientific objectives of this campaign are to constrain the black hole spin and to investigate the temporal variability of the magnetic field and plasma properties within the horizon-scale jet-launching region. In this work, we present theoretical predictions for the spectral index maps expected from these observations.
Our predictions are based on Magnetically Arrested Disk simulations performed with the \ebhlight~code and post-processed using the \ipole~general relativistic ray-tracing code. We investigate the temporal evolution of the spectral index for a range of electron distribution models and five different black hole spin values. The analysis includes both frequency bands currently present within the EHT and neighboring frequencies proposed for future extensions of the EHT array. In addition to extending previous studies to a broader range of observing frequencies, our work systematically investigates the time variability of the predicted spectral index maps.
In purely thermal electron models, the temporal variability of the spectral index maps closely traces variations in the magnetic field strength and electron temperature in the immediate vicinity of the event horizon. By contrast, models incorporating a non-thermal electron population exhibit substantially weaker spectral index variability, owing to the contribution of a power-law component with a fixed spectral slope. Our numerical results are consistent with theoretical expectations. Finally, a comparison with the recently measured integrated spectral index of the M87* enables us to place preliminary constraints on the model parameters. 
\end{abstract}




\section{Introduction}

The Event Horizon Telescope (EHT) has revolutionized black hole imaging over the past years. Through the use of linked radio telescopes across the world, it has managed to measure the size and shape of the emission regions around two supermassive black holes with some of the largest apparent event horizons, namely SgrA* at the center of the Milky Way galaxy and M87* at the center of the Virgo A galaxy, also known as the Messier 87 galaxy \citepalias{M87_PaperI:2019,M872018_PaperI:2024,SgrA_PaperI:2022,M87_2017_2021:2025}. Further imaging at a variety of wavelengths provides the possibility to gain a sense of the dynamic evolution and thermodynamics of black hole accretion flow. Starting in March 2026, EHT launched a two-month observing campaign to assemble the first time resolved sequence of images of M87* to measure the variability of the source on short time-scales to probe magnetic field dynamics, black hole spin, and plasma properties in the jet-launching regions at black hole event horizon scales \citepalias{M87_movie:2026}.

So far, the comparison between M87* models and observations made by EHT remains inconclusive \citepalias{M87_PaperI:2019}. There is, however, a preference for models using magnetically arrested disks (MADs; \citetalias{M87_PaperV:2019,M872018_PaperII:2025}), which assume that the magnetic field is dynamically dominant in the vicinity of the black hole and thus disrupts the accretion flow of the gas in the accretion disk as it establishes a maximum threshold for the magnetic flux \citep{igumenshchev:2003,narayan:2003,mckinney:2012}. The spin of the black hole is still largely unknown \citep[e.g.,][]{bernshteyn:2026} due to the biggest unsolved problem, namely that the electron temperatures, or more generally the electron distribution function, depend on microphysics which are not captured by the global simulations of MADs. The radiation from the accretion disk is caused by the accelerating electrons and the protons are primarily heated by processes in the disk \citep{abramowicz:2013}. Protons and electrons normally exchange energy through Coulomb collisions. However, these Coulomb collisions are very inefficient in systems like M87*, since the black hole is accreting at a very sub-Eddington accretion rate, meaning the heat in the protons cannot be efficiently transferred to the radiating electrons, possibly creating a so called two-temperature plasma \citep{shapiro:1976}. This means that other processes, like plasma waves and kinetic instabilities, need to be considered to provide a more efficient way to couple the protons and electrons \citep{abramowicz:2013}. All of this makes it very hard to accurately know how the electrons are heated by all these different processes and what their temperatures are exactly, so some assumptions have to be made about how hot the electrons are. It is also possible that the electrons are (partially) non-thermal in this type of accretion flow. 

EHT images of black holes are also taken in polarized light \citepalias{M87_PaperVII:2021}. Constraining models with the linear polarization of light efficiently limits the possible parameter space \citepalias{M87_PaperVII:2021}. However, the linear polarization is very sensitive to external/internal Faraday screen effects that rotate the polarization vectors (\citealt{moscibrodzka:2017}, \citetalias{M87_PaperVII:2021}), which could cause issues with the theoretical interpretations of the images. The other type of polarization, circular polarization, is very weak in M87* when compared to linear polarization \citepalias{M87_PaperIX:2023} and turned out 
not to be very useful in selecting the preferred models.

Multi-epoch and multifrequency observations on horizon scales are the next step towards a more robust understanding of accretion flow and jet launching in M87*. Variability measurements are rarely done (on long time-scales; \citetalias[see][]{M87_2017_2021:2025}); however, the first quasi-simultaneous measurements in multiple frequencies are already available (\citealt{lu:2023}, \citetalias{M872018_PaperI:2024}), so it is already possible to access the model constraining power of the multi-wavelength observations. 

This work focuses on multi-epoch and multifrequency modeling of the M87* horizon scale emissions. In particular, we generate and study the spectral index maps of the models, both for its values and its variability. Until now, M87* spectral index studies have been scarce, so these measurements could be useful for testing models and constraining possible combinations of model parameters. We compare our model observables to currently available real measurements by EHT and other interferometric networks. The sequences of images EHT is currently making of M87* also provide the possibility to examine the variability of the black hole at horizon scales, meaning that a variability study is also needed to compare to these movies EHT is making.

The article is organized as follows. In section~\ref{sec:methods}, we describe how M87* models are constructed and how they are compared to current observations. Readers familiar with GRMHD simulations and ray-tracing methods could skip reading this section, with the exception of subsection~\ref{sec:real_data}.
In section~\ref{sec:results}, we present results, explain them using simple theoretical considerations, and compare the results to the current observational constraints.
We discuss the results and conclude in section~\ref{sec:discussion}.

\section{Methods}\label{sec:methods}

\subsection{Accretion flow models}

\begin{figure}
\centering
  \centering
  Density $\rho$ (code units)
  \includegraphics[trim={0cm 0.3cm 0.8cm 0cm},clip,width=\linewidth]{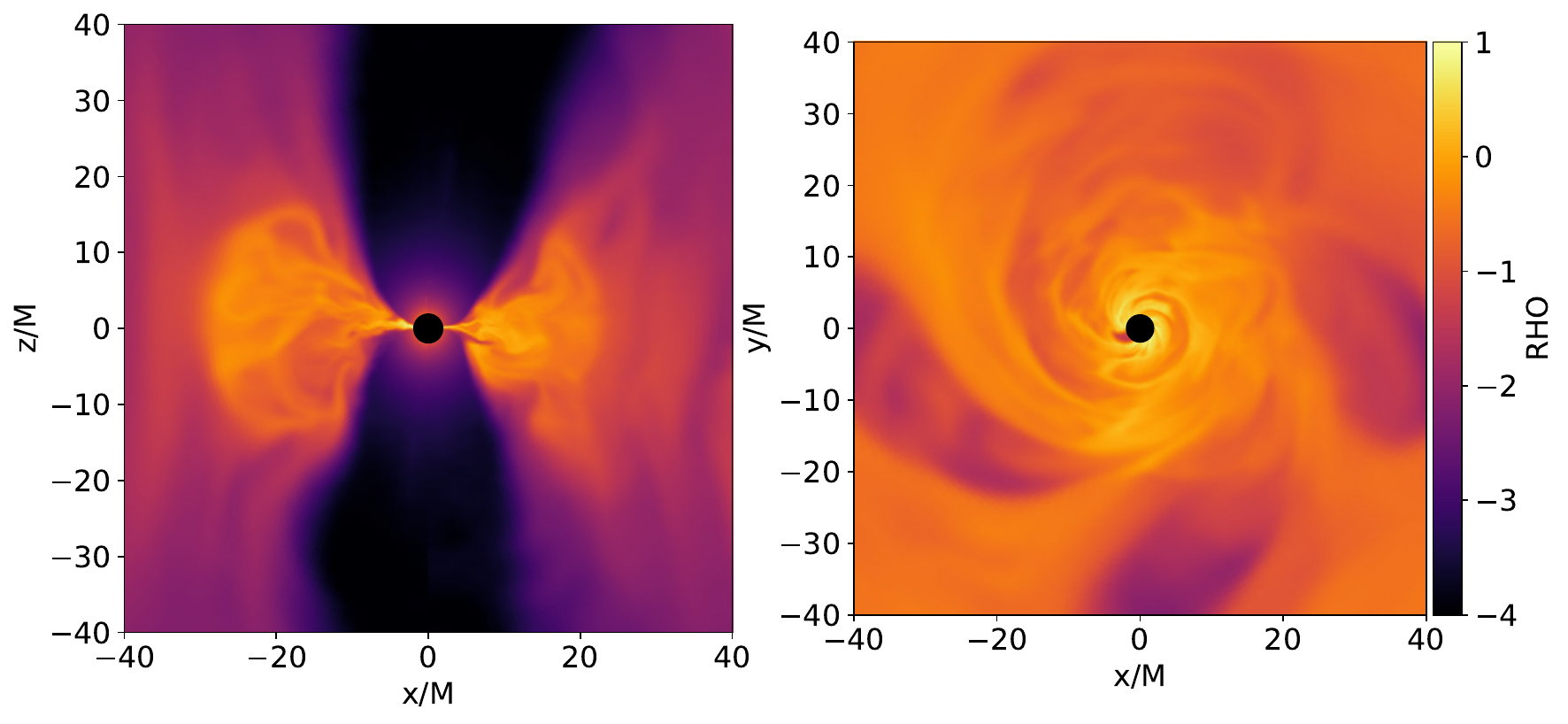}
  Pressure ratio $\beta$
  \includegraphics[trim={0cm 0.3cm 0.8cm 0cm},clip,width=\linewidth]{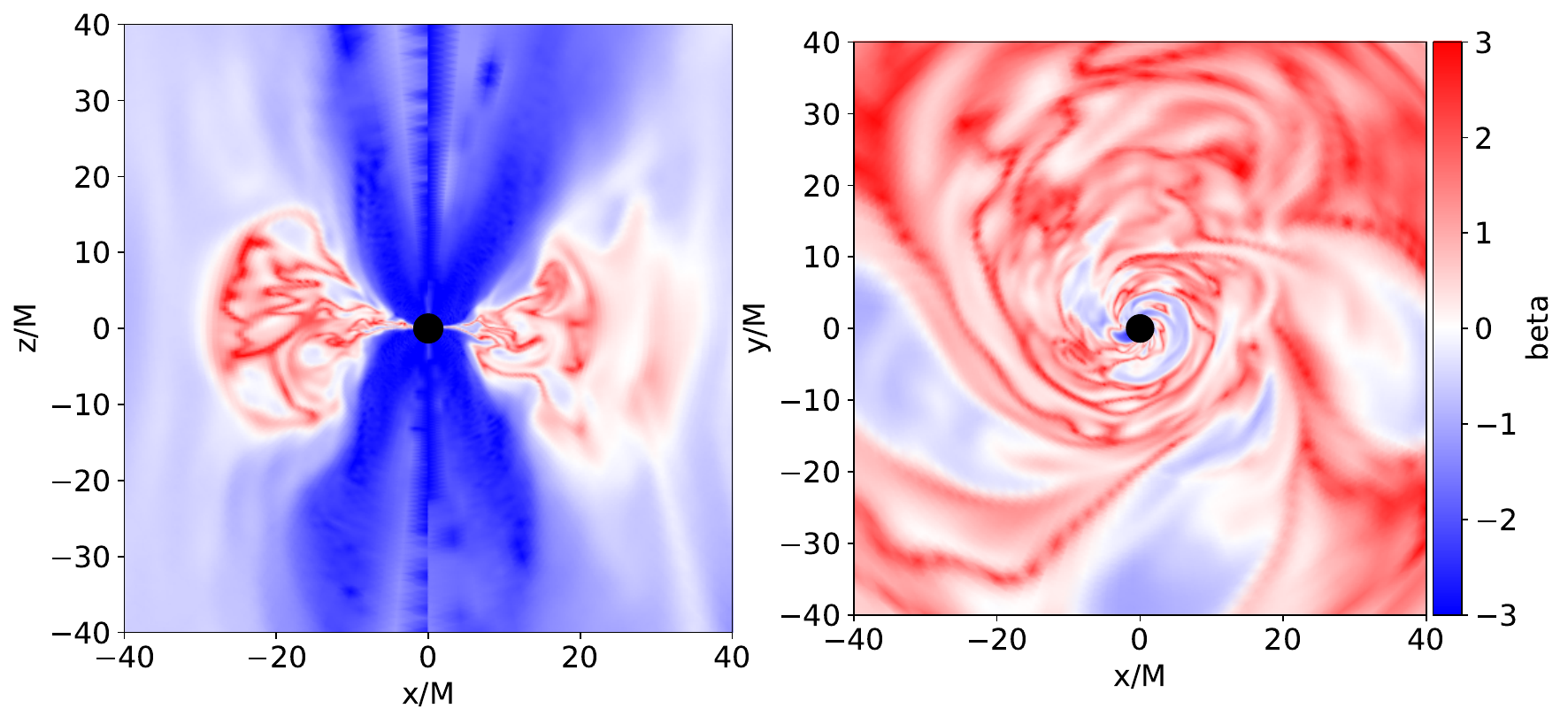}
\caption{Example of spatial distributions of quantities in the adopted MAD models \citep{moscibrodzka:2025}. The left and right panels show the meridional and equatorial planes with plasma density $\rho$ in dimensionless units and and gas to magnetic pressure ratio $\beta$ for a black hole spin $a_*~=~0.5$. Simulations for other black hole spins look similar.}
\label{fig:distribution_of_params_ebhlight_a0.5}
\end{figure}

We model M87* spectral slopes using a set of non-radiating 3-D GRMHD simulations of black hole accretion flows originally presented by \citet{moscibrodzka:2025}. These simulations used the \ebhlight~code\footnote{\url{https://github.com/afd-illinois/ebhlight}} \citep{ryan:2015}, which solves the General Relativistic (Radiation) magnetohydrodynamics equations (GR(R)MHD). The simulations assume particle number conservation and the magnetohydrodynamic stress-energy tensor to derive equations for mass, energy, and momentum conservation. \ebhlight~integrates these equations using a second order shock-capturing scheme. For more in-depth information about these equations and how they are exactly integrated, we refer the reader to the \ebhlight~code paper by \citet{ryan:2015}.
The adopted models are non-radiating MADs around a black hole with five different dimensionless spin parameters $a_*\equiv Jc / GM^2\in(-0.9375, -0.5, 0, 0.5, 0.9375)$. 
All simulations were first given some time to relax from initial conditions; $a_* \in (0, 0.5, 0.9375)$ models were given a settle-time of $20,000 M$ (where the time unit is $M \equiv \frac{GM}{c^3}$), and $a_* \in (-0.5, -0.9375)$ models were given a settle-time of $9,000 M$, after which snapshots across a time domain of $5,000 M$ were saved every $10 M$. 
\citet{moscibrodzka:2025} shows that at this stage, all models reach a steady state within the inner $20 M$ (where the length unit is $M \equiv \frac{GM}{c^2}$). Figure~\ref{fig:distribution_of_params_ebhlight_a0.5} shows a single frame snapshot of one of the models used in this work for both the density $\rho$ and the gas to magnetic pressure ratio $\beta \equiv \frac{P_{\rm gas}}{P_{\rm mag}}$.
Simulations were carried out with initial conditions (torus and its size; \citealt{fishbone:1976}) and grid resolutions comparable to or identical to the MAD simulations presented recently, e.g., in \citetalias{M872018_PaperII:2025}. 

\subsection{Synchrotron emission models}\label{sec:synchrotron_rad}

Next, \ipole~code\footnote{\url{https://github.com/moscibrodzka/ipole_lite}} \citep{moscibrodzka:2018} is used to post-process the GRMHD simulations to create model emission maps/images and spectral index maps at various frequencies. \ipole~ uses ray-tracing techniques along geodesics for covariant, polarized radiative transport. The code parallel transports wave vectors $k^\mu$ and the coherency tensors and evolves the Stokes parameters in the fluid frame under emission, absorption, and Faraday conversions. Integration of the transfer equation for many rays is carried out from the model towards a pinhole camera at a significant distance from the simulation origin. The governing equations for fully polarized radiative transfer in the frame co-moving with the fluid are
\begin{equation}
    \frac{d}{d\lambda} \begin{pmatrix}
    I_{\nu}\\
    Q_{\nu}\\
    U_{\nu}\\
    V_{\nu}
    \end{pmatrix} = \begin{pmatrix}
    j_{\nu, I}\\
    j_{\nu, Q}\\
    j_{\nu, U}\\
    j_{\nu, V}
    \end{pmatrix} - \begin{pmatrix}
    \alpha_{\nu, I} & \alpha_{\nu, Q} & \alpha_{\nu, U} & \alpha_{\nu, V}\\
    \alpha_{\nu, Q} & \alpha_{\nu, I} & \rho_{\nu, V} & -\rho_{\nu, U}\\
    \alpha_{\nu, U} & -\rho_{\nu, V} & \alpha_{\nu, I} & \rho_{\nu, Q}\\
    \alpha_{\nu, V} & \alpha_{\nu, U} & -\rho_{\nu, Q} & \alpha_{\nu, I}
    \end{pmatrix} \begin{pmatrix}
    I_{\nu}\\
    Q_{\nu}\\
    U_{\nu}\\
    V_{\nu}
    \end{pmatrix}
    \label{eq:radiative_transfer_eq_polarized}
\end{equation}
where
\begin{equation}
    \nu = -k^{\mu} u_{\mu}.
    \label{eq:ipole_eq5}
\end{equation}
is the photon frequency measured by an observer co-moving with the fluid four-velocity $u_{\mu}$ and $(I_\nu,Q_\nu,U_\nu,V_\nu)$ are the Stokes parameters along a ray paths parametrized by the affine parameter $\lambda$. In the transfer equations, $j$ is the synchrotron emissivity, $\alpha$ is the absorptivity, and $\rho$ is the rotativity, while Compton scatterings are neglected. For more information on how these equations are solved in the coordinate frame, including the appropriate treatment of the transport of the polarization vector in curved spacetime, we refer the reader to the \ipole~ code paper.

\begin{figure*}
\centering
  \includegraphics[trim={0cm 0cm 0cm 0cm},clip,width=0.6\textwidth]{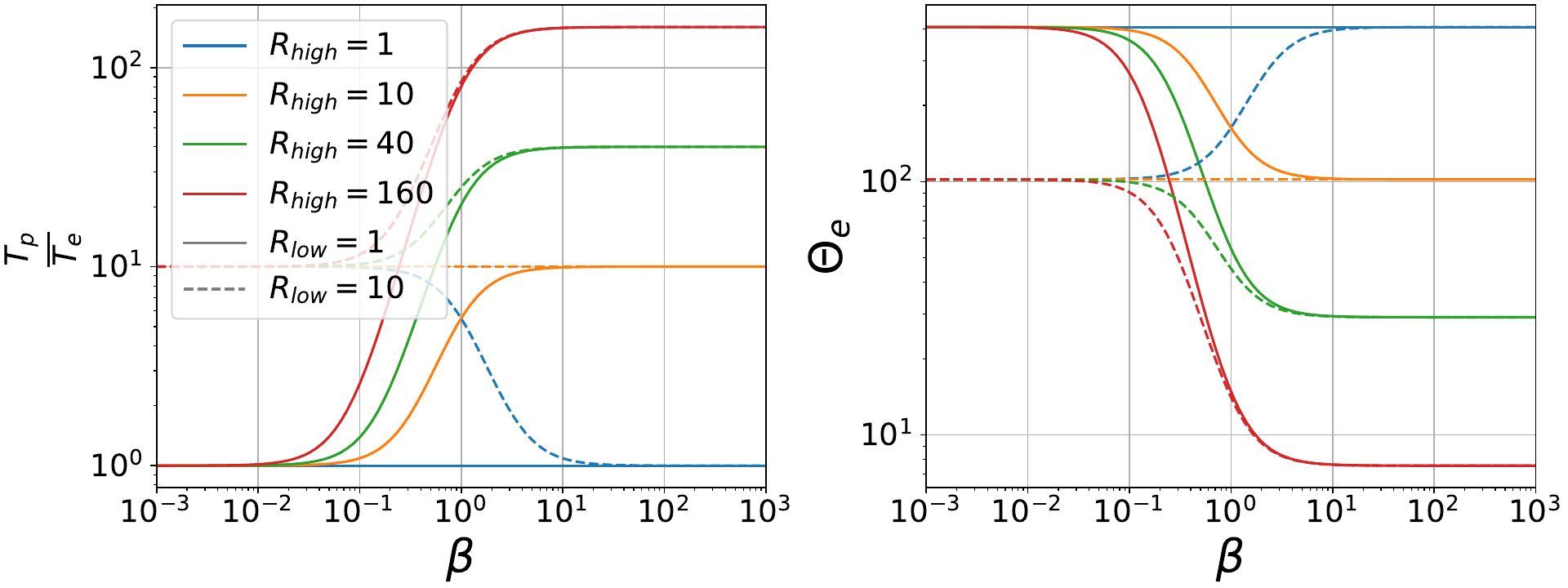}
  \includegraphics[trim={0cm 0cm 0cm 0cm},clip,width=0.3\textwidth]{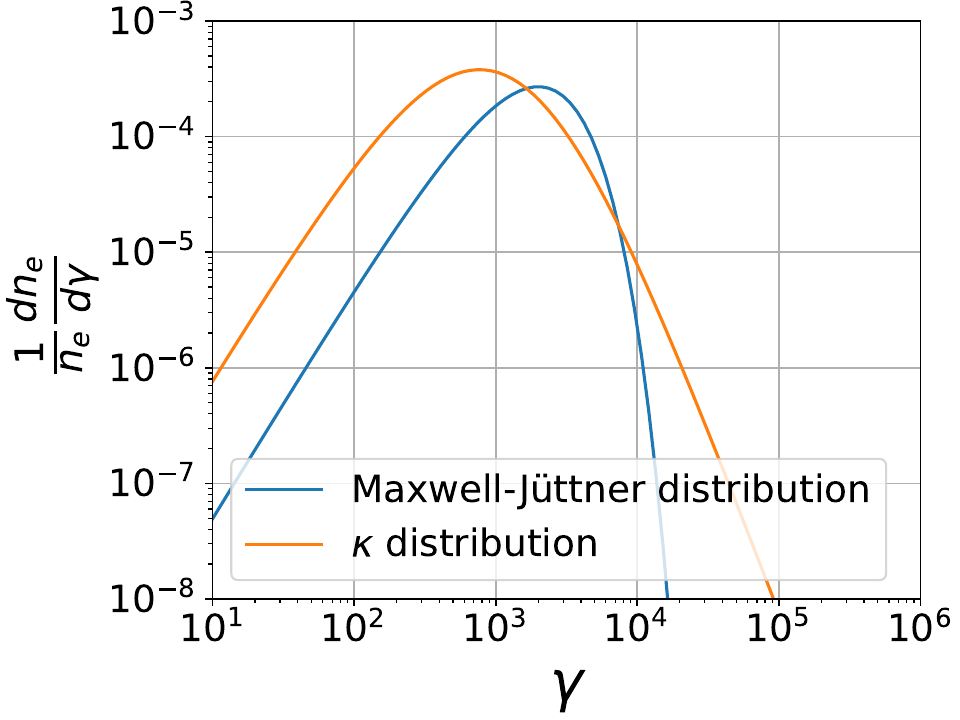}
  \caption{Visualization of the change in ratio of the electron temperature and proton temperature, the change in dimensionless plasma temperature $\Theta_{\rm e}$ with changing $\beta$ and an example for both a Maxwell-Jüttner distribution from Equation~\ref{eq:maxwell_juttner_dist} and a $\kappa$ distribution from Equation~\ref{eq:kappa_dist}. Notice that for $\beta \gg 1$ the ratio $\frac{T_{\rm p}}{T_{\rm e}}$ becomes $\rhigh$ and for $\beta \ll 1$ the ratio becomes $\rlow$. Electron temperature $\Theta_{\rm e}$ follows Equation~\ref{eq:theta_e} and thus becomes smaller for $\beta \gg 1$ when $\rhigh > \rlow$. The distributions in the right-most panel have $\Theta_{\rm e} = 1000$  and $\kappa = 4.25$ (where $w=\Theta_{\rm e} (\kappa-3)/\kappa\approx 300$ in case of the $\kappa$ distribution function). }\label{fig:ratio_thetae_visualizations}
\end{figure*}

The synchrotron emissivity coefficients for Stokes parameters are given by \citep{pandya:2016}:
\begin{equation}
    j_{\nu, S} = \begin{pmatrix}
    j_{I}\\
    j_{Q}\\
    j_{U}\\
    j_{V}
    \end{pmatrix} = \frac{2 \pi e^2 \nu^2 n_{\rm e}}{c} \int d^3 p f(p) \sum_{n=1}^{\infty} \delta(y_n)K_S,
\end{equation}
where $\delta$ is the Dirac delta function, and $S = (I, Q, U , V)$ is the Stokes vector.
\begin{equation}
  K_S = \begin{cases}
        M^2 J_n^2(z) + N^2 J_n'^2(z), & \text{if } S = I, \\
        M^2 J_n^2(z) - N^2 J_n'^2(z), & \text{if } S = Q, \\
        0, & \text{if } S = U, \\
        -2 M N J_n(z) J_n'(z), & \text{if } S = V
    \end{cases}
\end{equation}
and
\begin{equation}
  y_n = \frac{n \nu_{\rm cyc}}{\gamma} - \nu (1 - \beta \cos{\xi} \cos{\theta})
\end{equation}
are functions specific to a synchrotron process. Here, $M = \frac{\cos{\theta} - \beta \cos{\xi}}{\sin{\theta}}$, $N~=~\beta \sin{\xi}$, $z = \frac{\nu \gamma \beta \sin{\theta} \sin{\xi}}{\nu_{\rm cyc}}$, $J_n$ is the Bessel function of the first kind, $J_n'$ is its derivative, and $\nu_{\rm cyc} = \frac{e B}{2 \pi m_{\rm e} c}$ is the electron cyclotron frequency. The synchrotron emissivity is a function of $f(p) \equiv \frac{1}{n_{\rm e}} \frac{dn_{\rm e}}{d^3p}$, the distribution function per particle for a momentum $p$, which has to be assumed. For an isotropic distribution function, $f(p)$ can be defined via $f(\gamma) \equiv \frac{1}{n_{\rm e}} \frac{dn_{\rm e}}{d\gamma}$, which is the distribution function per particle for the electron Lorentz factor $\gamma$. Two commonly used models for $f(\gamma)$ are the thermal Maxwell-Jüttner distribution:
\begin{equation}
    f_{\rm th}(\gamma) = \frac{\gamma \sqrt{\gamma^2 - 1}}{\Theta_{\rm e} K_2(\frac{1}{\Theta_{\rm e}})}e^{\frac{-\gamma}{\Theta_{\rm e}}},
    \label{eq:maxwell_juttner_dist}
\end{equation}
where parameter $\Theta_{\rm e} = \frac{k_B T_{\rm e}}{m_{\rm e} c^2}$ is the dimensionless plasma temperature, and the non-thermal $\kappa$ distribution function \citep{xiao:2006}:
\begin{equation}
    f_{\kappa}(\gamma) = N \gamma \sqrt{\gamma^2 - 1} \left( 1 + \frac{\gamma - 1}{\kappa w} \right)^{-(\kappa + 1)},
    \label{eq:kappa_dist}
\end{equation}
where $w,\kappa$ are parameters and $N$ is a normalization constant. The $\kappa$ function smoothly connects a thermal core with a power-law distribution function. The parameter $w$, which describes the position of the thermal core as a function of $\gamma$, can be anything, but in this work, we assume that it is scaled through $w = \Theta_{\rm e} \frac{\kappa - 3}{\kappa}$. With this assumption, the total energy of the $\kappa$ distribution function will be equivalent to the energy of the thermal distribution function with $\Theta_{\rm e}$. This assumption will allow us to model the $w$ parameter in the same way as $\Theta_{\rm e}$ in thermal models.
$\kappa$ index is the power-law parameter that controls the slope of the distribution function at higher energies; for $\kappa \rightarrow \infty$, the $\kappa$ distribution function collapses to the Maxwell-Jüttner distribution function. The value of the parameter $\kappa$ is typically unknown and must be assumed (a condition must be satisfied that $\kappa>3$ for $w>0$). Synchrotron emissivities $j$, corresponding absorptivities $\alpha$, and rotativities $\rho$ for the $\kappa$ distribution are implemented in the \ipole~code using the stochastic average concept introduced by \citet{moscibrodzka:2024} (see also \citealt{vanduren:2025}). In this method, any non-thermal distribution functions and their transfer coefficients are constructed as a sum of the thermal components. Emissivities, absorptivities, and rotativities for thermal distribution functions, on the other hand, are taken from \citet{pandya:2016}.

In this work, both distribution functions (Equations~\ref{eq:maxwell_juttner_dist} and~\ref{eq:kappa_dist}) are used when computing emission from GRMHD models, and since in both cases the electron temperature $\Theta_{\rm e}$ is used as an energy parameter, we must make an assumption for $\Theta_{\rm e}$. In strongly sub-Eddington sources such as M87* the electron temperature $T_{\rm e}$ likely differs from the proton temperature $T_{\rm p}$ that dominates the accretion flow dynamics \citep{shapiro:1976}.
Although simulations presented in \citet{moscibrodzka:2025} passively evolve the parameter $\Theta_{\rm e}$ assuming a specific prescription for electron heating \citep{kawazura:2019}, in this work we follow a more general approach adopted by \citetalias{M872018_PaperII:2025} (and references therein). In this work, the temperature ratio between protons and electrons depends on the parameter $\beta$.
Following \citet{moscibrodzka:2016}, the ratio between the electron temperature and the proton temperature is
\begin{equation}
    R(\beta) \equiv \frac{T_{\rm p}}{T_{\rm e}} = \rhigh \frac{\beta^2}{1 + \beta^2} + \rlow \frac{1}{1 + \beta^2},
    \label{eq:rhigh_rlow}
\end{equation}
where $\rhigh$ represents the ratio for the regions where $\beta \gg 1$, and $\rlow$ represents the ratio for the regions where $\beta \ll 1$. The dimensionless plasma temperature can then be calculated using
\begin{equation}
    \Theta_{\rm e} = \frac{u}{\rho} \frac{m_{\rm p}}{m_{\rm e}} \frac{(\gamma_{\rm p} - 1)(\gamma_{\rm e} - 1)}{(\gamma_{\rm e} - 1)R(\beta) + (\gamma_{\rm p} - 1)}
    \label{eq:theta_e}
\end{equation}
where $\rho$ equals the rest-mass density, $u$ is the gas total internal energy, and the adiabatic indices are $\gamma_{\rm e} = \frac{4}{3}$ and $\gamma_{\rm p} = \frac{5}{3}$, as assumed by \citet{moscibrodzka:2025}. Figure~\ref{fig:ratio_thetae_visualizations} shows visualizations of $\frac{T_{\rm p}}{T_{\rm e}}$ and $\Theta_{\rm e}$ as functions of $\beta$, together with an example of a Maxwell-Jüttner distribution from Equation~\ref{eq:maxwell_juttner_dist} and a $\kappa$ distribution from Equation~\ref{eq:kappa_dist}.

\subsection{Feature extraction and model comparison}

\subsubsection{Future and current observational data}\label{sec:real_data}

Our work focuses on modeling the observed spectral index of M87* emission on the black hole horizon scales (so called compact flux). Our main goal is to make predictions for future multifrequency and/or time-resolved observations by EHT that could span from $86 \unit{GHz}$ to $345 \unit{GHz}$ \citep[e.g.,][and references therein]{zhao:2025}. 
Nevertheless, M87* has already been observed quasi-simultaneously at $86\unit{GHz}$ and $\sim230\unit{GHz}$ in 2018. The 2018 measurement allows us, for the first time, to put preliminary constraints on the accretion model parameters.

The 2018 observations at $86\unit{GHz}$ were made by the Global Millimeter VLBI Array (GMVA) with participation from the Atacama Large Millimeter Array (ALMA) \citep{lu:2023}. 
The net measured flux at $86 \unit{GHz}$ is taken from their Table~S2. 
Depending on the assumed model, the net compact flux (i.e., excluding the extended jet emission that we do not model) ranges between $562.1 \pm 1.3 \unit{mJy}$ and $591.0 \pm 2.1 \unit{mJy}$, so the chosen value for the measured flux at $\nu = 86 \unit{GHz}$ is the average of the two extrema: $F_{86 \unit{GHz}} = 577 \pm 17 \unit{mJy}$. 
The flux measured at $\sim230 \unit{GHz}$ is taken from Figure~7 in \citetalias{M87_2017_2021:2025} publication, which, among other things, contains the compact flux M87 * measured by EHT in 2018 (a few days before the observation of 86\unit{GHz}. At $\nu = 230 \unit{GHz}$, $F_{230 \unit{GHz}} = 0.43_{-0.03}^{+0.29}$. Using these values for compact fluxes gives us the measured net spectral index of the compact emission region: $\alpha_{I, 86-230 \unit{GHz}} = -0.30_{-0.11}^{+0.56}$.

Apart from the spectral index, we also model the sizes of the emitting regions at different frequencies as the size of the image is the second observable which depends on the optical thickness of the emitting region.
Both mentioned GMVA and EHT images of M87* show rings. The observed sizes of these rings are ${\rm FWHM} = 64_{-8}^{+4} ~\upmu {\rm as}$ at $\nu = 86 \unit{GHz}$ \citep{lu:2023} and ${\rm FWHM} = 42 \pm 3 ~\upmu {\rm as}$ at $\nu\sim230 \unit{GHz}$ \citepalias{M872018_PaperI:2024}.
However, the measured ring sizes depend on the nominal resolutions of both VLBI arrays and on the types of many imaging algorithms used in these works. To be independent of these two issues, we will compare our model sizes to M87* size estimates based on pre-imaging considerations. Based on Gaussian model fits directly to the $\sim230\unit{GHz}$ EHT visibility data in the $(u,v)$ space on intermediate baselines, M87* has ${\rm FWHM}=39-98 ~\upmu {\rm as}$ \citepalias[see Appendix F1 in][]{M872018_PaperI:2024}. Using the Gaussian fit model to all $(u,v)$ GMVA data, the pre-imaging size of M87* at $86\unit{GHz}$ is ${\rm FWHM}=70.7\pm0.2 ~\upmu {\rm as}$ \citep[see Table S2 in][]{lu:2023}. The fit not only uses all $(u,v)$ data but also has unrealistically small errorbars, hence, we use only the $\sim 230\unit{GHz}$ size constraint. We explain how these Gaussian sizes are compared to our typically non-Gaussian GRMHD images in the following subsection~\ref{sec:model_comparison}. However, we first turn to listing the model parameters. 

\subsubsection{Model parameters and finding $\munit$}

\begin{table*}
\centering
\caption{The values for $\munit$ and $\langle \dot{M} \rangle$ used for different combinations of $(a_*, \rlow, \rhigh)$ for the Maxwell-Jüttner/$\kappa$ distribution functions.}
\label{tab:munits_maxwell_juttner}
\begin{tabular}{|c|cc|ccccc|}
\hline
\multirow{2}{*}{} & \multicolumn{2}{c|}{\textbf{}} & \multicolumn{5}{c|}{\textbf{$a_*$}} \\ \cline{2-8} 
 & \multicolumn{1}{c|}{\textbf{$\rlow$}} & \textbf{$\rhigh$} & \multicolumn{1}{c|}{\textbf{$-0.9375$}} & \multicolumn{1}{c|}{\textbf{$-0.5$}} & \multicolumn{1}{c|}{\textbf{$0$}} & \multicolumn{1}{c|}{\textbf{$0.5$}} & \textbf{$0.9375$} \\ \hline
\multirow{8}{*}{\textbf{$\munit/10^{25}$}} & \multicolumn{1}{c|}{\multirow{4}{*}{\textbf{$1$}}} & \textbf{$1$} & \multicolumn{1}{c|}{$0.573/0.589$} & \multicolumn{1}{c|}{$1.43/1.42 $} & \multicolumn{1}{c|}{$1.10/1.11 $} & \multicolumn{1}{c|}{$0.844/0.86$} & $0.518/0.546 $ \\ 
 & \multicolumn{1}{c|}{} & \textbf{$10$} & \multicolumn{1}{c|}{$0.854/0.867$} & \multicolumn{1}{c|}{$2.21/2.12 $} & \multicolumn{1}{c|}{$1.88/1.79$} & \multicolumn{1}{c|}{$1.49/1.41 $} & $0.836/0.851$ \\ 
 & \multicolumn{1}{c|}{} & \textbf{$40$} & \multicolumn{1}{c|}{$1.34/1.33 $} & \multicolumn{1}{c|}{$3.38/3.16 $} & \multicolumn{1}{c|}{$3.02/2.78 $} & \multicolumn{1}{c|}{$2.50/2.25 $} & $1.36/1.35$ \\ 
 & \multicolumn{1}{c|}{} & \textbf{$160$} & \multicolumn{1}{c|}{$2.62/2.39$} & \multicolumn{1}{c|}{$6.20/5.41 $} & \multicolumn{1}{c|}{$5.38/4.67$} & \multicolumn{1}{c|}{$4.61/3.9$} & $2.65/2.44$ \\ \cline{2-8} 
 & \multicolumn{1}{c|}{\multirow{4}{*}{\textbf{$10$}}} & \textbf{$1$} & \multicolumn{1}{c|}{$1.06/1.01 $} & \multicolumn{1}{c|}{$2.71/2.51 $} & \multicolumn{1}{c|}{$1.85/1.76 $} & \multicolumn{1}{c|}{$1.41/1.37$} & $0.879/0.865$ \\ 
 & \multicolumn{1}{c|}{} & \textbf{$10$} & \multicolumn{1}{c|}{$1.80/1.58$} & \multicolumn{1}{c|}{$5.18/4.14$} & \multicolumn{1}{c|}{$4.16/3.24 $} & \multicolumn{1}{c|}{$3.17/2.5$} & $1.66/1.47 $ \\ 
 & \multicolumn{1}{c|}{} & \textbf{$40$} & \multicolumn{1}{c|}{$3.09/2.45$} & \multicolumn{1}{c|}{$9.22/6.4 $} & \multicolumn{1}{c|}{$8.55/5.51$} & \multicolumn{1}{c|}{$6.72/4.33$} & $3.15/2.42$ \\ 
 & \multicolumn{1}{c|}{} & \textbf{$160$} & \multicolumn{1}{c|}{$6.11/4.14 $} & \multicolumn{1}{c|}{$17.2/10.3 $} & \multicolumn{1}{c|}{$16.7/9.28 $} & \multicolumn{1}{c|}{$13.7/7.55 $} & $6.45/4.24$ \\ \hline
\multirow{8}{*}{\textbf{\begin{tabular}[c]{@{}c@{}}$\langle \dot{M} \rangle /10^{-4}$\\ $( M_{\odot}/\rm yr)$\end{tabular}}} & \multicolumn{1}{c|}{\multirow{4}{*}{\textbf{$1$}}} & \textbf{$1$} & \multicolumn{1}{c|}{$0.842/0.866$} & \multicolumn{1}{c|}{$1.26/1.25$} & \multicolumn{1}{c|}{$0.984/0.993$} & \multicolumn{1}{c|}{$0.728/0.742$} & $0.441/0.465 $ \\ 
 & \multicolumn{1}{c|}{} & \textbf{$10$} & \multicolumn{1}{c|}{$1.26/1.27 $} & \multicolumn{1}{c|}{$1.95/1.87 $} & \multicolumn{1}{c|}{$1.68/1.6 $} & \multicolumn{1}{c|}{$1.29/1.22 $} & $0.712/0.725$ \\ 
 & \multicolumn{1}{c|}{} & \textbf{$40$} & \multicolumn{1}{c|}{$1.97/1.95 $} & \multicolumn{1}{c|}{$2.98/2.79 $} & \multicolumn{1}{c|}{$2.70/2.49$} & \multicolumn{1}{c|}{$2.16/1.94$} & $1.16/1.15$ \\ 
 & \multicolumn{1}{c|}{} & \textbf{$160$} & \multicolumn{1}{c|}{$3.85/3.51 $} & \multicolumn{1}{c|}{$5.47/4.77$} & \multicolumn{1}{c|}{$4.81/4.18 $} & \multicolumn{1}{c|}{$3.98/3.36$} & $2.26/2.08$ \\ \cline{2-8} 
 & \multicolumn{1}{c|}{\multirow{4}{*}{\textbf{$10$}}} & \textbf{$1$} & \multicolumn{1}{c|}{$1.56/1.48$} & \multicolumn{1}{c|}{$2.39/2.21$} & \multicolumn{1}{c|}{$1.66/1.57$} & \multicolumn{1}{c|}{$1.22/1.18$} & $0.749/0.737 $ \\ 
 & \multicolumn{1}{c|}{} & \textbf{$10$} & \multicolumn{1}{c|}{$2.65/2.32 $} & \multicolumn{1}{c|}{$4.57/3.65 $} & \multicolumn{1}{c|}{$3.72/2.9 $} & \multicolumn{1}{c|}{$2.73/2.16 $} & $1.41/1.25 $ \\ 
 & \multicolumn{1}{c|}{} & \textbf{$40$} & \multicolumn{1}{c|}{$4.54/3.6 $} & \multicolumn{1}{c|}{$8.13/5.64 $} & \multicolumn{1}{c|}{$7.65/4.93 $} & \multicolumn{1}{c|}{$5.80/3.74 $} & $2.68/2.06 $ \\ 
 & \multicolumn{1}{c|}{} & \textbf{$160$} & \multicolumn{1}{c|}{$8.98/6.09$} & \multicolumn{1}{c|}{$15.2/9.08$} & \multicolumn{1}{c|}{$14.9/8.3$} & \multicolumn{1}{c|}{$11.8/6.51$} & $5.50/3.61$ \\ \hline
\end{tabular}
\end{table*}

Throughout the work, we assume that the mass of M87* is $M = 6.5 \times 10^9 M_{\odot}$ at a distance of $D = 16.8 \unit{Mpc}$ (\citealt{bird:2010}, \citetalias{M872018_PaperI:2024}). $M$ and $D$ allow us not only to scale the dimensionless GRMHD simulations to M87* but also to scale any angular size of the source to linear sizes via the following:
\begin{equation}
    {\rm FWHM}_M = \frac{{\rm FWHM}_{{\rm as}}}{3600} \frac{\pi}{180 \degree} \frac{c^2 D}{G M}.
\end{equation}

Along with the MAD snapshot from \ebhlight~ (500 time-snapshots for all five spins, thus already providing 2500 data files for each combination of the other parameters),
\ipole~ takes a few parameters, namely: the observing viewing angle $i$, the observing frequency $\nu$, values for $\rlow$ and $\rhigh$ in Equation~\ref{eq:rhigh_rlow}, a mass and magnetic field strength scaling constant $\munit$, and the magnetization cut parameter $\sigma_{\rm cut}\equiv B^2/\rho$, where B is the magnetic field strength and $\rho$ is plasma rest-mass density, which is used to suppress emission from highly magnetized regions where the GRMHD simulations are no longer reliable as a fluid model. All models assume $\sigma_{\rm cut}=1$.

The M87* jet viewing angle is fixed to $i=160 \degree$ \citep{walker:2018}. Our simulations assume that the accretion flow is perpendicular to the jet axis, which means that the accretion flow viewing angle is nearly face-on and rotates clock-wise in the sky.

To cover a wide range of frequencies for the spectral index maps, $\nu$ is chosen to be $\nu \in (86, 213, 215, 227, 229, 345) \unit{GHz}$, 
since $\nu \in (86, \sim230) \unit{GHz}$ has been measured previously by GMVA and EHT high bands, respectively, and $\nu=345 \unit{GHz}$ provides a prediction of what images in that frequency might look like. 
The intermediate frequencies $\nu \in (213, 215, 227, 229) \unit{GHz}$ are the four EHT intra bands in which EHT is creating the movie of M87*. In the near future, 
the spectral index will be measured between the low and high bands. Notice that there is almost no difference between \ipole~images separated by 1\unit{GHz}; we do not calculate spectral indices between closely spaced frequencies because their real measured uncertainties may be large.

The electron energy parameters $\rlow$ and $\rhigh$ are chosen as $\rlow \in (1, 10)$ and $\rhigh \in (1, 10, 40, 160)$. The values for $\rhigh$ have been tested before assuming $\rlow = 1$ and a thermal electron distribution function, as well as smaller scale tests with non-thermal $\kappa$ distribution functions \citep{ricarte:2022}. Here we assume a family of non-thermal models with $\rlow \in (1, 10)$, $\rhigh \in (1, 10, 40, 160)$ and constant $\kappa=4.25$. The value of the parameter $\kappa$ could be  motivated by observations of the M87 jet at 22 and 43 GHz. \citet{ro:2023} shows that the spectral index of the M87 jet roughly asymptotically reaches $\alpha_{22-43\unit{GHz}}=-1$ at a distance of  $\sim2000M$ from the black hole (see Figure 2 in \citealt{ro:2023}). 
Assuming that at these scales/frequencies, we see mostly the power-law contribution to the total flux $\alpha_{\rm I}=(2-\kappa)/2$ \citep{rybicki_lightman:2004} gives us $\kappa\sim 4$. In general, the parameter $\kappa$ itself might be different from $\sim4$ or even a variable \citep[see e.g.,][]{ball:2018,davelaar:2019,meringolo:2023}. It is preferable to treat this parameter as an unconstrained quantity. 

Finally, the mass scaling parameter $\munit$ is not easy to estimate, since increasing the value of this scalar increases the mass accretion rate through $\dot{M}_{\rm physical} = \munit \dot{M}_{\rm dimentionless}$, as well as the strength of the magnetic fields, and thus also increases the emitted flux. The appropriate values for $\munit$ must be found empirically. This scalar can be determined by requiring that the total flux of the image at a frequency of $\sim230 \unit{GHz}$ has an average flux of around $0.5 \unit{Jy}$ (see subsection~\ref{sec:real_data}). This was accomplished by computing the average flux over $50$ snapshots that are equally separated in time for two different values of $\munit$. Since the total flux strictly increases with $\munit$, it is then possible to find a value of this $\munit$ by fitting an increasing function to these generated points $(M_{\rm unit,1}, F_1)$ and $(M_{\rm unit,2}, F_2)$, and saving the value of $M_{\rm unit}$ for which $F = 0.5 \unit{Jy}$. This point has to be closer to the real value than the previous two; therefore, using the new point $(M_{\rm unit,3}, F_3)$ as either the new highest or new lowest, depending on their relative positions, and repeating the process, we eventually end up with the desired value. The function used to fit the aforementioned points in our bisection scheme is: $F \propto \munit^{3/2}$.  
All final values of $\munit$ for all possible sets of parameters 
reproduce $0.5 \unit{Jy}$ within $2\%$ errorbar. They can be found in Table~\ref{tab:munits_maxwell_juttner}, along with their respective values for the time-averaged mass accretion rates $\langle \dot{M} \rangle$. Notice that both $\munit$ and $\langle \dot{M} \rangle$ increase when increasing $\rhigh$ and $\rlow$, which is consistent with the expectations, since larger values for $R \equiv T_{\rm p}/T_{\rm e}$ mean that the electrons are colder compared to the protons, which in turn means they are less radiative. Also notice that the order for the spins from highest to lowest for both $\munit$ and $\langle \dot{M} \rangle$ is $a_* = -0.5$, $a_* = 0$, $a_* = 0.5$, $a_* = -0.9375$, and $a_* = 0.9375$, of which the last two are very close together for $\munit$ and even flip their order at $\rhigh \in (40, 160)$ for the thermal emission model and at $\rhigh = 160$ for the non-thermal emission model.

\subsubsection{Modeling image sizes, net and resolved spectral indices}\label{sec:model_comparison}

The sizes of the emitting regions in our library of images are computed using image second moments. These are defined as the eigenvalues of the image covariance matrix. The second moments in the image domain map to Gaussian sizes fitted to the visibilities in the $(u,v)$ domain on short baselines \citep{issaoun:2019}. If either of the two (major or minor) image second moments falls within the observed range, the model is deemed consistent with the observations. This is a standard procedure used for model comparisons with EHT data \citepalias[see e.g.,][]{SgrA_PaperV:2022}.

The spectral index maps are defined by means of
\begin{equation}
    \alpha_{I, \nu_1-\nu_2} = \frac{\Delta \ln F_{\nu}}{\Delta \ln \nu} = \frac{\ln \left( \frac{F_{\nu_1}}{F_{\nu_2}} \right)}{\ln \left( \frac{\nu_1}{\nu_2} \right)}.
    \label{eq:alpha_I_nu1_nu2}
\end{equation}
The calculation is carried out for each pixel in the image pair, resulting in a map. Averaging over these maps for a set of parameters provides the time-averaged spectral index map along with its standard deviation.

The net spectral index is defined via Equation~\ref{eq:alpha_I_nu1_nu2}, assuming that $F_{\nu_1}$ and $F_{\nu_2}$ represent total fluxes rather than single pixel fluxes. The net spectral indices are then plotted against the size of the emitting region and compared with the observational constraints. 

Finally, azimuthally averaged spectral index profiles are created by taking the spectral index maps and plotting the azimuthally averaged values for $\alpha_I$. 

\section{Results}\label{sec:results}

\begin{figure*}
  \centering
  \includegraphics[trim={5cm 2.5cm 2cm 0cm},clip,width=0.45\textwidth]{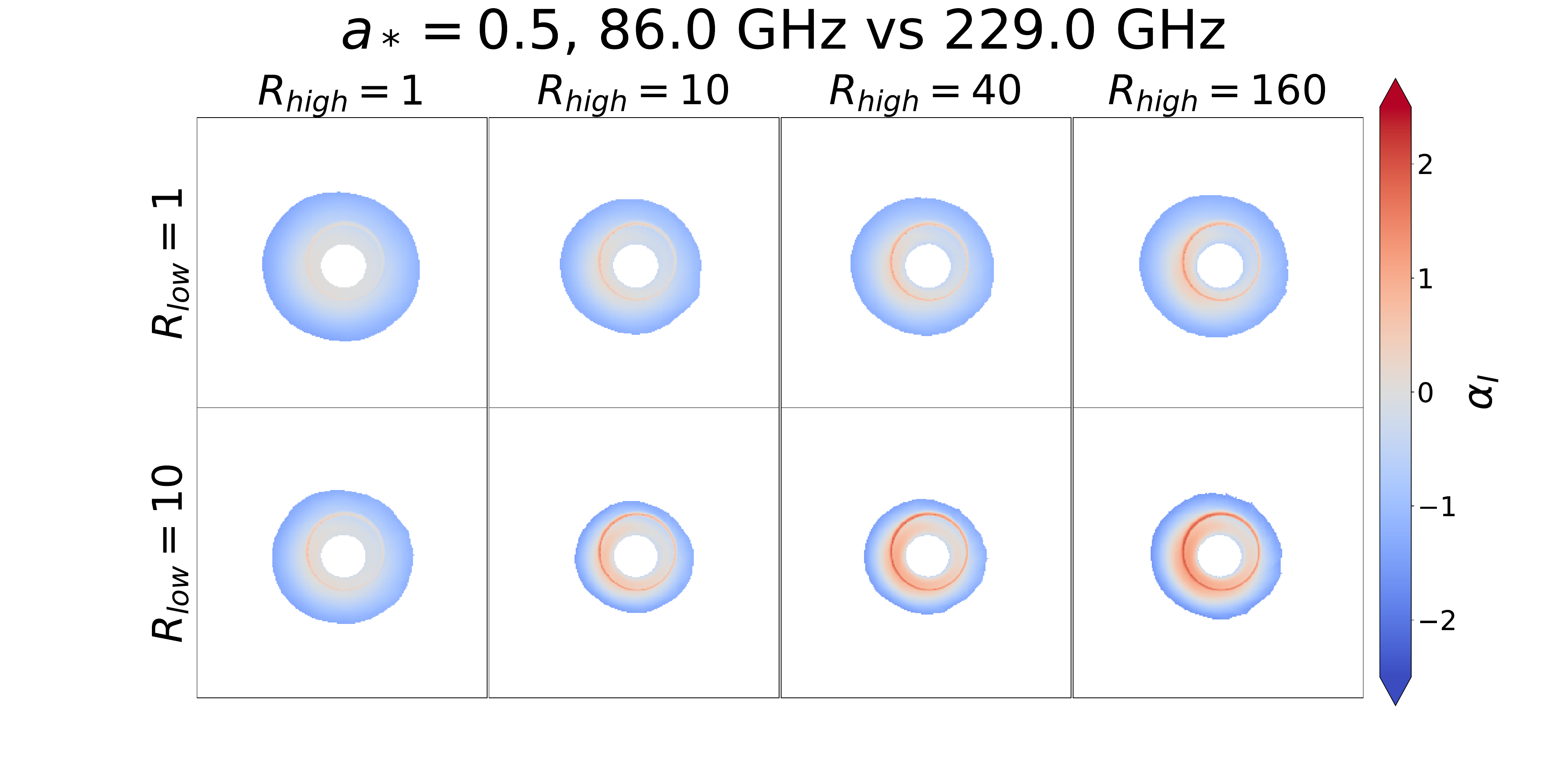}
  \includegraphics[trim={5cm 2.5cm 2cm 0cm},clip,width=0.45\textwidth]{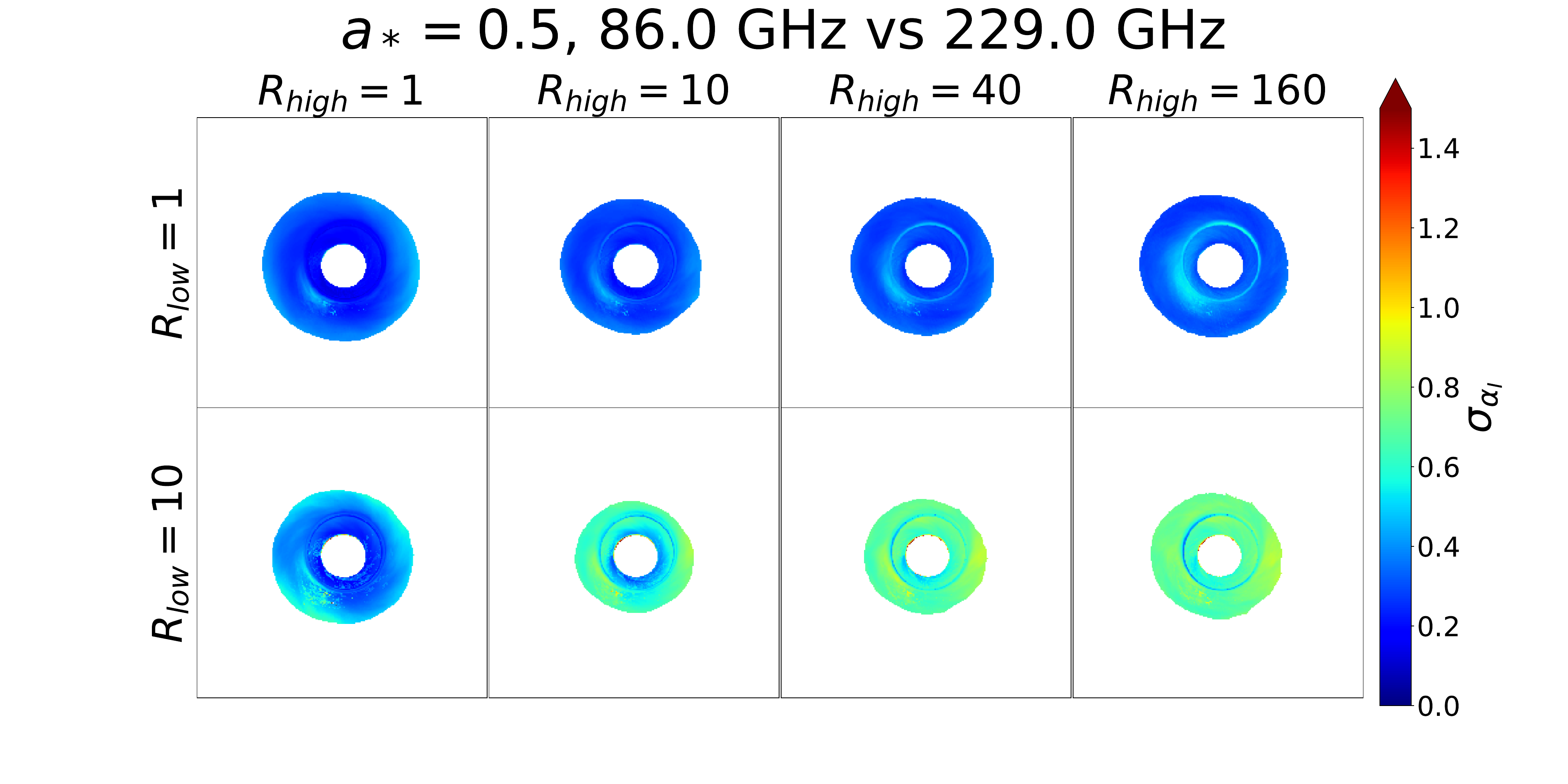}\\
 \includegraphics[trim={5cm 2.5cm 2cm 0cm},clip,width=0.45\textwidth]{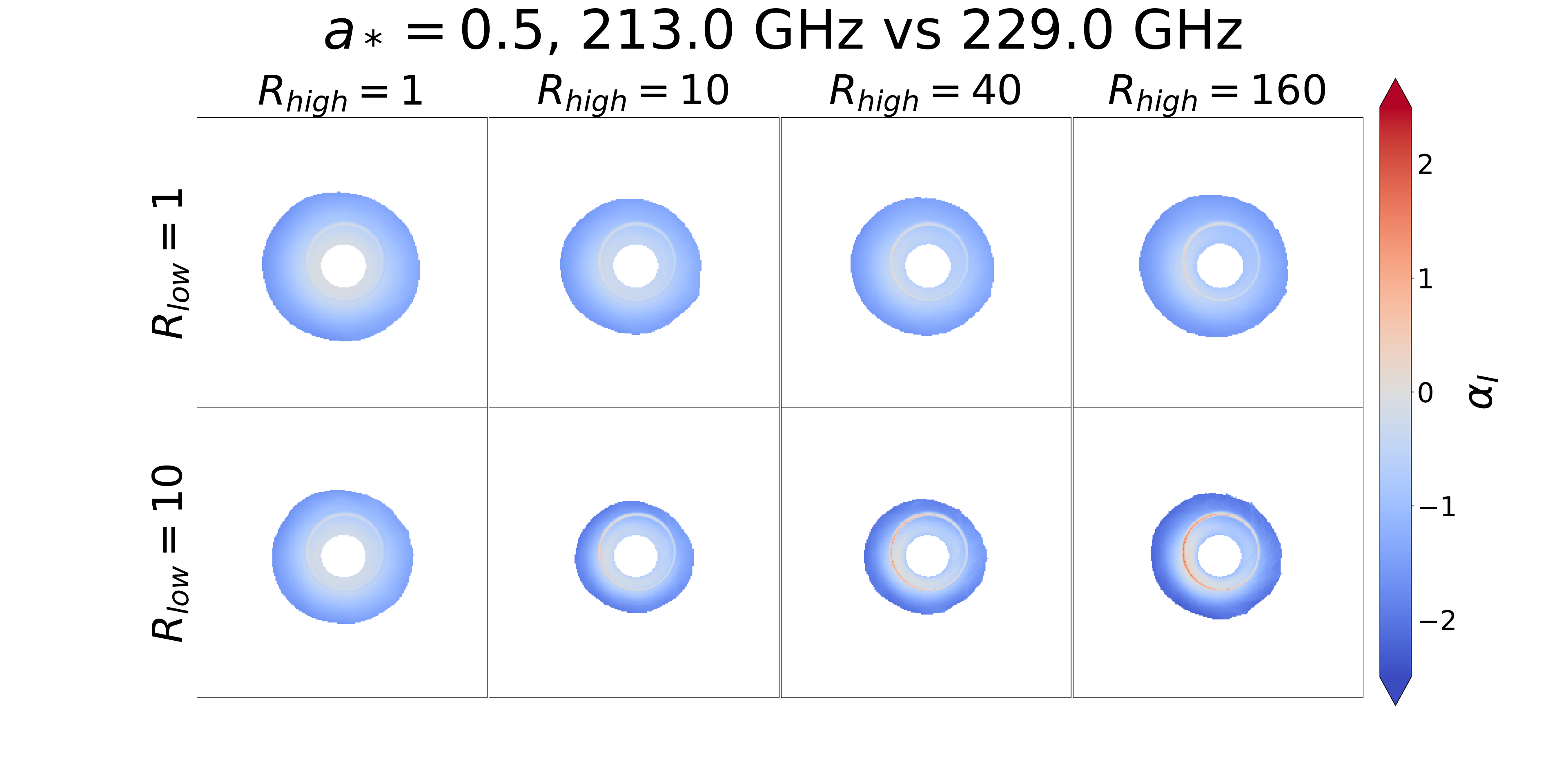}
  \includegraphics[trim={5cm 2.5cm 2cm 0cm},clip,width=0.45\textwidth]{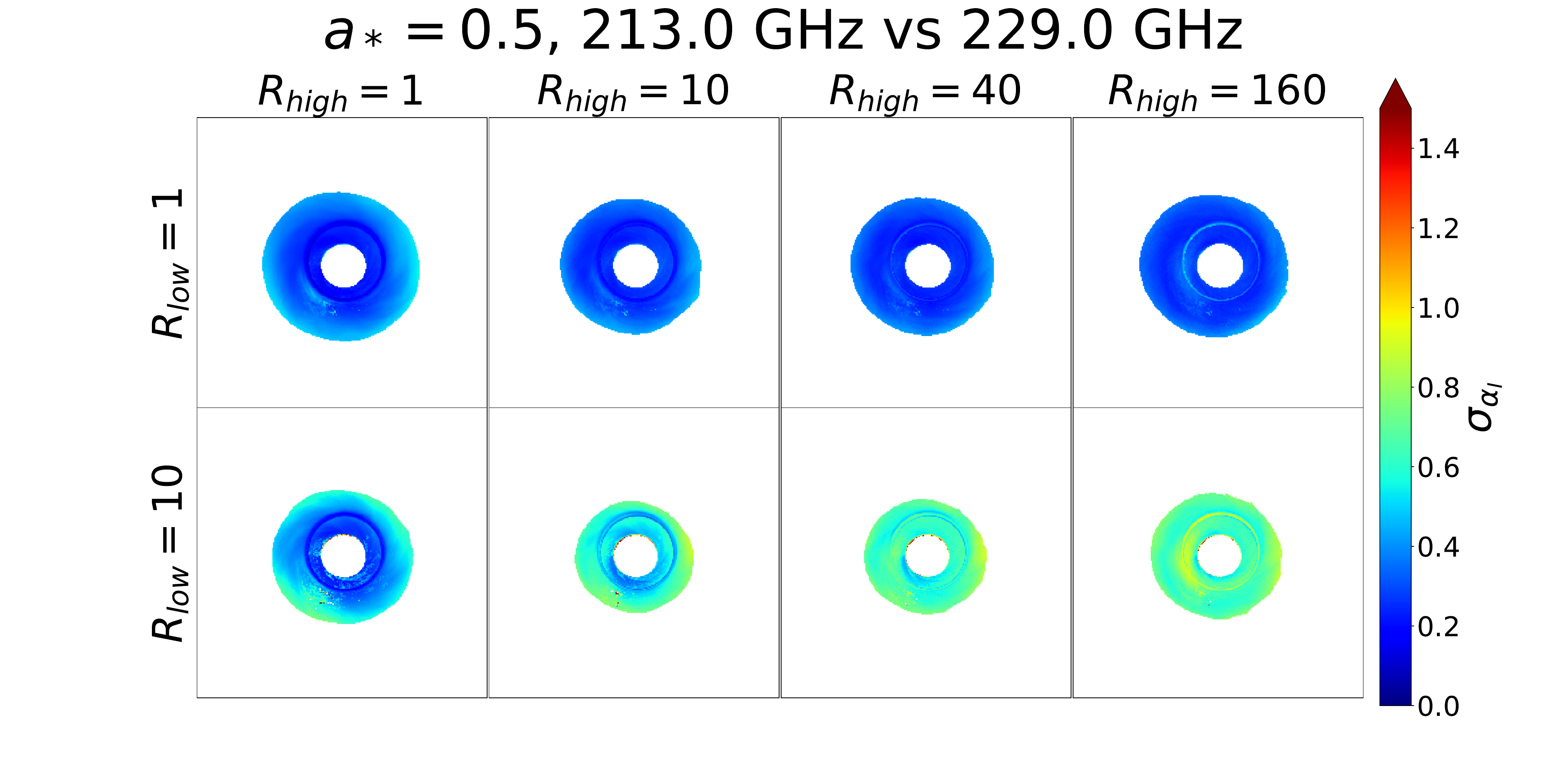}\\  
  \includegraphics[trim={5cm 2.5cm 2cm 0cm},clip,width=0.45\textwidth]{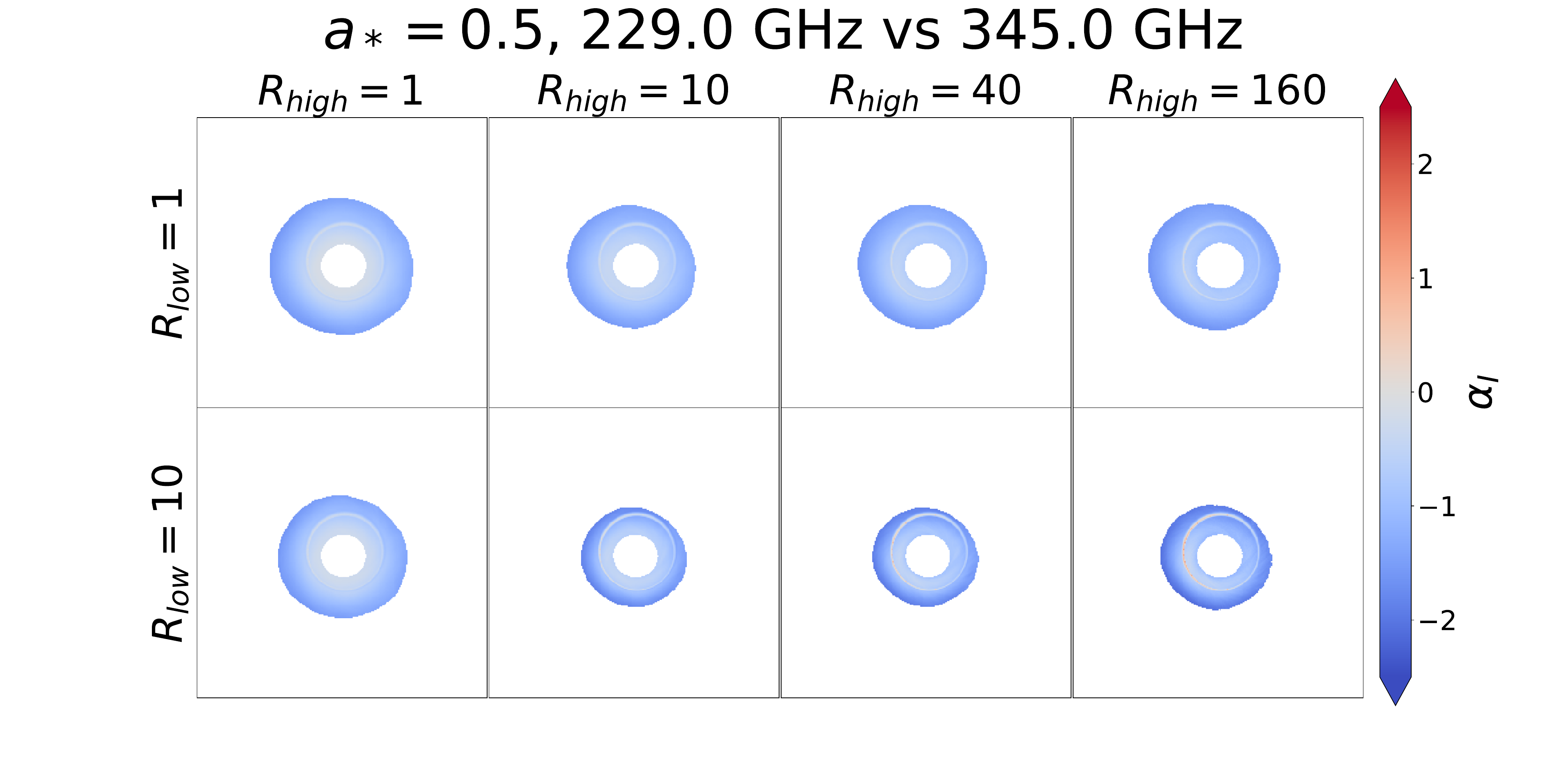}
  \includegraphics[trim={5cm 2.5cm 2cm 0cm},clip,width=0.45\textwidth]{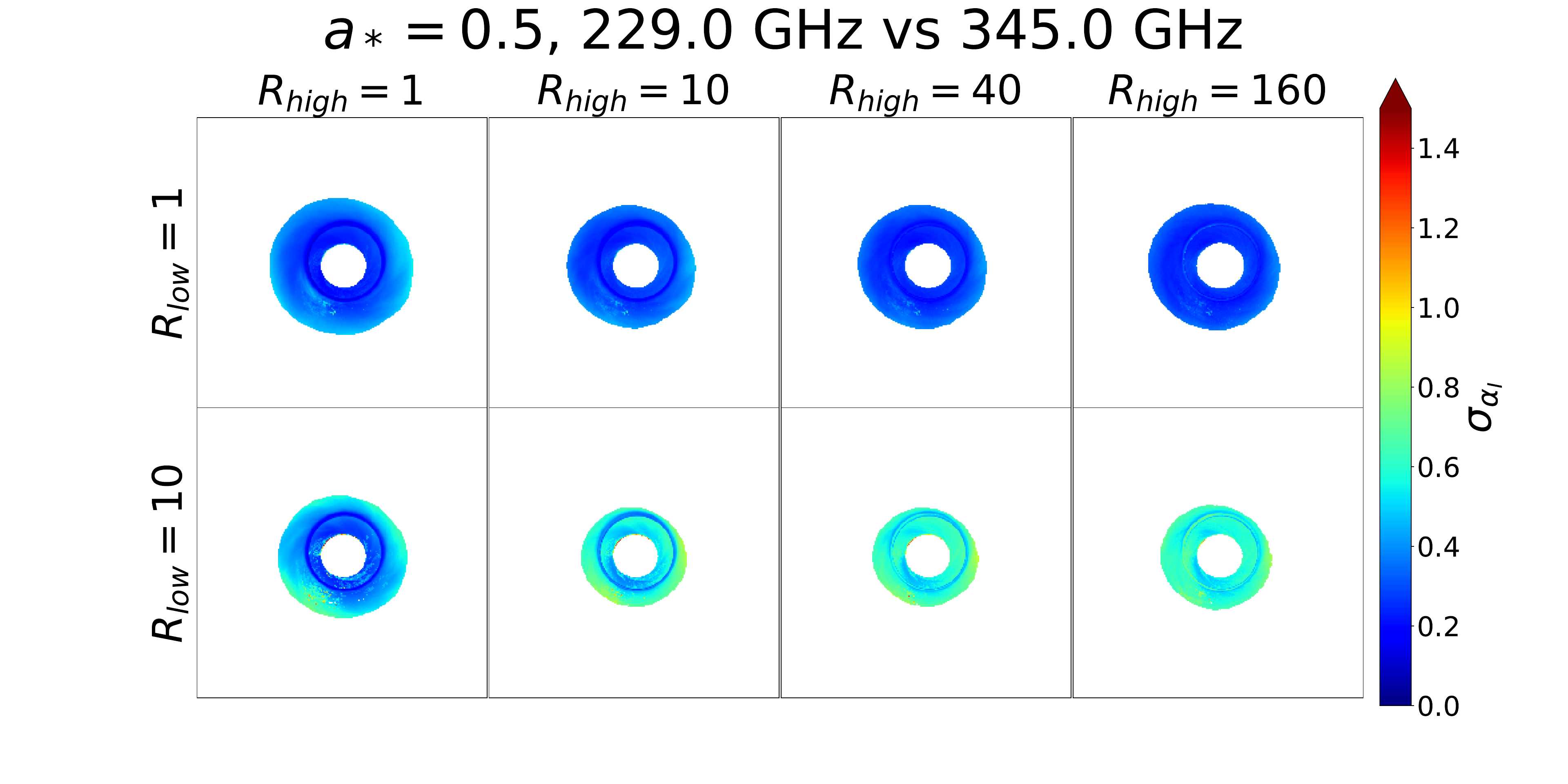}
\caption{Spectral indices of thermal models. Left panels show unblurred, time-averaged spectral index maps for all combinations of parameters $\rlow \in (1, 10)$ and $\rhigh \in (1, 10, 40, 160)$ and frequencies $86-345 \unit{GHz}$ (from top to bottom) for a single black hole spin $a_*=0.5$. Right panels show the corresponding standard deviation which indicates the variability of a model. Field of view of each image is $40\times40$M with resolution $256\times256$ pixels. Only the regions where $I > 0.01 I_{\rm max}$ are shown.}
\label{fig:rhigh_rlow_time_averaged_a0.5}
\end{figure*}

\begin{figure*}
  \centering
  \includegraphics[trim={5cm 2.5cm 2cm 0cm},clip,width=0.45\textwidth]{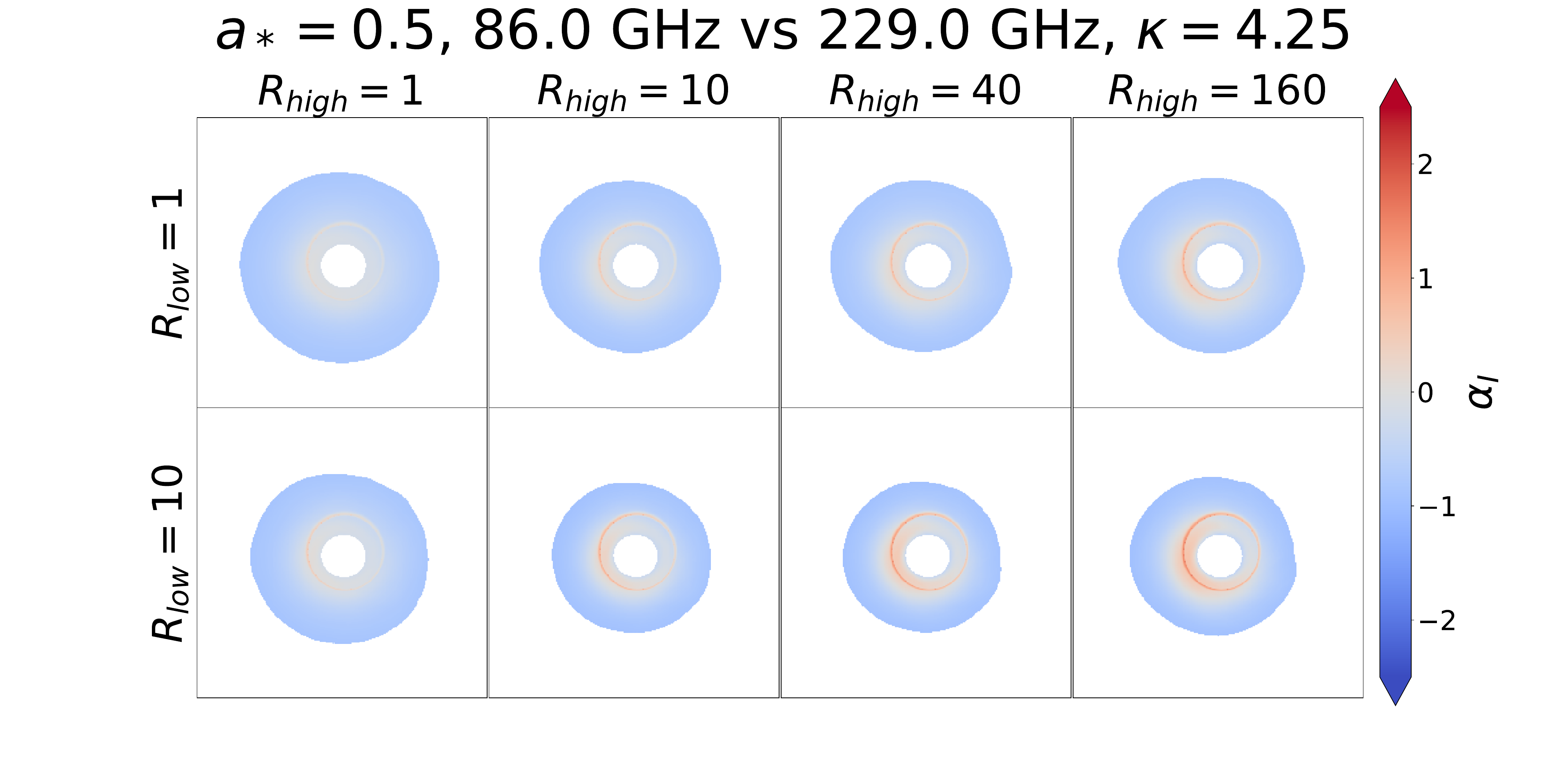}
  \includegraphics[trim={5cm 2.5cm 2cm 0cm},clip,width=0.45\textwidth]{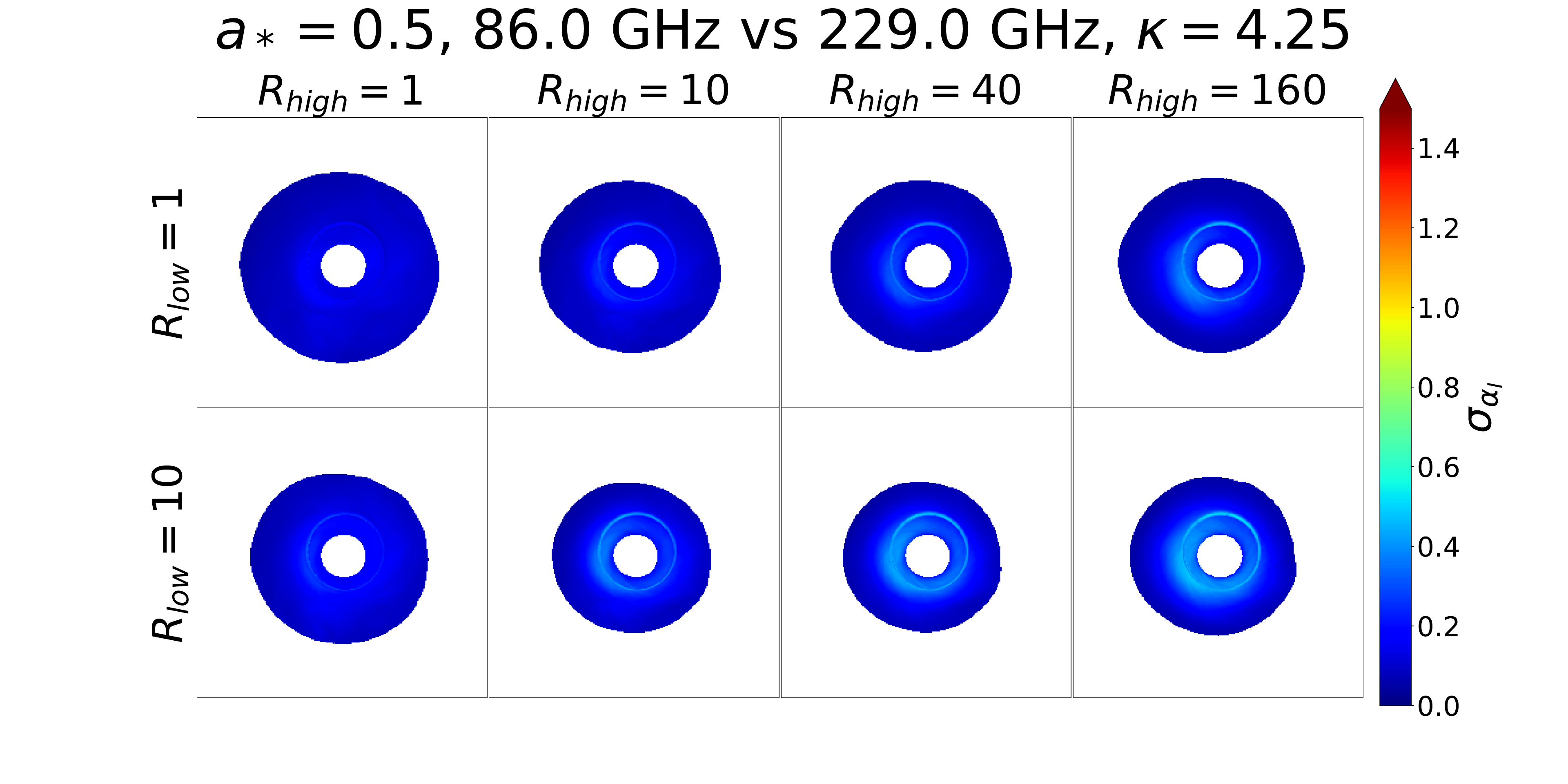}\\
\includegraphics[trim={5cm 2.5cm 2cm 0cm},clip,width=0.45\textwidth]{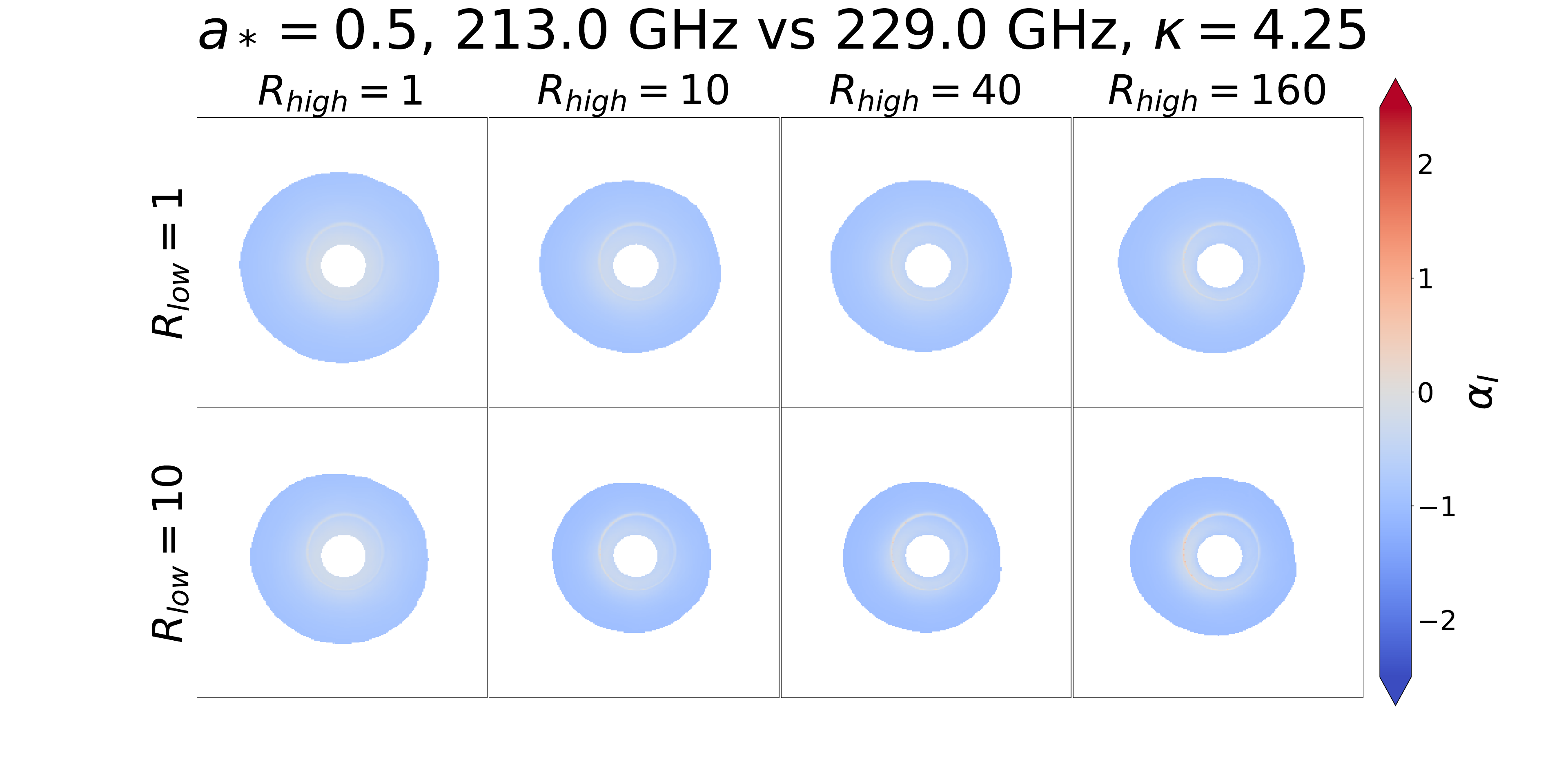}
  \includegraphics[trim={5cm 2.5cm 2cm 0cm},clip,width=0.45\textwidth]{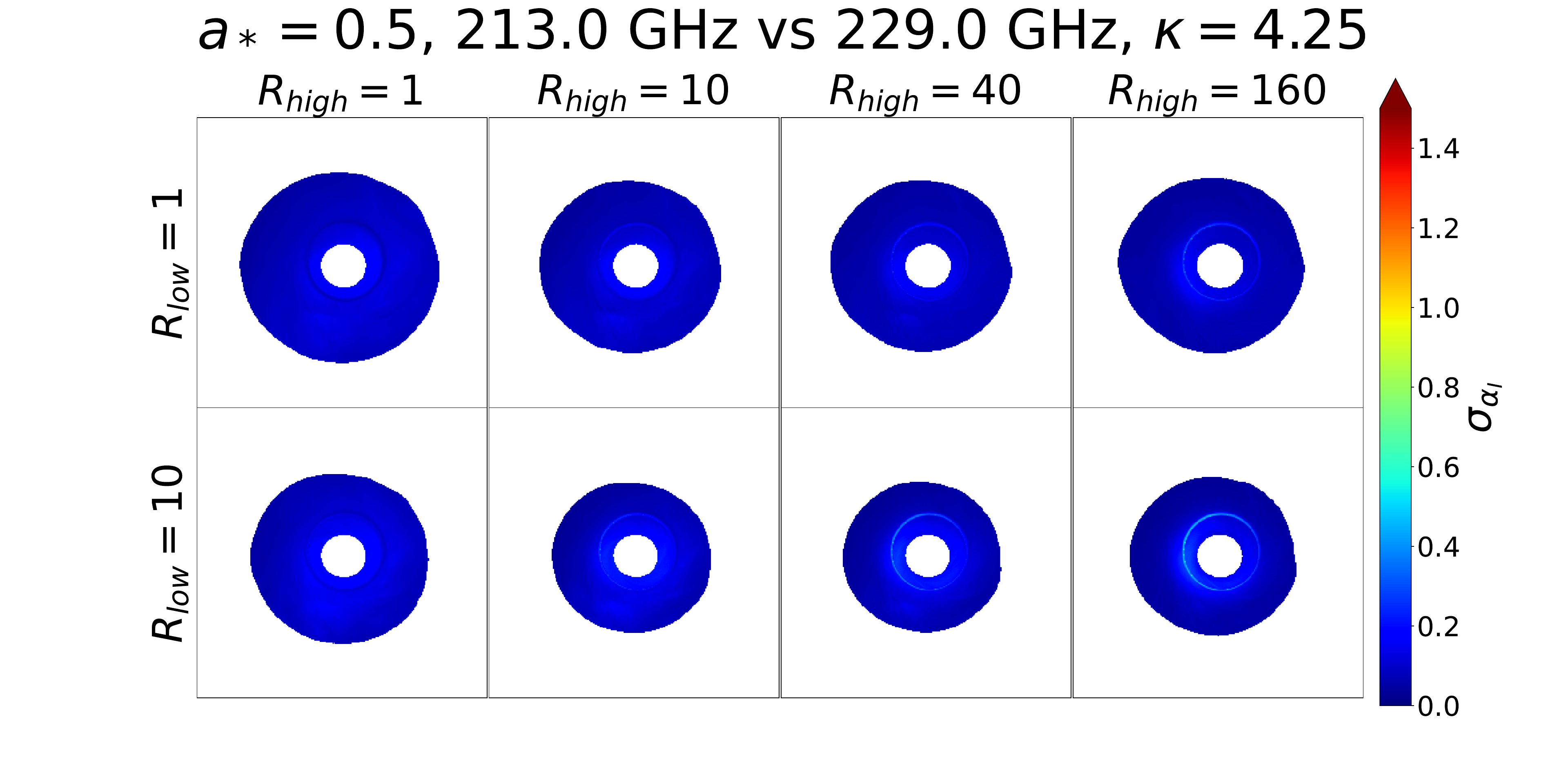}\\
  \includegraphics[trim={5cm 2.5cm 2cm 0cm},clip,width=0.45\textwidth]{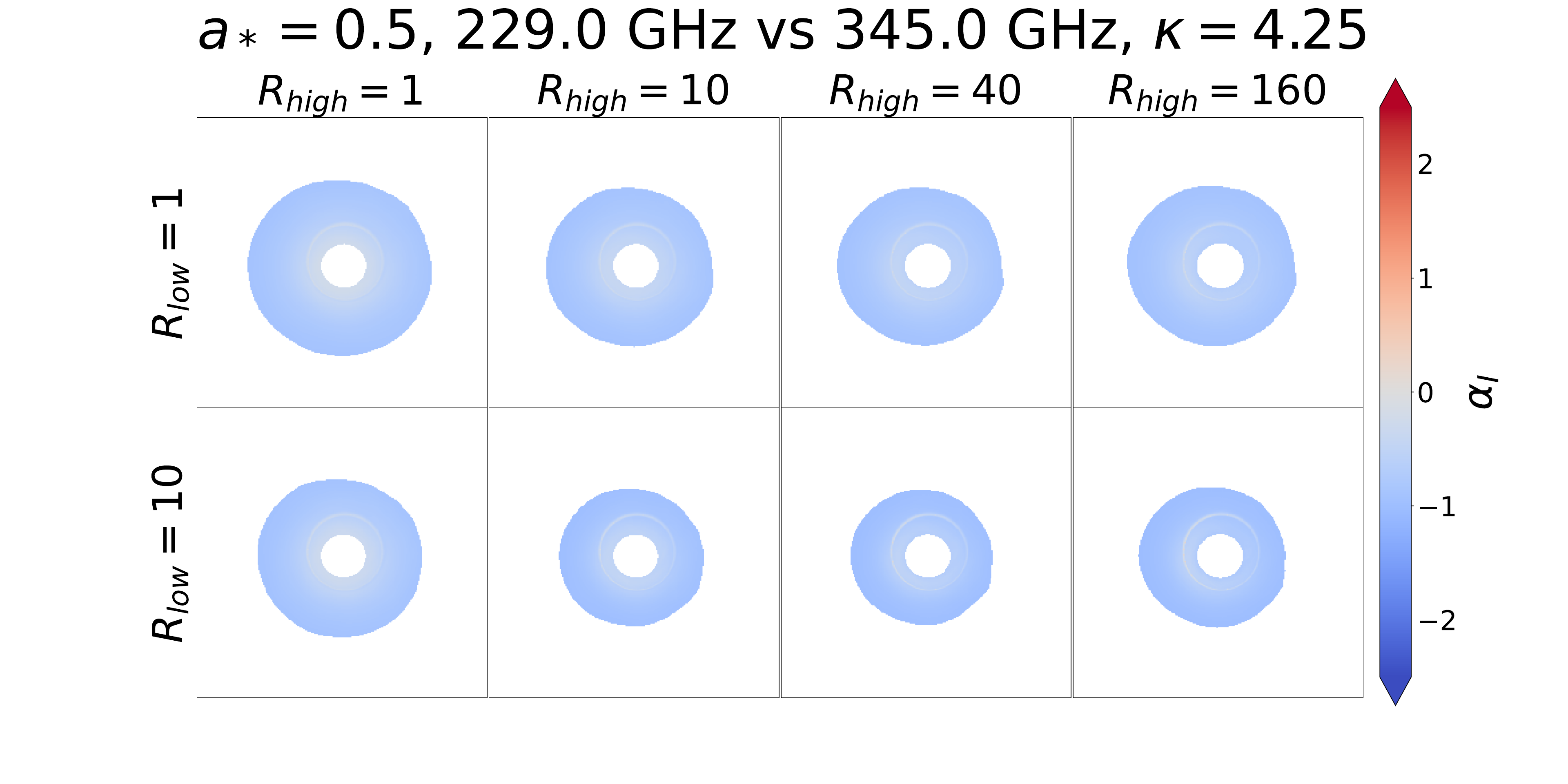}
  \includegraphics[trim={5cm 2.5cm 2cm 0cm},clip,width=0.45\textwidth]{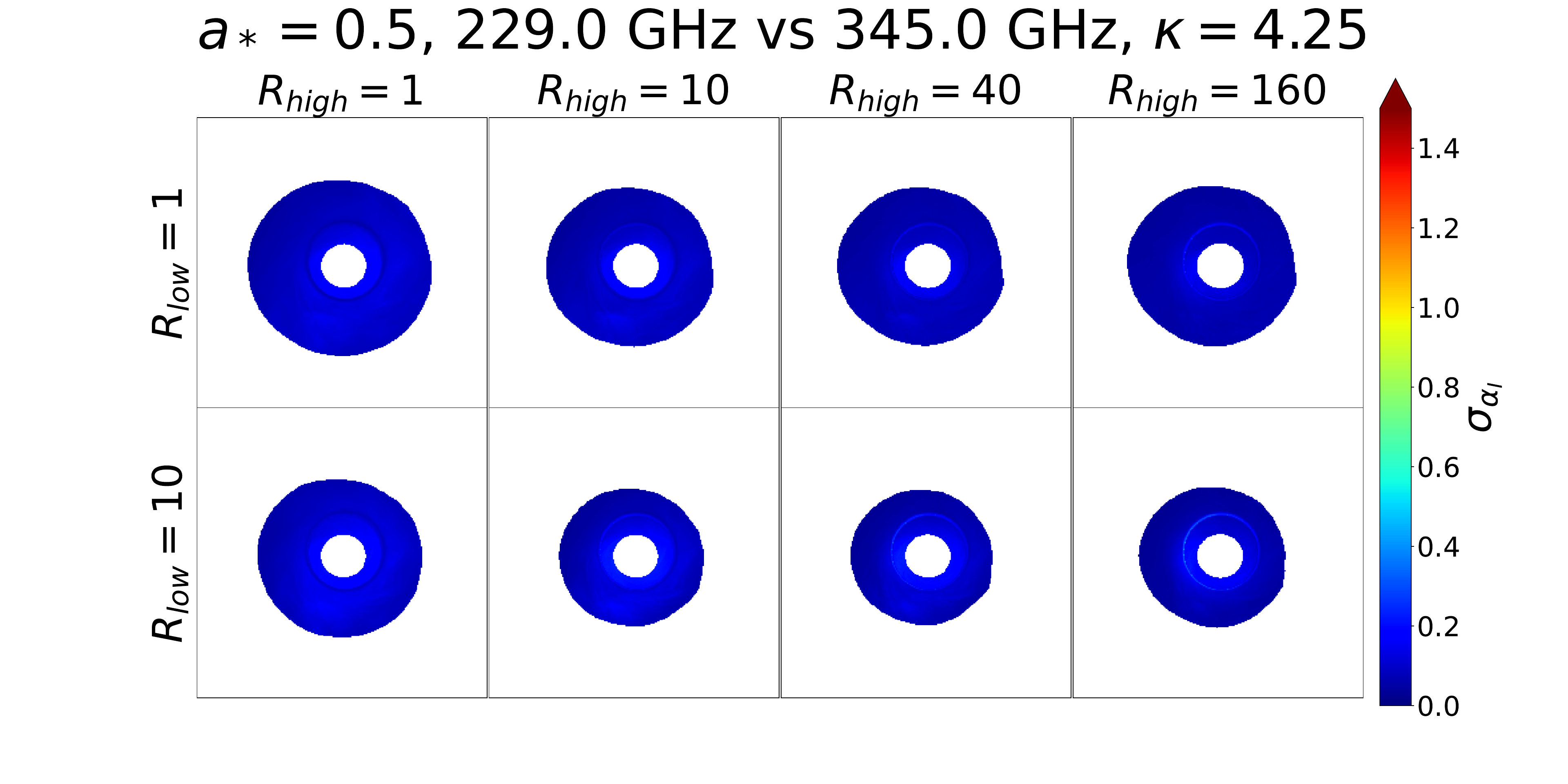}
\caption{The same as Figure~\ref{fig:rhigh_rlow_time_averaged_a0.5} but for non-thermal models.}
\label{fig:rhigh_rlow_time_averaged_kappa_a0.5}
\end{figure*}

\subsection{Model validation and total intensity variability}

Our library of M87* model images contains 240,000 
snapshots with a resolution of $256\times256$ pixels. To ensure that all models are computed without random numerical errors, we first examine our entire data set for any unphysical behavior. 

The multifrequency light-curves for all possible combinations of parameters $(a_*,\rlow,\rhigh)$ are shown in Appendix~\ref{sec:app_intensity_over_time} (Figures~\ref{fig:intensity_over_time_a-0.94}-~\ref{fig:intensity_over_time_a0.94}). None of these light-curves show fluxes reaching unphysically large values, so the result of the post-processing with \ipole is reliable. It is also important to note that, although the average total flux at $\sim230\unit{GHz}$ is $0.5 \unit{Jy}$, the flux varies significantly over time, sometimes increasing to as high as $1.8 \unit{Jy}$ or decreasing to as low as $0.1 \unit{Jy}$. Significant variability is observed at all modeled frequencies.

The snapshot intensity maps for each frequency, each pair of ($\rlow$,$\rhigh$), and the example $a_*=-0.5$ can be found in Appendix~\ref{sec:app_intensity_over_time} (Figures~\ref{fig:intensity_map_a-0.5}-~\ref{fig:intensity_map_kappa_a-0.5}). Notice that as the frequency increases, the size of the emission region decreases, as the outer regions of the emission region diminish in flux faster than the inner regions. The photon ring is the slowest to decrease in flux (as it typically has higher optical depth), which means that at sufficiently high frequencies, the size of the emission region — for any $a_*$, $\rlow$, and $\rhigh$ — should converge to the same value, namely that of the photon ring. It should then also become nearly circularly symmetric, indicating that the minor and major axes should converge as well.

After the spectral indices $\alpha_I$ for all pixels and all images were computed using Equation~\ref{eq:alpha_I_nu1_nu2} they are weighted with the intensity $I_{\nu}$ via $\alpha_{I, {\rm weighted}} = \int \alpha_I(x,y) I(x,y) dx dy/\int I(x,y) dx dy$ where $x,y$ are the image coordinates. Additional figures in Appendix~\ref{sec:app_intensity_over_time} (Figures~\ref{fig:intensity_weighted_spectral_index_a-0.94}-~\ref{fig:intensity_weighted_spectral_index_a0.94}) show the intensity $I_{\nu}$ weighted spectral index as a function of time. None of the figures have the weighted spectral index approaching infinity at any time; hence, the pixels of the spectral index maps presented in the next subsection are all valid pixels. 

All these sanity checks are also performed for all considered non-thermal models.

\subsection{Spectral index maps}

Figures~\ref{fig:rhigh_rlow_time_averaged_a0.5} and \ref{fig:rhigh_rlow_time_averaged_kappa_a0.5} display the time-averaged spectral index maps  for a chosen black hole spin ($a_*=0.5$) and all possible combinations of parameters $\rlow$ and $\rhigh$. Spectral indices are computed between: $86$ and $229 \unit{GHz}$ (GMVA-EHT frequencies);
$213$ and $229 \unit{GHz}$ (intra EHT frequencies);
$229$ and $345 \unit{GHz}$ (forthcoming EHT capability).
The standard deviation maps reflect the variability of the spectral index in each model. 
Spectral index maps for other black hole spins are shown in the Appendix~\ref{sec:other_spins} (Figures~\ref{fig:rhigh_rlow_time_averaged_a-0.94}-~\ref{fig:rhigh_rlow_time_averaged_a0.94}) and they appear similar.

From $86 - 229 \unit{GHz}$ through $229-345 \unit{GHz}$, the spectral indices in the maps decrease, regardless of the exact combination of $\rlow$ and $\rhigh$ parameters. This is expected, as all thermal and non-thermal models become optically thinner with increasing frequency. With the exception of the $86-229\unit{GHz}$ range, the averaged spectral slopes of all our MAD simulations are predominantly negative. 
Measuring the spectral slope variability at any of the considered frequency ranges could be critical for distinguishing between thermal vs. non-thermal or thermal $\rlow=1$ vs. $\rlow=10$ models.
Overall, thermal models with $\rlow = 1$ are significantly less variable than those with $\rlow = 10$, regardless of the frequency range. All non-thermal models are significantly less variable even when compared to the thermal $\rlow=1$ models. 

An important feature of all black hole images is the photon ring - a thin ring of lensed emission at the edge of the black hole shadow. In our models, this feature 
typically has a higher spectral index than the rest of the emission region for all black hole spins, $\rlow$, $\rhigh$, and frequencies, including both thermal and non-thermal models. While most of the emission region is negative, with $\alpha_I=-1$ and sometimes even $-2$, the photon ring $\alpha_I$ is mostly around zero, or even positive. 
For example, the spectral index of the photon ring at $86 - 229 \unit{GHz}$ is mostly positive, between $0$ and $1$ for $\rlow = 1$, and it can increase up to $2$ for $\rlow = 10$.  The spectral index of the photon ring at $213-229\unit{GHz}$ and $229-345 \unit{GHz}$ is generally around $0$ or lower, with the only exception occurring at $\rlow = 10$ and $\rhigh \in (40, 160)$ for all black hole spins; yet, it still does not exceed $0.5$. 
For each respective $\rlow$, the spectral index of the photon ring also increases as $\rhigh$ increases. This trend also seems to be independent of black hole spin. 
Our results suggest that on average the photon ring has a higher turnover frequency than the rest of the emission region, and this turnover frequency for the thermal/non-thermal
models is typically located at frequencies higher than the $86-229\unit{GHz}$ band.
%
The turn-over frequency for the rest of the emission region appears to be at lower frequencies than any of the examined ones, since the spectral indices are already negative at the $86 - 229 \unit{GHz}$ band.
Interestingly, the photon ring spectral index has a distinct variability pattern that often differs from the rest of the image. In some particular cases, e.g., in thermal $\rlow=1$ and $\rhigh>10$ models at $86-229\unit{GHz}$ bands, the $\alpha_I$ of the photon ring is more variable than the rest of the emission region. In other cases, e.g., in thermal models with $\rlow=10$ and $\rhigh>1$ the $\alpha_I$ of the photon ring is less variable then the rest of the emission region.

All maps shown in Figures~\ref{fig:rhigh_rlow_time_averaged_a0.5} and \ref{fig:rhigh_rlow_time_averaged_kappa_a0.5} show the distribution of the spectral index in regions where the intensity $I > 0.01  I_{\rm max}$. It is evident that the thermal models with $\rlow=1$ are slightly ($1-44$ percent)
more extended compared to the models with $\rlow=10$. The non-thermal models are also
more extended ($5-62$ percent) compared to all their thermal counterparts with the same ($\rlow,\rhigh$) parameters. 

Finally, it is worth mentioning that all our findings apply to other black hole spins, with the caveat that as the spin increases (whether negative or positive), the azimuthal asymmetry of the spectral index in the photon ring also increases. This asymmetry is similar to the asymmetry observed in total intensity images used to estimate the spin of the M87* black hole \citep{bernshteyn:2026}. Spectral index maps for other black hole spins are shown in Appendix~\ref{sec:other_spins}.

\subsection{Spectral index profiles}

\begin{figure}
    \centering
    \includegraphics[trim={0cm 0cm 0cm 1.0cm},clip,width=0.48\textwidth]{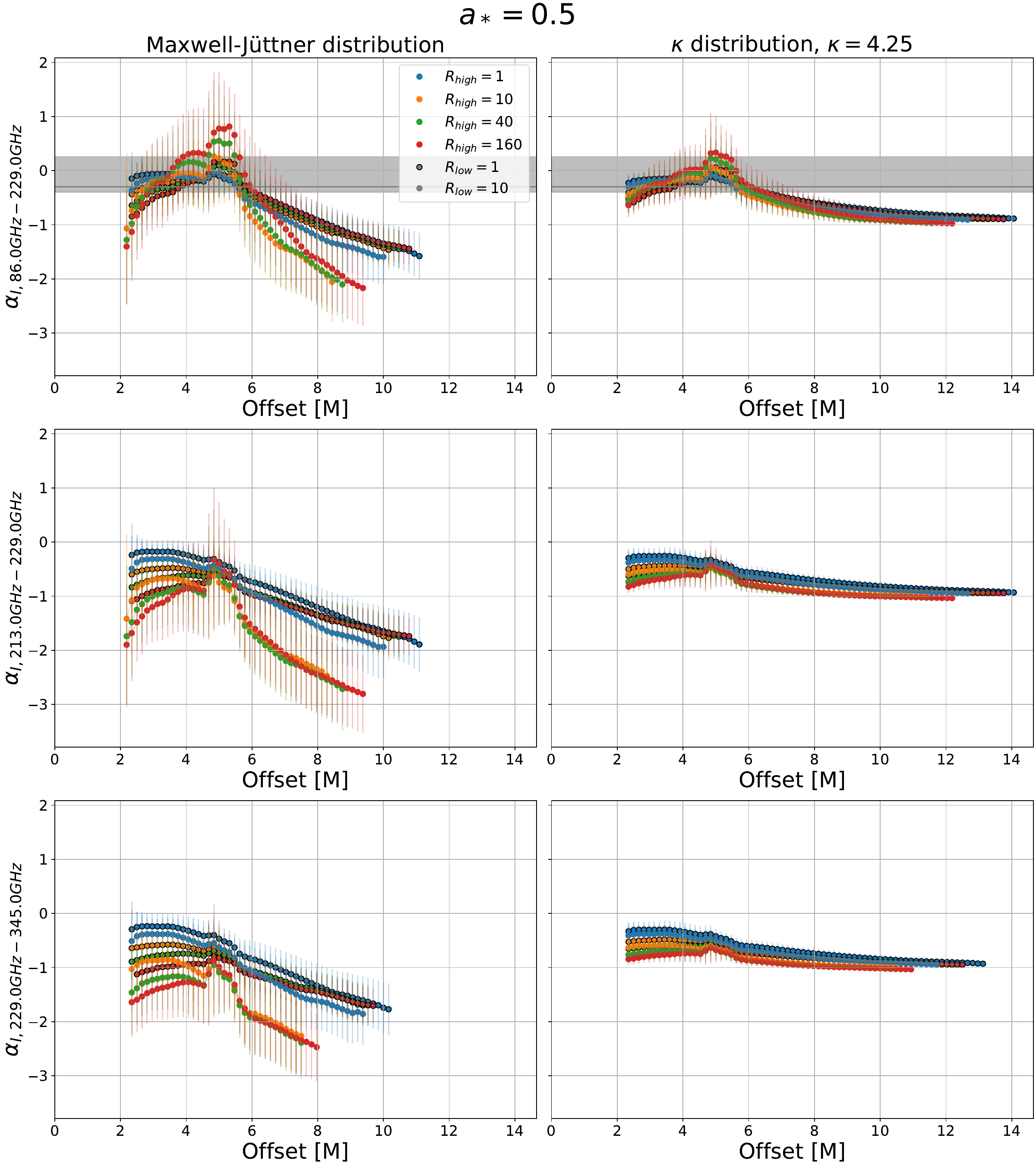}
    \caption{Azimuthal averages of all spectral index maps presented in Figures~\ref{fig:rhigh_rlow_time_averaged_a0.5} and~\ref{fig:rhigh_rlow_time_averaged_kappa_a0.5}. The left panels display models assuming the thermal Maxwell-Jüttner distribution while the right panels display the non-thermal $\kappa$ distribution cases.  Only the regions where $I > 0.01 \cdot I_{\rm max}$ are shown. The gray band indicates the observational constraint for the net spectral index (see Section~\ref{sec:unresolved_spectral_indices}).}
    \label{fig:gradients_kappa_a0.5}
\end{figure}

Figure~\ref{fig:gradients_kappa_a0.5} displays the azimuthal average profile of the spectral index maps shown in Figures~\ref{fig:rhigh_rlow_time_averaged_a0.5} and~\ref{fig:rhigh_rlow_time_averaged_kappa_a0.5}. 
Figure~\ref{fig:gradients_kappa_a0.5} supports all statements from the previous subsection but also reveals more detailed  characteristics. A generic characteristic of all models is that the averaged spectral index always declines with the radial distance from the position of the photon ring (located approximately at $5M$). Moreover, this decline is faster for thermal models compared to non-thermal models. Non-thermal models tend to asymptote to $\alpha_I=-1$ at larger distances, independently of the frequency band. 
Within the thermal models, models with $\rlow=10$ (with the exception of the $\rhigh=1$ model) show a steeper gradient towards more negative spectral indices compared to models with $\rlow=1$. In extreme cases, the thermal models can even decrease to $\alpha_I=-3$ in the less luminous outer regions.

If the electron energies near the M87* horizon are distributed according to a thermal distribution function, it will be easier to distinguish the thermal parameters $\rlow,\rhigh$ at higher frequency bands ($213-229\unit{GHz}$, $229-345\unit{GHz}$) rather than in lower bands ($86-229\unit{GHz}$). While in the lower bands, the spectral slope values tend to overlap in the brightest part of the image, in the higher bands these profiles tend to separate better. However, the variability level may be a better indicator of exact parameters $\rlow,\rhigh$. 
If the electron energies near the M87* horizon are distributed according to a non-thermal distribution function with the same power-law slope, the $\rlow,\rhigh$ parameters of this distribution function are difficult to distinguish. In the two higher frequency bands, all spectral index profiles fit within the 0 to $-1$ range.

\subsection{Physical explanation of the spectral indices}

\begin{figure*}
    \centering
    \includegraphics[trim={0cm 0cm 0cm 0cm},clip,width=1.0\textwidth]{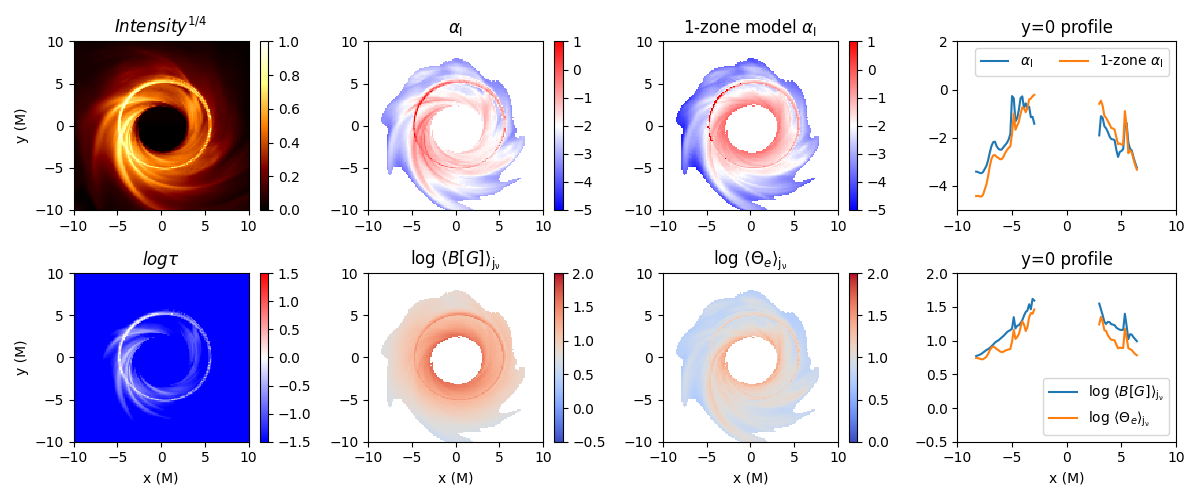}
    \caption{Explanation of the observed spectral index based on a single snapshot of MAD model with parameters $(a_*,\rlow,\rhigh)=(0.5,10,40)$. Top panels from left to right: total intensity at $\nu=229\unit{GHz}$ (using non-linear scale), exact spectral index between $213-229\unit{GHz}$, spectral index predicted by the one-zone model of \citet{ricarte:2022}, comparison of the exact and model spectral index profiles. Bottom panels from left to right: optical thickness of the simulation, synchrotron emissivity averaged magnetic field strength, synchrotron emissivity averaged electron temperature, profiles of the latter two as a function of image offset. All quantities in the bottom panels are shown in logarithmic scale.}
    \label{fig:spectral_slope}
\end{figure*}

When considering the synchrotron radiation from thermal and non-thermal distributions of electrons, the turn-over frequency where the spectrum transitions from optically thick to optically thin is $\nu_{\rm crit}\propto \Theta_e^2 B$ \citep{pandya:2016}. 
Taking the electron temperature parametrization from Equation~\ref{eq:theta_e}, we expect that $\nu_{\rm crit} \propto \rhigh^{-2} \rlow^{-2} M_{\rm unit}^{1/2}$. At fixed frequency, the spectral index increases when $\nu_{\rm crit}$ increases. At a fixed frequency, models with higher $\rhigh$ should, therefore, have a more negative spectral index. This is approximately what is visible in the azimuthally averaged profiles in Figure~\ref{fig:gradients_kappa_a0.5},
except for the regions with a strong contribution from the photon ring and outer regions where all profiles converge to the same spectral index for a fixed $\rlow$.
Notice that the thermal models with $\rlow=10$ have a more negative spectral index in these outer regions compared to the models with $\rlow=1$, which is also in agreement with the theoretical expectation. In all our cases, the dependence of the spectral slope on the electron temperature is always slightly contaminated by a weaker dependence on the magnetic field strength, which is proportional to $M_{\rm unit}^{1/2}$. In addition to this, the models are variable. Hence, although the theoretical trend is valid, it may not always be evident in the time-averaged models.  

The regions farther away from the image center emit photons with frequencies that become larger than the local $\nu_{\rm crit}$. This is because both $\Theta_{\rm e}$ and B decrease with distance from the black hole. 

In case of the thermal models, in the outer regions $\alpha_I\rightarrow -\infty$, as is evident in our Figure~\ref{fig:gradients_kappa_a0.5}. 
It would be interesting to understand the exact physical reason behind the gradient of the spectral index in thermal models. We employ a one-zone model of \citet{ricarte:2022} to investigate this issue. Their one-zone model predicting the spectral index maps needs three quantities as a function of image pixel: $\tau_\nu$ - the optical thickness of matter along the line of sight, $ \left< B \right>_{\rm j_{\nu,inv}}$ - the average magnetic field along the line of sight weighted by plasma emissivity, and $ \left< \Theta_{\rm e} \right>_{\rm j_{\nu,inv}}$ - the average electron temperature along the line of sight weighted by plasma emissivity. The predicted spectral index is then computed from the formal solutions of the (unpolarized) radiative transfer equation in a uniform slab: $\alpha \equiv dlnj_\nu/dln\nu$ (for $\tau_\nu <1$ pixels) or $\alpha \equiv dlnj_\nu/dln\nu-dln\alpha_\nu/dln\nu$ (for $\tau_\nu>1$ pixels), where $j_\nu$ and $\alpha_\nu$ are the synchrotron emissivity and absorptivity, respectively. Plasma density $n_e$ does not affect these derivatives, so the resulting $\alpha_{\rm I}$ is typically a function of $B$ and $\Theta_{\rm e}$ averaged over the most emitting regions (notice that $\tau_\nu$ does depend on $n_e$).
Using a single GRMHD snapshot with $(a_*,\rlow,\rhigh)=(0.5,10,40)$ and frequency band $213-229\unit{GHz}$, Figure~\ref{fig:spectral_slope} demonstrates that the spectral index typically follows fluctuations in $\Theta_{\rm e}$ and $B$. 
The spectral index radial decrease corresponds to the decline of both $\left<B\right>_{j_\nu}$ and $\left <\Theta_e\right>_{j_\nu}$ as a function of the offset of the image center. In models with $\rlow=10$ the decrease in $\left <\Theta_e\right>_{j_\nu}$ is slightly steeper compared to $\rlow=1$ models resulting in a steeper spectral index gradient. The one-zone model also shows us that the spectral indices in models with $\rlow=10$ vary more due to larger variations in temperatures in the emitting region compared to $\rlow=1$ models. 

In the non-thermal models in the outer regions, for all frequency ranges $\alpha_I \rightarrow \sim -1$, as expected. Moving away from the photon ring location plasma becomes increasingly optically thinner and the spectral index asymptotes to $(2-\kappa)/2=-1.125$ \citep{rybicki_lightman:2004}, independently of the local $w$ (parametrized with $\Theta_{\rm e}$) and $B$. Our assumption of the constant parameter $\kappa$ contributes to the stability of the spectral index in all non-thermal models. 
 

\subsection{Net spectral indices and current observational constraints}\label{sec:unresolved_spectral_indices}

\begin{figure*}
    \centering
    \includegraphics[trim={0cm 0cm 0cm 0cm},clip,width=\textwidth,height=0.8\textheight,keepaspectratio]{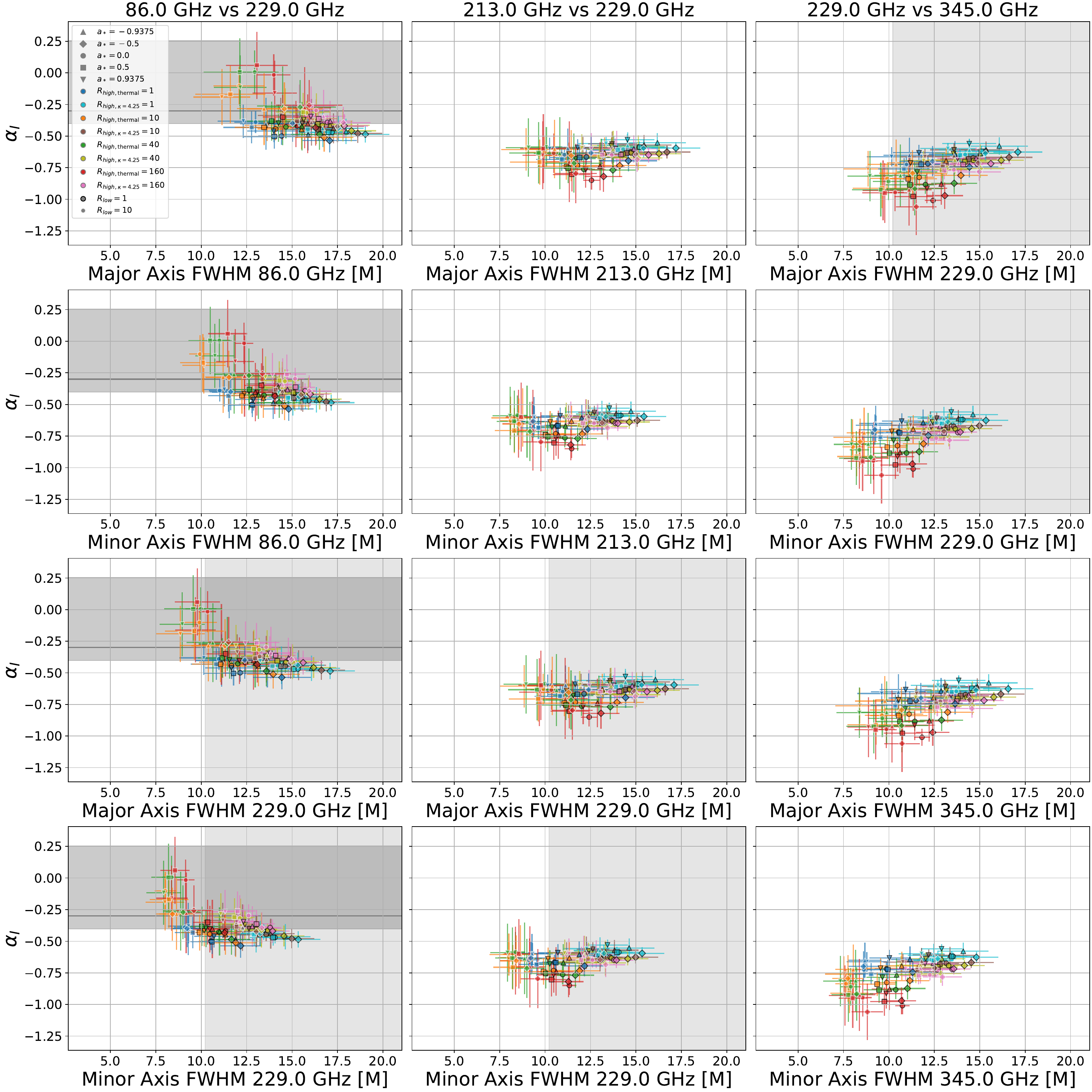}
    \caption{The time-averaged net spectral indices as a function of the time-averaged sizes of the emission region calculated for all possible combinations of parameters $(a_*, \rlow, \rhigh)$ and several observing frequencies. Errorbars indicate the standard deviation reflecting each model's variability. The light gray areas indicate 2018 observational constraints 
    from \citet{lu:2023} and \citetalias{M872018_PaperI:2024}. See Section~\ref{sec:real_data} for more details.}
    \label{fig:emission_vs_si_avg}
\end{figure*}

We currently do not yet have observational constraints on the time- and spatially-resolved spectral index. However, it would be interesting to compare the models to the observational data based on image sizes and the integrated spectral index of the images. 
Figure~\ref{fig:emission_vs_si_avg} shows the net spectral indices against the image size for all $(a_*,\rlow,\rhigh)$ in both thermal and non-thermal models. 

For $86-229\unit{GHz}$ band, the figure shows a trend that larger sized images have a lower net spectral index. In this band, $\rlow = 1$ parameter makes up larger images with lower net spectral index and $\rlow = 10$ makes the images smaller with higher net spectral index. This is also true for non-thermal models, although the difference between spectral indices is smaller in this case. 
This observation could be explained by the fact that for larger images, the net emission is more influenced by the outer emission region, which has negative values for the spectral index, resulting in the net spectral index of the image also being more negative.

Examining higher frequency cases, $213-229\unit{GHz}$ and
$229-345 \unit{GHz}$, the net spectral indices have shifted down into the negative range. Within these bands, there is not a single model that has a net positive spectral index, as expected. Larger images also tend to have roughly the same net spectral index, while smaller images tend to have a more spread out variety of spectral indices.
In both high frequency ranges, there seem to be very few
differences between the image sizes, as the intensity becomes dominated by the photon ring, which depends solely on the distance-to-mass ratio of the black hole. 

When shifting from $86-229\unit{GHz}$ to $ 229-345 \unit{GHz}$, 
spectral index decrease is more significant for models with $\rhigh \in (10, 40, 160)$ compared to $\rhigh = 1$. Higher $\rhigh$ also means a higher net spectral index that is less spread out across the domain of the size at $86 \unit{GHz}$ versus $229 \unit{GHz}$, while for $229 \unit{GHz}$ versus $345 \unit{GHz}$, a higher $\rhigh$ means a lower net spectral index and all $\rhigh$ have roughly the same spread across the domain of the size. The only exception to this is $\rhigh = 1$, which is pretty concise regarding the spread across the domain of the size.

Figure~\ref{fig:emission_vs_si_avg} also contains the measured sizes for M87* at $229 \unit{GHz}$ in the form of light gray areas and the measured net spectral index of M87* for frequencies of $86-229 \unit{GHz}$ in the form of darker gray areas (see section~\ref{sec:real_data}).
Many model averages fall within the allowed range or 
are permitted due to the large error bars associated with model variability. There are however a few models which are already 
inconsistent with currently available single-epoch observational data, those are: thermal models with $(a_*, \rlow, \rhigh)=(0.9375,10,10$/$40)$ due to too compact emission region size and both thermal and non-thermal models with $(a_*, \rlow, \rhigh)=(-0.5$/$0$/$0.5, 1, 1)$ and non-thermal models with $(a_*, \rlow, \rhigh)=(-0.5, 1, 10)$ due to too negative net spectral indices.


\section{Discussions}\label{sec:discussion}

The expected spectral index maps of M87* near the horizon region have been computed using a set of MAD simulations. The parameters surveyed for this include black hole spins $a_* \in (-0.9375, -0.5, 0, 0.5, 0.9375)$, electron parameters $\rlow \in (1, 10)$ and $\rhigh \in (1, 10, 40, 160)$ for both a thermal Maxwell-Jüttner electron distribution and a non-thermal $\kappa$ electron distribution with constant $\kappa = 4.25$. These parameters were surveyed at frequencies $\nu \in (86, 213, 216, 227, 229, 345) \unit{GHz}$. The observed spectral indices can be explained by simple theoretical considerations. Spectral indices and image sizes have also been computed and used to compare with available multifrequency constraints based on measured data of M87* by EHT and other interferometric networks such as GMVA, and to make predictions for even higher frequencies. 

Our results for the spectral indices with $\rlow = 1$ at $213-229\unit{GHz}$ agree with the results obtained by \citet{ricarte:2022}. Figure~12 in their paper shows that for MAD simulations with thermal electrons, spectral indices for all $\rhigh$ and black hole spins are roughly around $\alpha_{\rm I} \sim -0.8$. This  agrees with our results, as can be seen in our Figure~\ref{fig:emission_vs_si_avg}. Any small differences could be explained by their sample images being much smaller than the one used in this work, as they used 11 snapshots per model, while we used 500 snapshots per model. We post-processed many more files per model; hence, our estimated errorbars for the variability are much more reliable. Our results for $229-345\unit{GHz}$ are also consistent with those found by \citet{ricarte:2022}. 
The $\rlow = 10$ models are not examined in \citet{ricarte:2022}. We have also recreated the single-zone model presented in their work and arrived at similar conclusion that, in the case of thermal models, the resolved spectral index gradient is caused by the decrease of both magnetic field strength and electron temperature in the emitting regions. 
When comparing the thermal spectral index maps for M87* from \citet{ricarte:2022} Figures 6 and 7 with their counterparts from Figures~\ref{fig:rhigh_rlow_time_averaged_a0.5} and \ref{fig:rhigh_rlow_time_averaged_kappa_a0.5}, it is noticeable that both maps agree that the photon ring and its near vicinity have a spectral index of around zero. There is also very little difference between the spectral indices of the photon ring with a thermal electron distribution and those with a $\kappa$ distribution for $\kappa \in (5.0, 7.0)$, something that was also found using Figures~\ref{fig:rhigh_rlow_time_averaged_kappa_a0.5} with $\kappa = 4.25$. Do notice, however, that there is a difference between thermal and $\kappa = 3.5$ according to \citet{ricarte:2022}, so this also indicates that the parameter space for electron distributions should be examined further. 

There is a variety of electron models that one could further consider. They can be divided into three groups:
i) purely thermal models, ii) models with non-thermal components, iii) anisotropic models. 
Regarding i), in this work, we use Equation~\ref{eq:rhigh_rlow} as a prescription for the ratio between the electron temperature and the proton temperature. However, this is not the only prescription one can use. The so called critical-$\beta$ model \citep{anantua:2023} is an alternative electron-to-proton temperature ratio model described as
\begin{equation}
    \frac{T_{\rm p}}{T_{\rm e}} = \frac{1}{f} e^{\frac{\beta}{\beta_{\rm c}}} - 1
\end{equation}
where $f \in (0,1]$ determines the slope of the function and $\beta_{\rm c}$ determines the transition between electron domination and proton domination. Using this to compute the dimensionless plasma temperature $\Theta_{\rm e}$ could lead to different results; thus, this may also be an interesting topic to examine. Two-temperature GRMHD simulations also belong to category i), where electron temperatures are modeled alongside GRMHD simulations and where the parameter is the electron heating function \citep{chael:2025}. 
Regarding category ii), non-thermal models similar to ours but with a variable $\kappa$ parameter have been considered \citep{davelaar:2019,fromm:2022,davelaar:2023}. 
Finally, in relation to category iii), \citet{galishnikova:2023} uses the anisotropic electron distribution function:
\begin{equation}
  f (\gamma, \xi) \equiv \frac{1}{n_{\rm e}} \frac{dn_{\rm e}}{d\gamma d\cos{\xi}} = \frac{\sqrt{\eta}}{4 \pi m^3 c^3 \epsilon_{\perp} K_2(\frac{1}{\epsilon_{\perp}})} e^{-\frac{\gamma}{\epsilon_{\perp}} \sqrt{1 + (\eta - 1) \cos^2{\xi}}},
  \label{eq:anisotropic_dist}
\end{equation}
where $\eta$ is the measure of anisotropy (notice that for $\eta = 1$, Equation~\ref{eq:anisotropic_dist} simplifies to an isotropic relativistic Maxwell electron distribution function) and $\epsilon_{\perp} \equiv \frac{k T_{\perp, \rm e}}{m_{\rm e} c^2}$ is the dimensionless perpendicular electron temperature. This electron distribution function allows for different temperatures both parallel and perpendicular to the local magnetic field lines, providing yet another possible parameter space that could be explored.
Anizotropic models have also been developed for distribution functions with a non-thermal component \citep{tsunetoe:2025}. 

Many of these electron models have been implemented in GRMHD simulations of MADs to model emission from M87* near horizon region (e.g., \citealt{cruzosorio:2022,cruzosorio:2026}). However, with the exception of \citet{ricarte:2022}, none of the other works makes explicit predictions of the resolved, near horizon, compact flux spectral index.
We therefore limit the discussion of the results to this work.

We also compared our models with the measured net spectral index and size of M87*. Our result for the excluded thermal models using the net spectral index complements the results obtained by \citetalias{M872018_PaperII:2025}. Table~3 of their paper shows passing and non-passing MAD thermal models 
equivalent to models considered in the present work. Based on joint 
multi-year variability (in single frequency band) thermal MAD simulations with $(a_*,\rlow,\rhigh)=(0.9375,10,10$/$40)$ and $(-0.5$/$0$/$0.5,1,1)$ often pass EHT variability constraints. Our study excludes these models.

Our results show that we could further severely limit the parameter space of models using time variability measurements like the ones from the observing movie campaign by EHT. An increased value of $\rlow$ was shown to make the spectral index more variable. This would mean that a movie of the spectral slope could possibly discard a significant amount of models. Notice, however, that the spectral indices are also less variable for both $\rlow$ when assuming the $\kappa$ electron distribution function with $\kappa = 4.25$. Low variability in $\alpha_{\rm I}$ could also indicate a non-thermal distribution function. Surveying through parameters of the non-thermal models needs to be optimized as it is currently extremely time-consuming. 



On the observational side, the compact flux in EHT observations should be more precisely estimated, since it will be crucial for measuring the spectral indices. 

Our simulations show that the photon ring has a visibly higher spectral index compared to the rest of the emission region, suggesting that the photon ring has a higher critical frequency $\nu_{\rm crit}$ (or higher $\tau$) indicating stronger magnetic fields and higher electron temperatures. The variability pattern of the spectral index in the photon ring merits a dedicated study. Here we limit our discussion of these effects due to the fast-light approximation used in our ray-tracing computations, which should be relaxed for photon ring variability studies. The current EHT configuration is also not able to resolve this feature of the image, but future space-based VLBI arrays might be able to do it.





\section*{Acknowledgements}

Regarding GRMHD simulations,
we gratefully acknowledge the HPC RIVR
consortium (www.hpc-rivr.si) and EuroHPC JU (eurohpc-ju.europa.eu) for funding this research by providing the computing resources of the HPC system Vega at the
Institute of Information Science (www.izum.si).
Ray-tracing simulations were carried out at Coma Cluster at Radboud University.

\section*{Data Availability}

GRMHD simulations used in this work are taken from \citet{moscibrodzka:2025} and have been generated by the publically available code \ebhlight~\citep{ryan:2015} and are reproducible. All images produced in this work using \ipole~\citep{moscibrodzka:2018} code will be made publically available on Zenodo in fits format once this work is peer-reviewed.



\bibliographystyle{mnras}
\bibliography{main} 




\appendix

\section{Multifrequency light curves, intensity maps and intensity weighted spectral index}\label{sec:app_intensity_over_time}
\begin{figure*}
  \centering
  \includegraphics[trim={0cm 0cm 0cm 0cm},clip,width=0.9\linewidth]{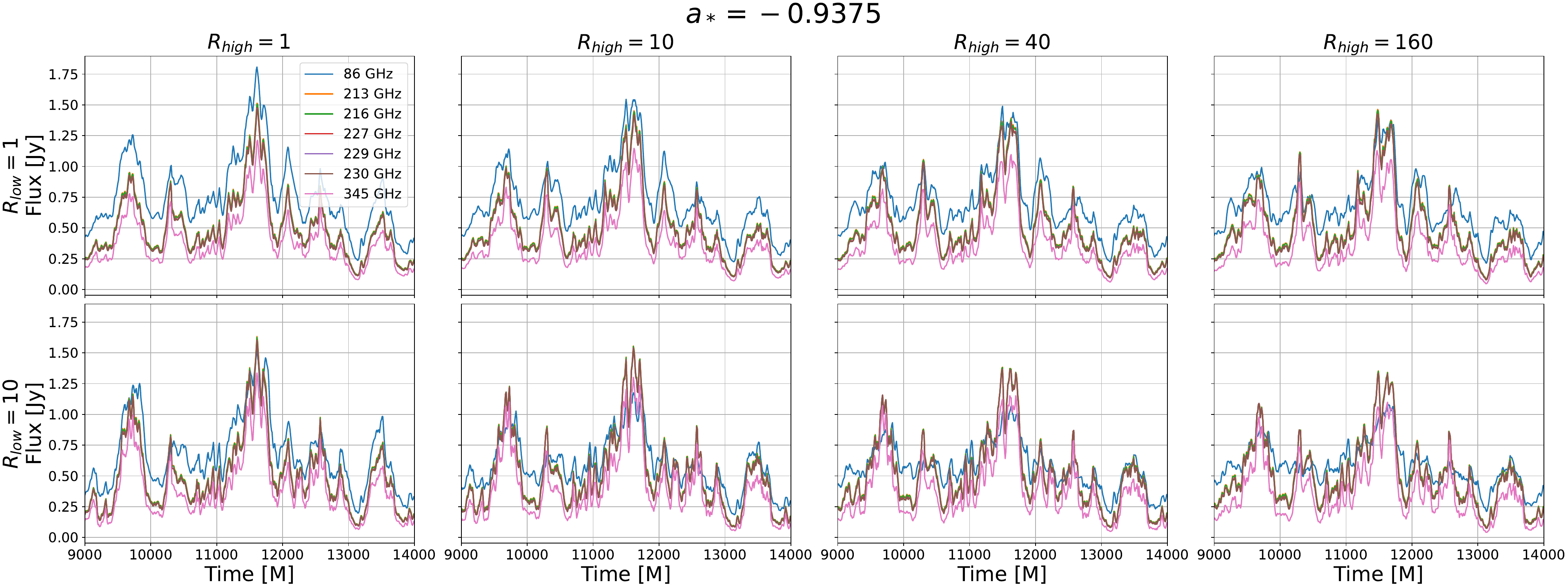}
\caption{Multifrequency light curves for $\nu \in (86, 213, 216, 227, 229, 230, 345 ) \unit{GHz}$, black hole spin $a_* = -0.9375$ and parameters $\rhigh \in (1, 10, 40, 160)$ and $\rlow \in (1, 10)$. The time is in units of $M$, where $1 M \approx 8.896 \unit{hr}$ for M87*. Notice that none of the graphs go to infinity along the time interval, indicating that the data is processed correctly.}
\label{fig:intensity_over_time_a-0.94}
\end{figure*}

\begin{figure*}
  \centering
  \includegraphics[trim={0cm 0cm 0cm 0cm},clip,width=0.9\linewidth]{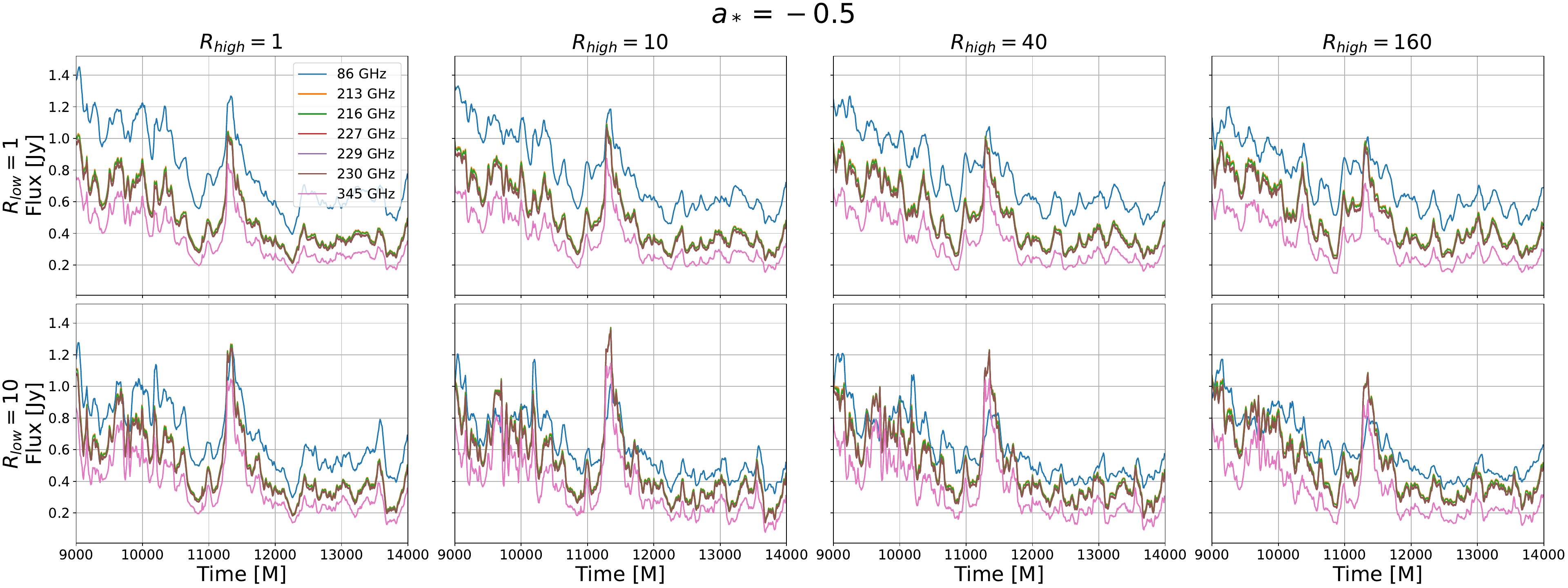}
\caption{Same as in Figure~\ref{fig:intensity_over_time_a-0.94} but for $a_*=-0.5$}
\label{fig:intensity_over_time_a-0.5}
\end{figure*}

\begin{figure*}
  \centering
  \includegraphics[trim={0cm 0cm 0cm 0cm},clip,width=0.9\linewidth]{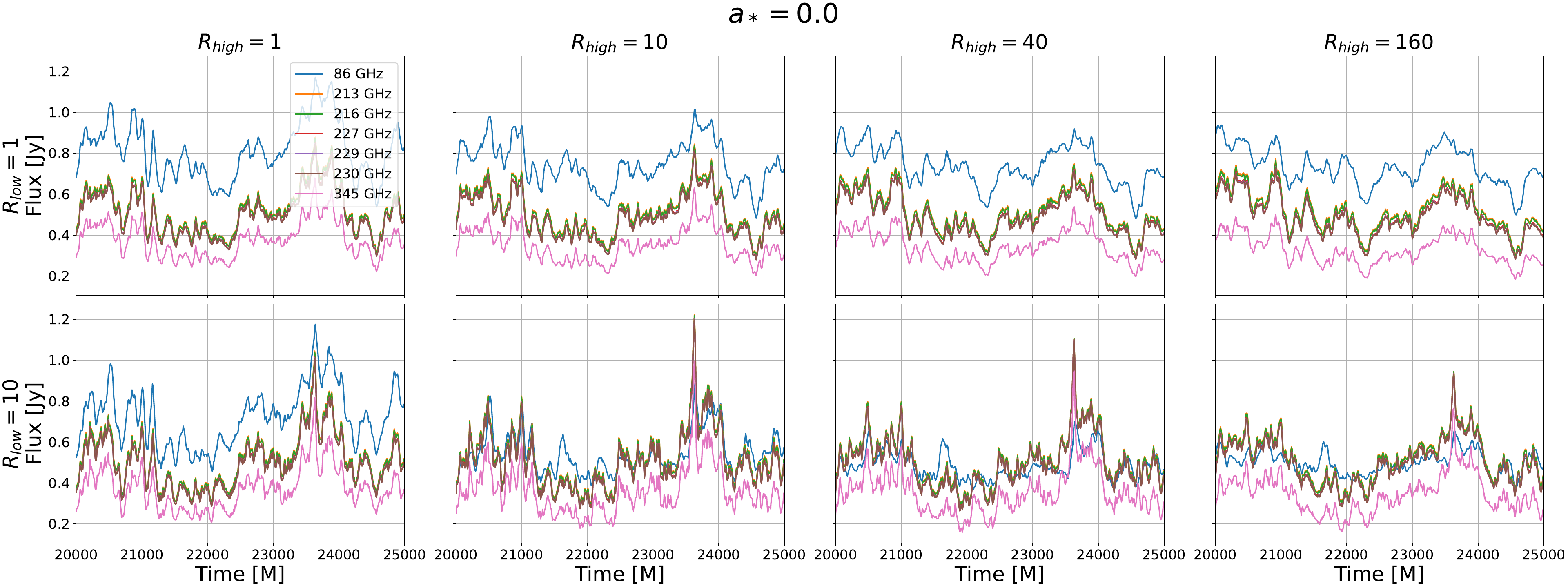}
\caption{The same as in Figure~\ref{fig:intensity_over_time_a-0.94} but for $a_* = 0$.}
\label{fig:intensity_over_time_a0.0}
\end{figure*}

\begin{figure*}
  \centering
  \includegraphics[trim={0cm 0cm 0cm 0cm},clip,width=0.9\linewidth]{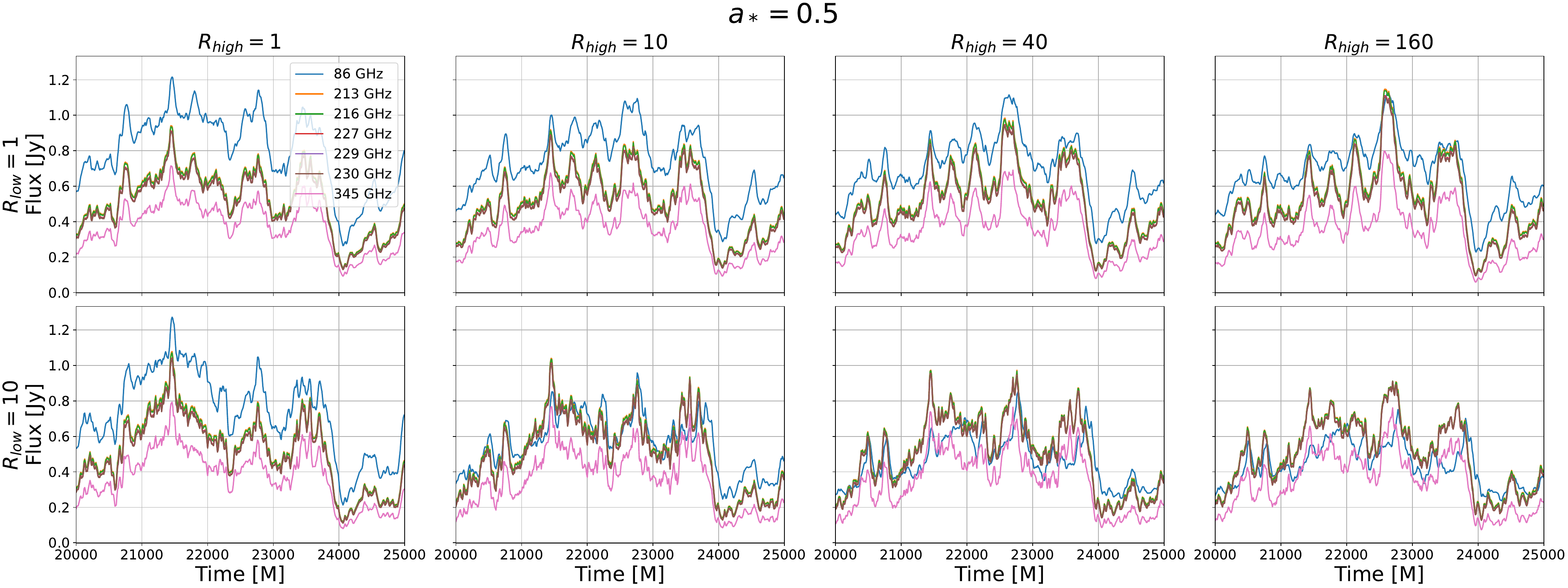}
\caption{The same as in Figure~\ref{fig:intensity_over_time_a-0.94} but for $a_* = 0.5$.}
\label{fig:intensity_over_time_a0.5}
\end{figure*}

\begin{figure*}
  \centering
  \includegraphics[trim={0cm 0cm 0cm 0cm},clip,width=0.9\linewidth]{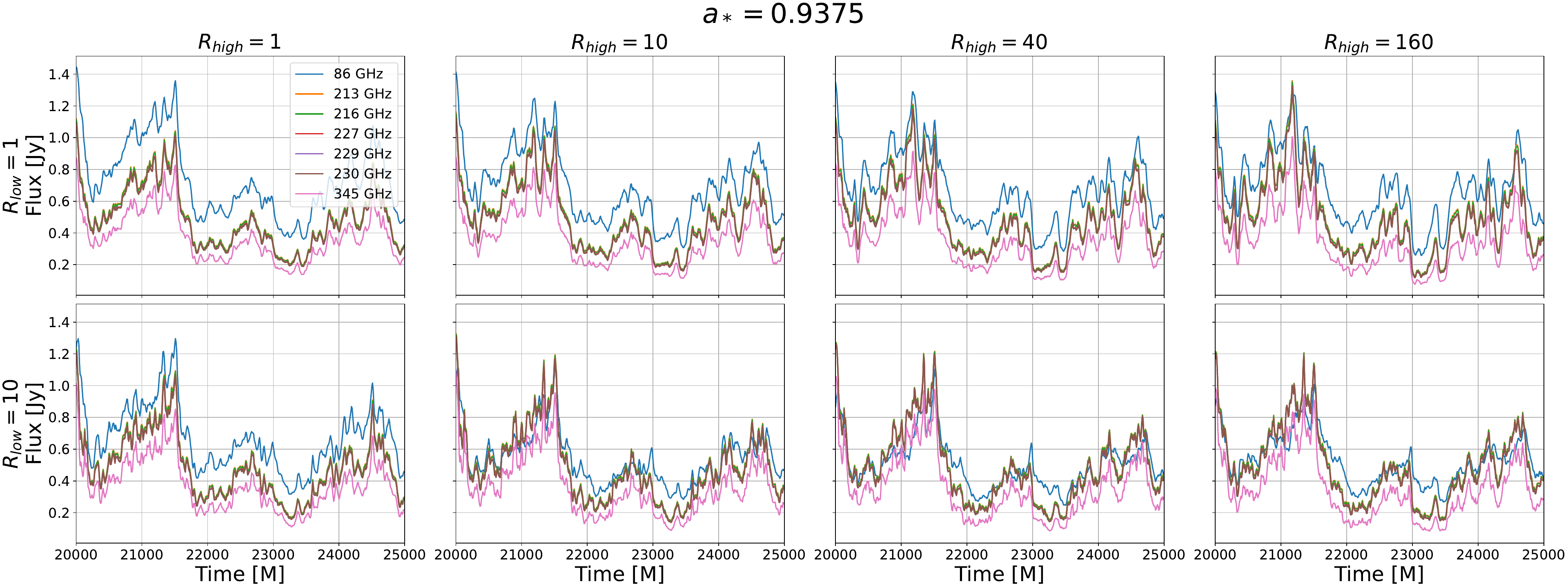}
\caption{The same as in Figure~\ref{fig:intensity_over_time_a-0.94} but for $a_* = 0.9375$.}
\label{fig:intensity_over_time_a0.94}
\end{figure*}

\begin{figure*}
  \centering
  \includegraphics[trim={1.7cm 0.6cm 0cm 0cm},clip,width=0.45\linewidth]{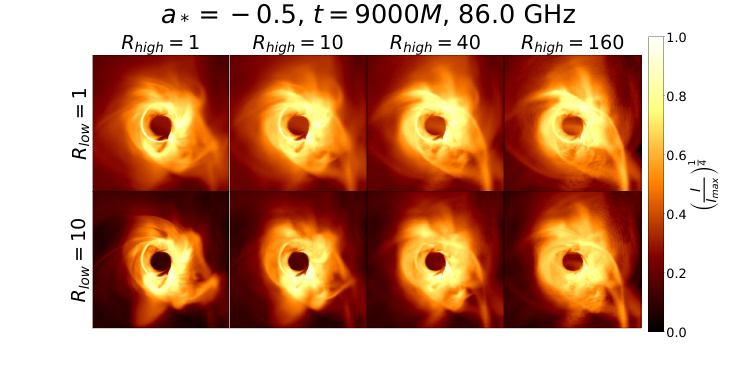}
  \includegraphics[trim={1.7cm 0.6cm 0cm 0cm},clip,width=0.45\linewidth]{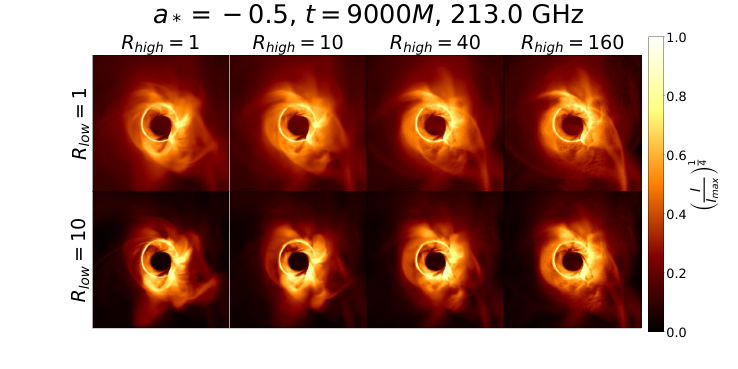}\\
  \includegraphics[trim={1.7cm 0.6cm 0cm 0cm},clip,width=0.45\linewidth]{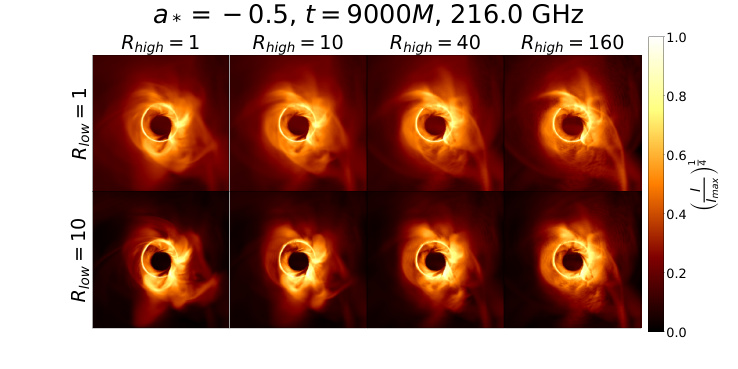}
  \includegraphics[trim={1.7cm 0.6cm 0cm 0cm},clip,width=0.45\linewidth]{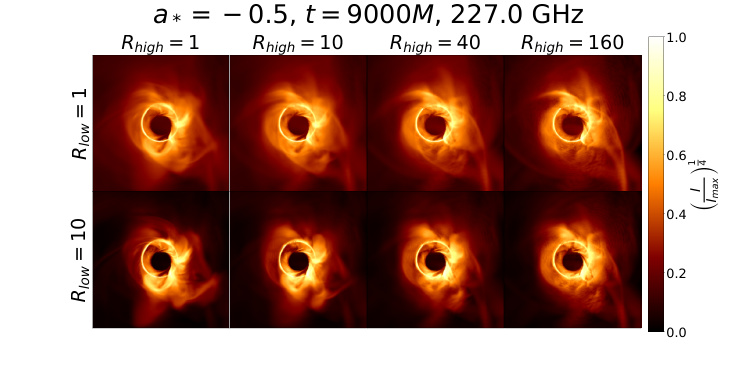}\\
  \includegraphics[trim={1.7cm 0.6cm 0cm 0cm},clip,width=0.45\linewidth]{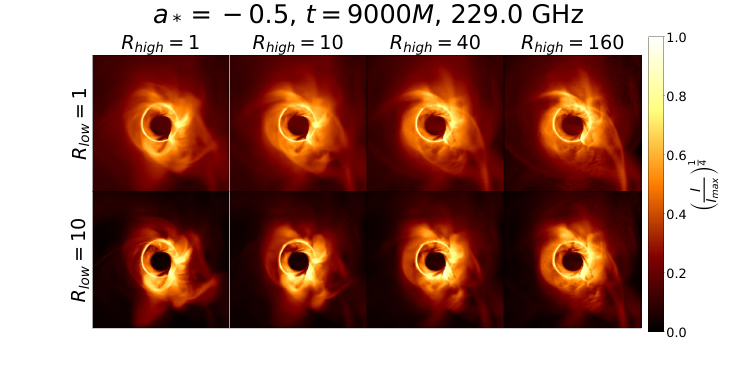}
    \includegraphics[trim={1.7cm 0.6cm 0cm 0cm},clip,width=.45\textwidth]{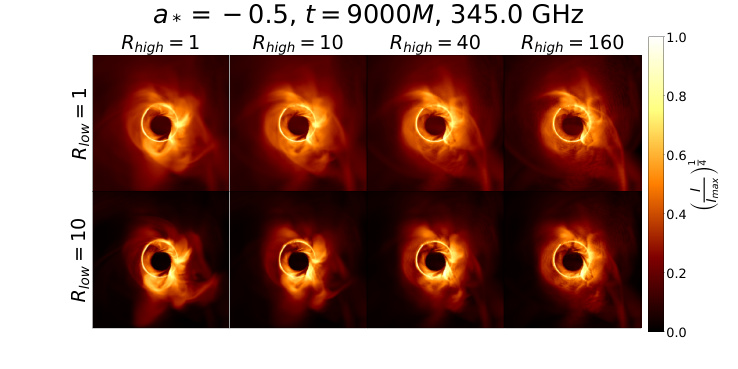}
     \caption{Example intensity maps for frequencies $\nu \in (86, 213, 216, 227, 229, 345 ) \unit{GHz}$, black hole spin $a_* = -0.5$ 
     and parameters $\rhigh \in (1, 10, 40, 160)$ and $\rlow \in (1, 10)$ at $9000 M$. The intensity has been normalized and the color scales with $I^{\frac{1}{4}}$ to increase contrast at the higher frequencies.}\label{fig:intensity_map_a-0.5}
\end{figure*}

\begin{figure*}
  \centering
  \includegraphics[trim={1.7cm 0.6cm 0cm 0cm},clip,width=0.45\linewidth]{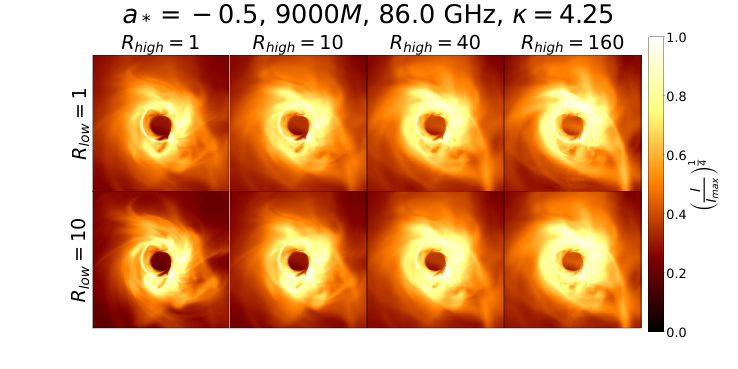}
  \includegraphics[trim={1.7cm 0.6cm 0cm 0cm},clip,width=0.45\linewidth]{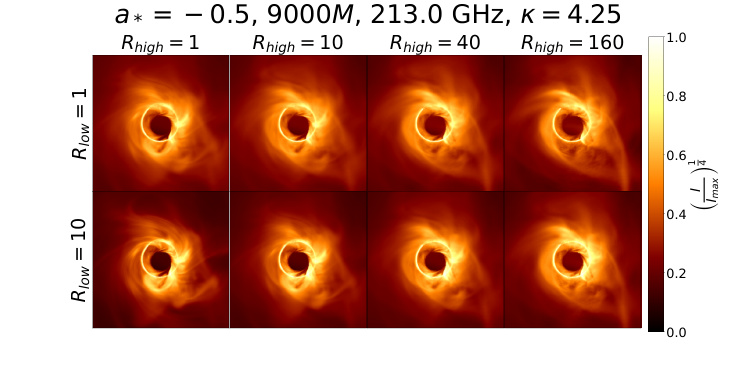}\\
  \includegraphics[trim={1.7cm 0.6cm 0cm 0cm},clip,width=0.45\linewidth]{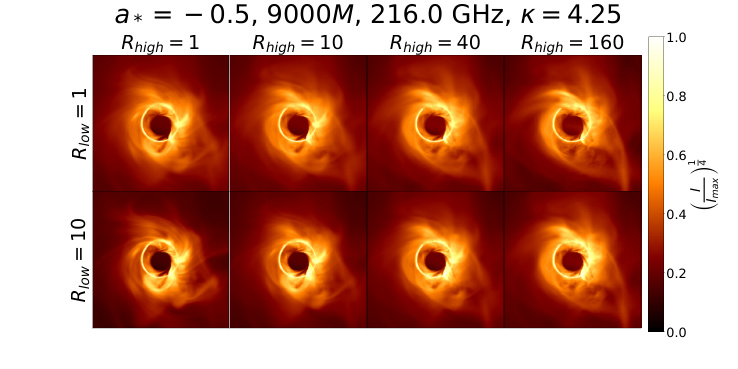}
  \includegraphics[trim={1.7cm 0.6cm 0cm 0cm},clip,width=0.45\linewidth]{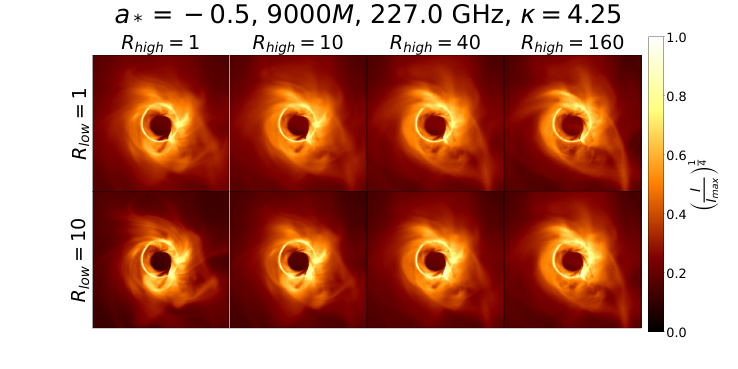}\\
  \includegraphics[trim={1.7cm 0.6cm 0cm 0cm},clip,width=0.45\linewidth]{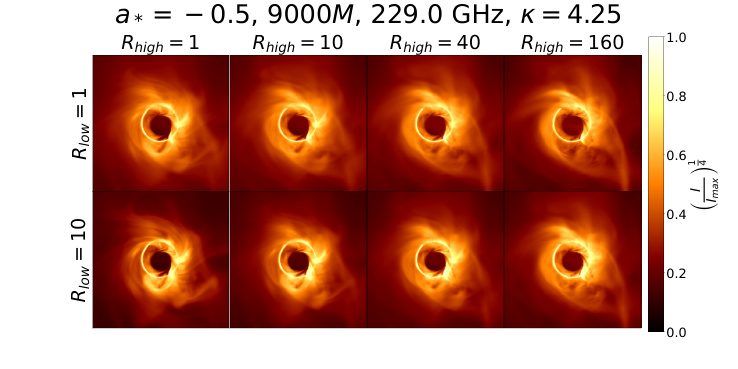}
    \includegraphics[trim={1.7cm 0.6cm 0cm 0cm},clip,width=.45\textwidth]{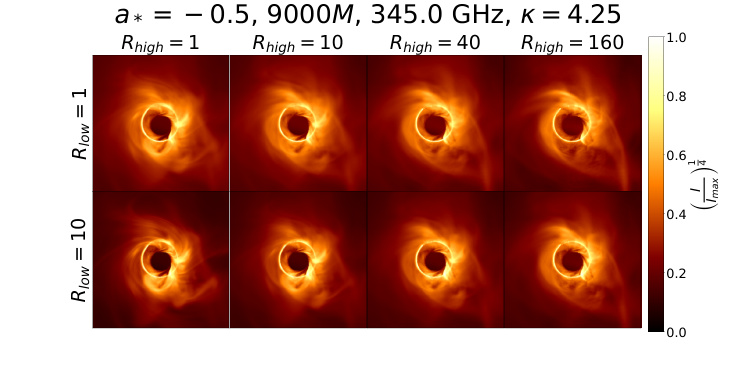}
     \caption{The same as Figure \ref{fig:intensity_map_a-0.5} but for non-thermal models.}\label{fig:intensity_map_kappa_a-0.5}
\end{figure*}

\begin{figure*}
  \centering
  \includegraphics[trim={0cm 0cm 0cm 0cm},clip,width=0.9\linewidth]{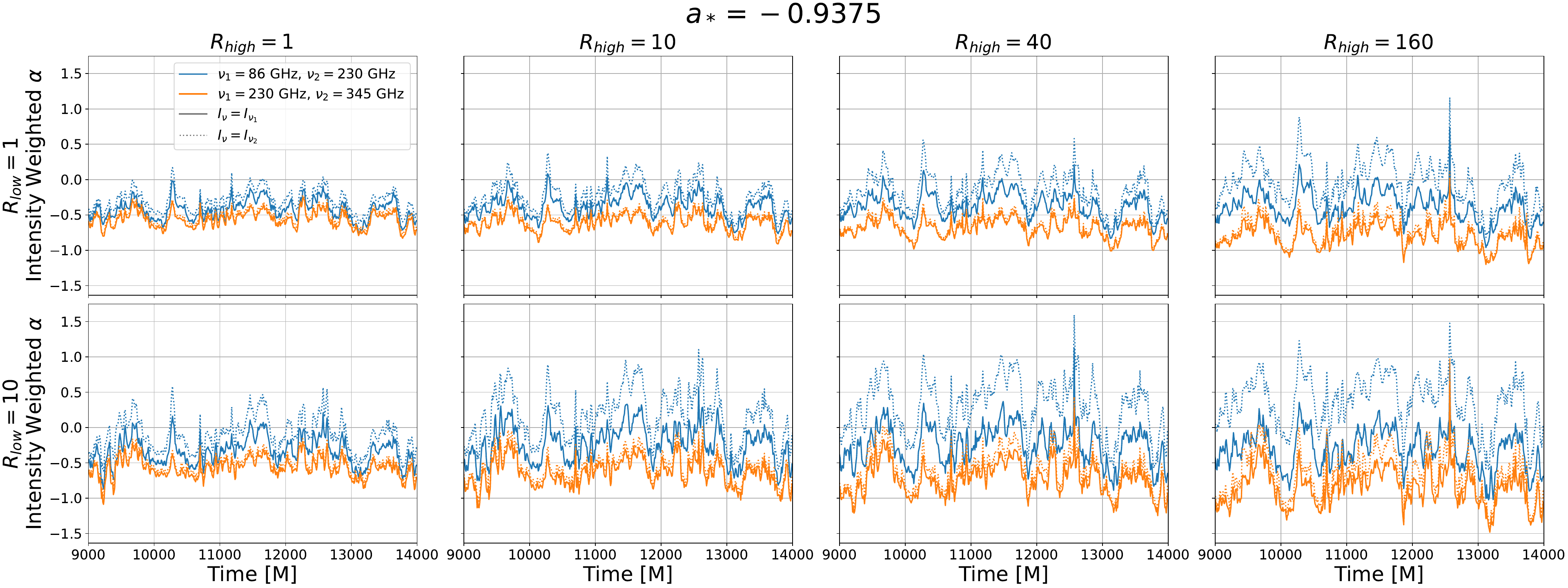}
\caption{Intensity weighted spectral index over time for the frequencies $\nu \in (86, 230, 345 ) \unit{GHz}$, black hole spin $a_* = -0.9375$ and parameters $\rhigh \in (1, 10, 40, 160)$ and $\rlow \in (1, 10)$. The time is in units of $M$, so for M87*, $1 M \approx 8.896 \unit{hr}$. Notice that none of the graphs go to infinity along the time interval, indicating that the spectral indices were computed correctly.}
\label{fig:intensity_weighted_spectral_index_a-0.94}
\end{figure*}

\begin{figure*}
  \centering
  \includegraphics[trim={0cm 0cm 0cm 0cm},clip,width=0.9\linewidth]{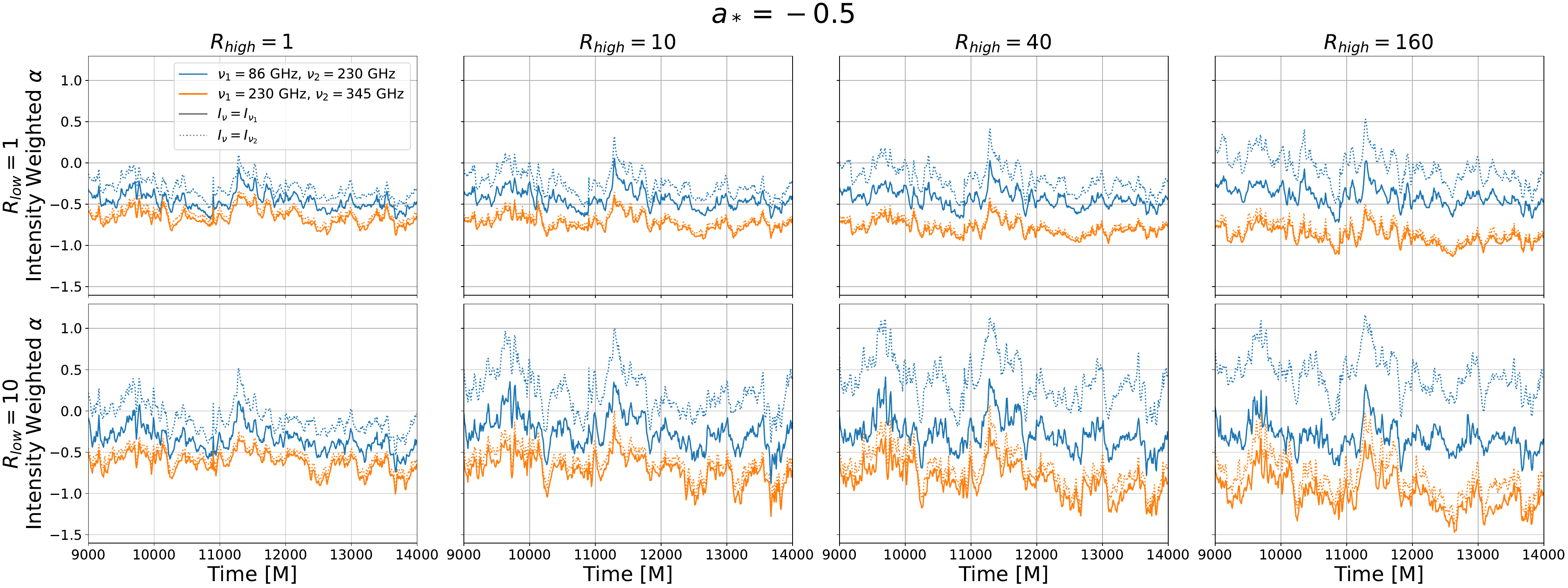}
\caption{The same as in Figure~\ref{fig:intensity_weighted_spectral_index_a-0.94} but for $a_* = -0.5$.}
\label{fig:intensity_weighted_spectral_index_a-0.5}
\end{figure*}

\begin{figure*}
  \centering
  \includegraphics[trim={0cm 0cm 0cm 0cm},clip,width=0.9\linewidth]{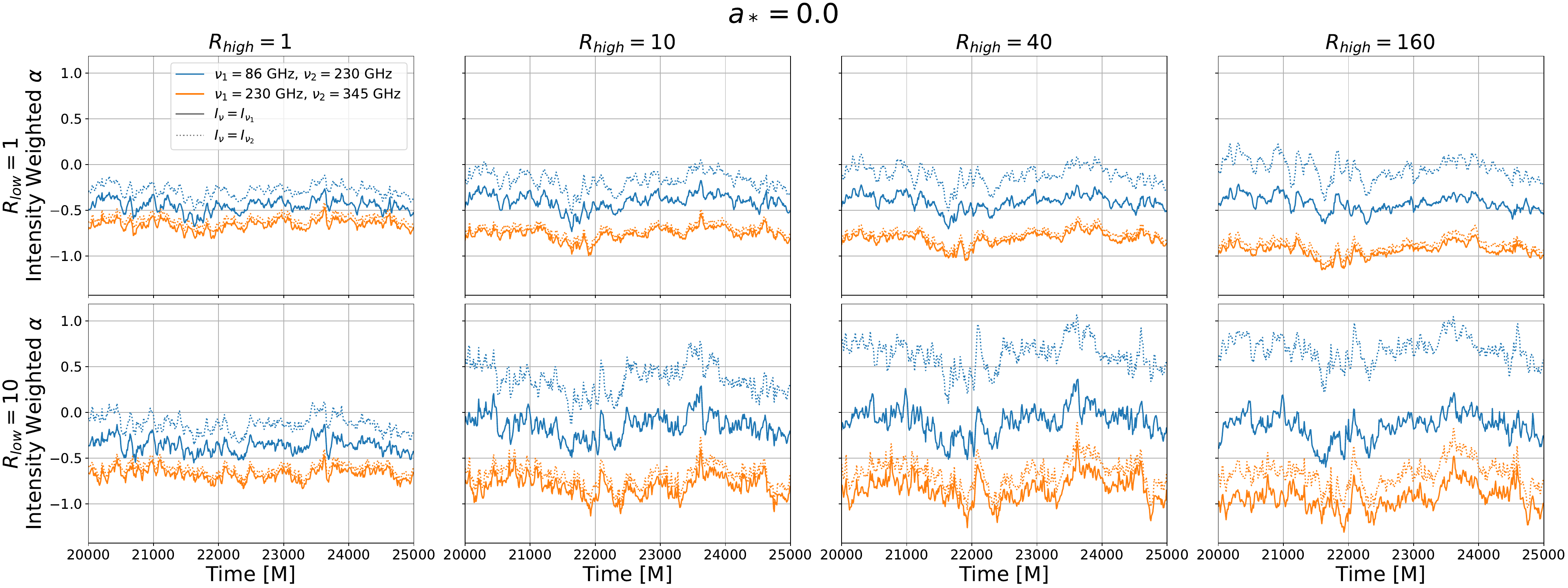}
\caption{The same as in Figure~\ref{fig:intensity_weighted_spectral_index_a-0.94} but for $a_* = 0$.}\label{fig:intensity_weighted_spectral_index_a0.0}
\end{figure*}

\begin{figure*}
  \centering
  \includegraphics[trim={0cm 0cm 0cm 0cm},clip,width=0.9\linewidth]{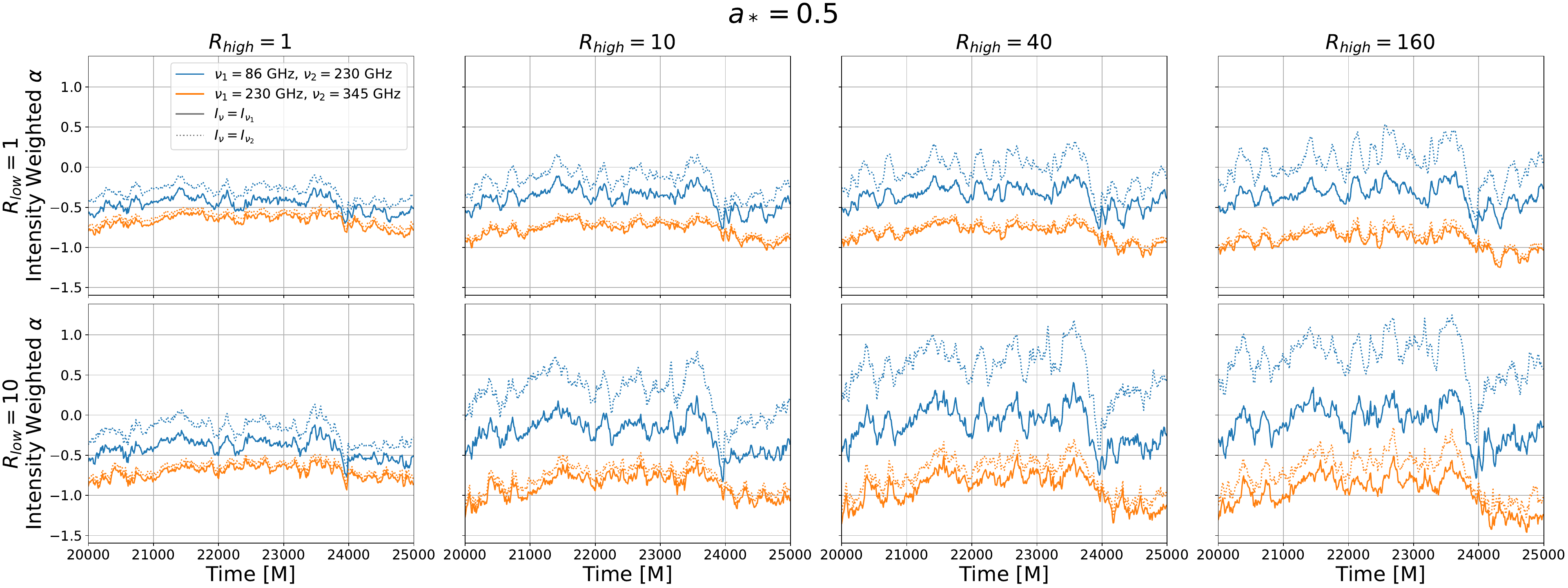}
\caption{The same as in Figure~\ref{fig:intensity_weighted_spectral_index_a-0.94} but for $a_* = 0.5$.}
\label{fig:intensity_weighted_spectral_index_a0.5}
\end{figure*}

\begin{figure*}
  \centering
  \includegraphics[trim={0cm 0cm 0cm 0cm},clip,width=0.9\linewidth]{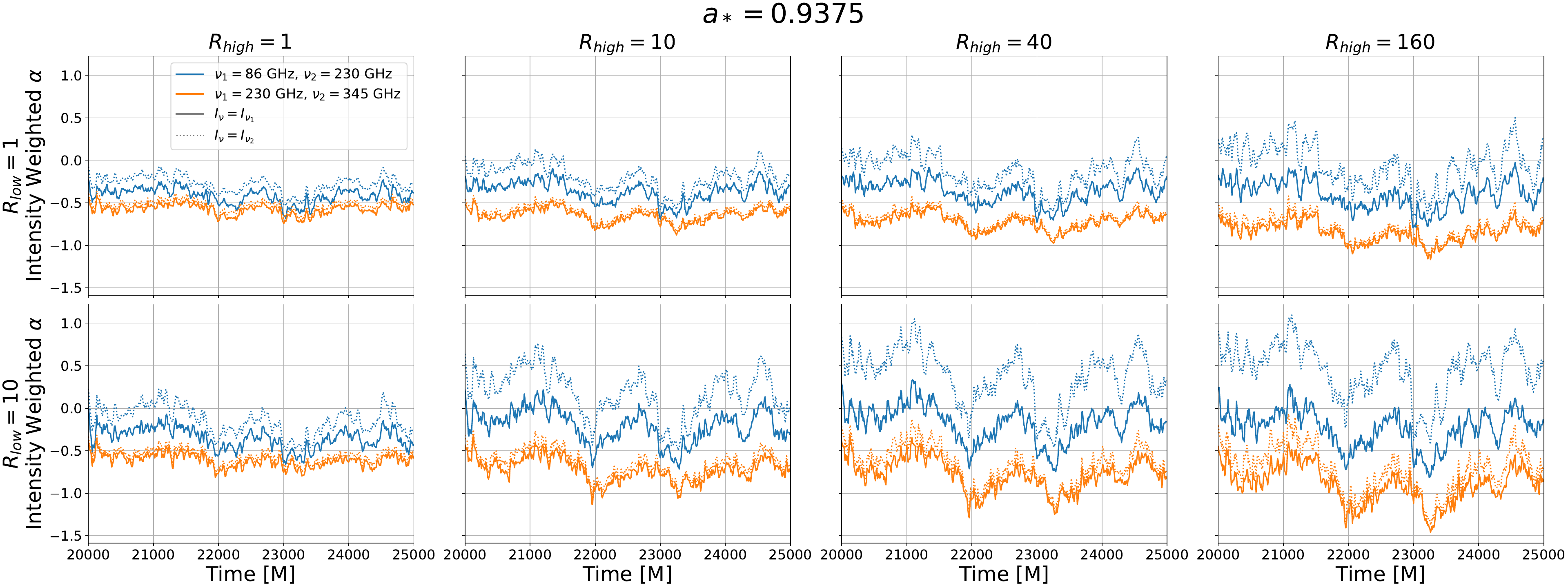}
\caption{The same as  in \ref{fig:intensity_weighted_spectral_index_a-0.94} but for  $a_* = 0.9375$.}
\label{fig:intensity_weighted_spectral_index_a0.94}
\end{figure*}

\section{Spectral index maps for remaining black hole spins}\label{sec:other_spins}

\begin{figure*}
  \centering
  \includegraphics[trim={5cm 2.5cm 2cm 0cm},clip,width=0.39\textwidth]{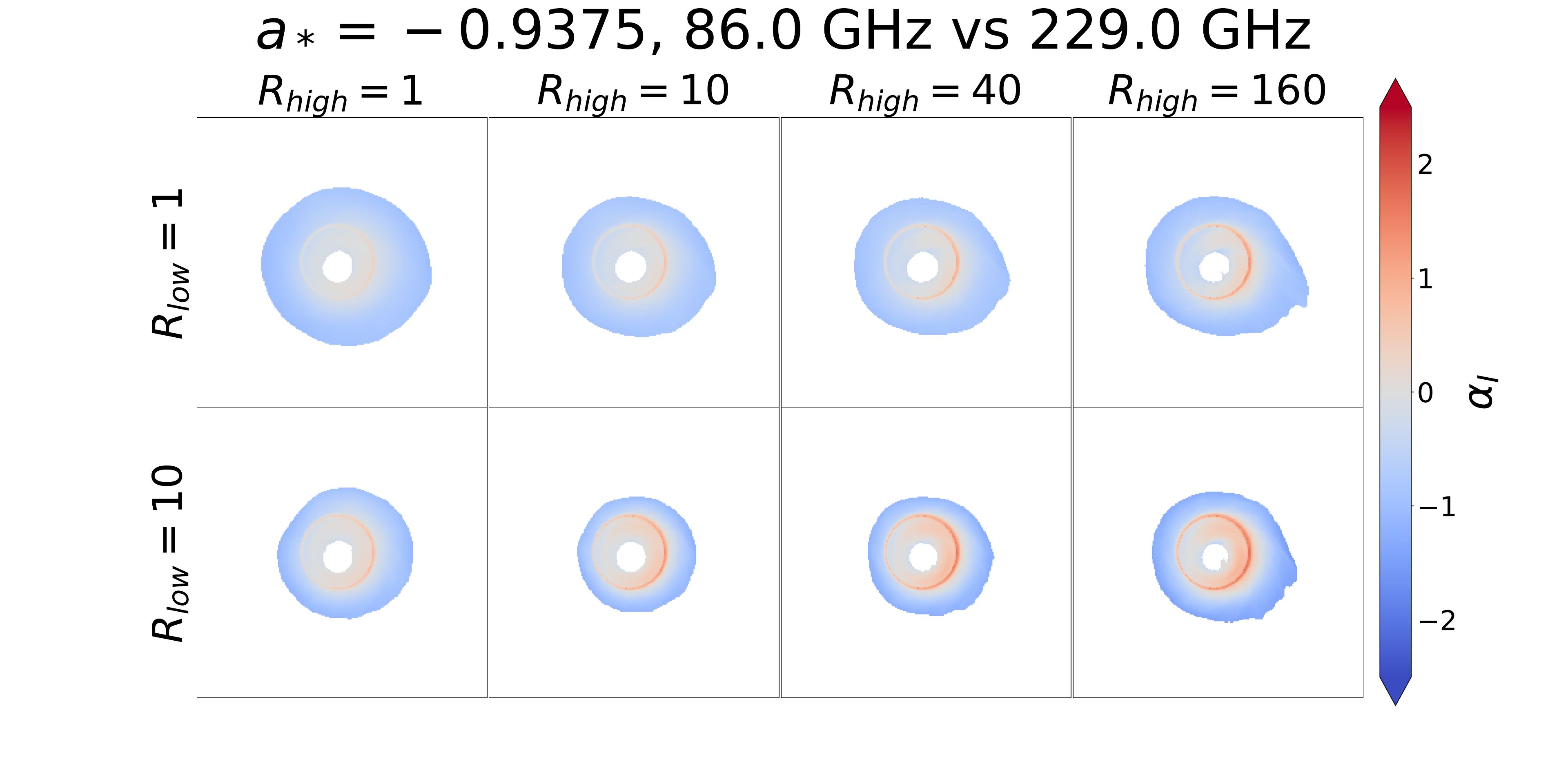}
  \includegraphics[trim={5cm 2.5cm 2cm 0cm},clip,width=0.39\textwidth]{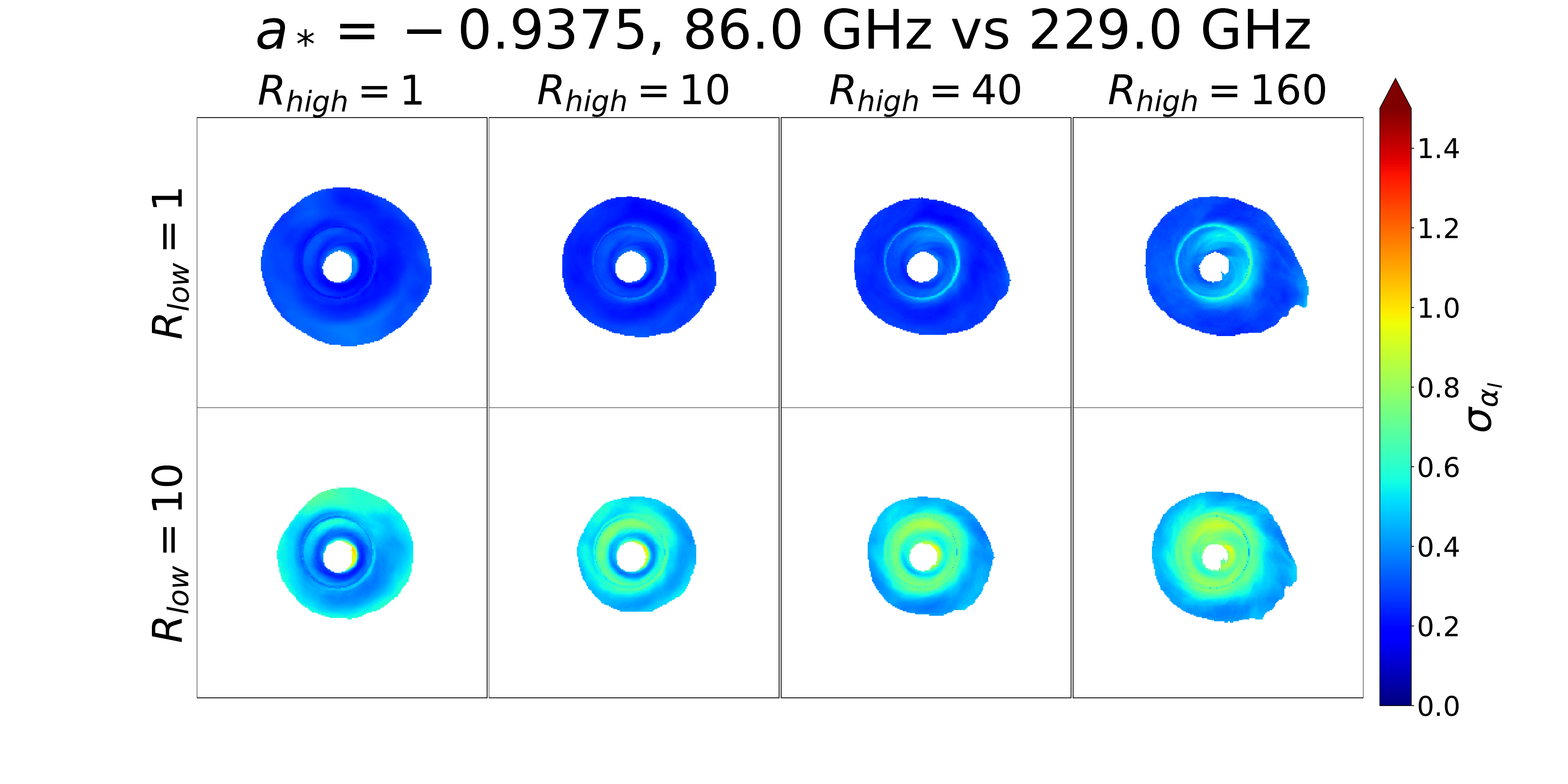}\\
 \includegraphics[trim={5cm 2.5cm 2cm 0cm},clip,width=0.39\textwidth]{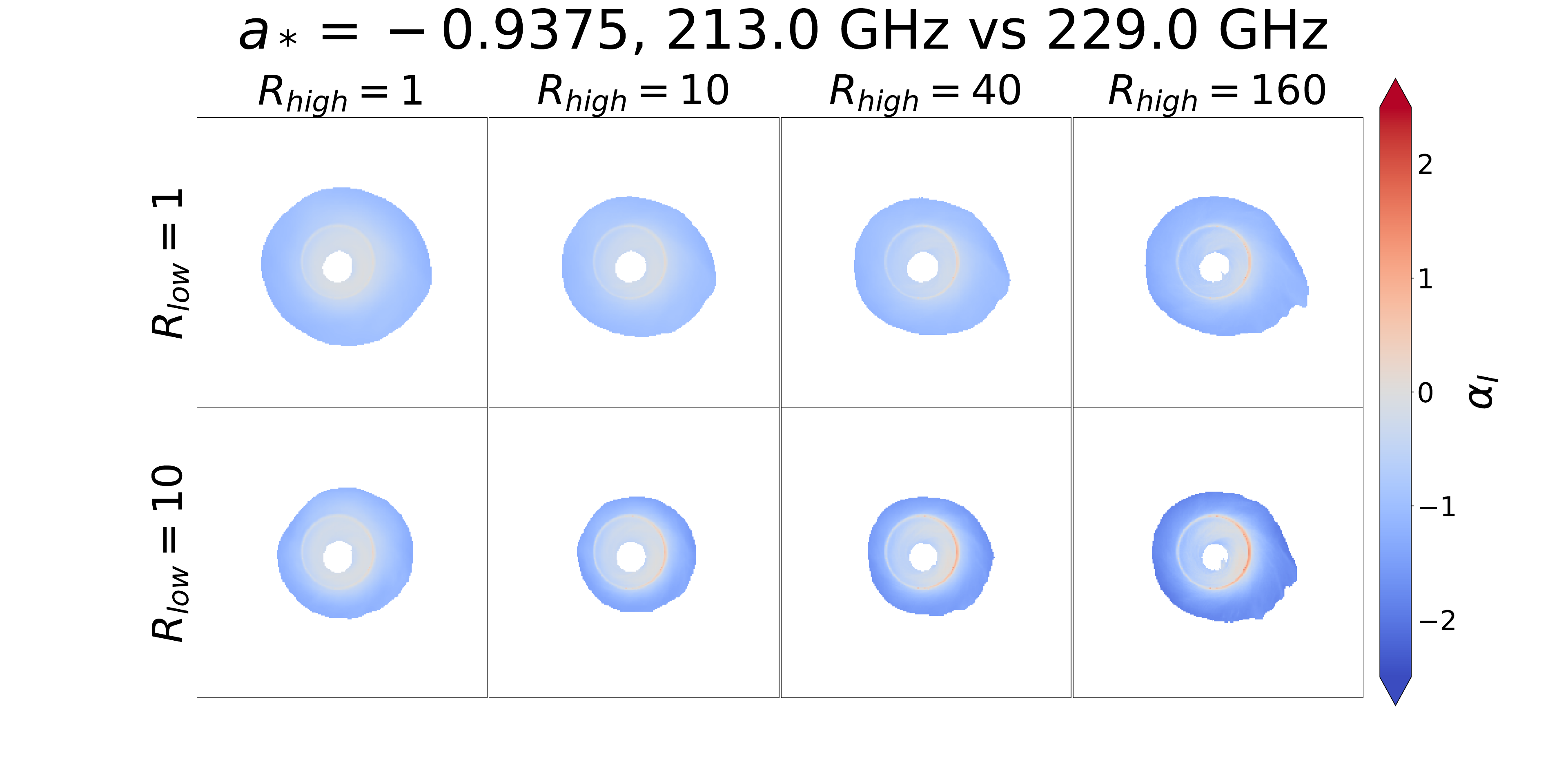}
  \includegraphics[trim={5cm 2.5cm 2cm 0cm},clip,width=0.39\textwidth]{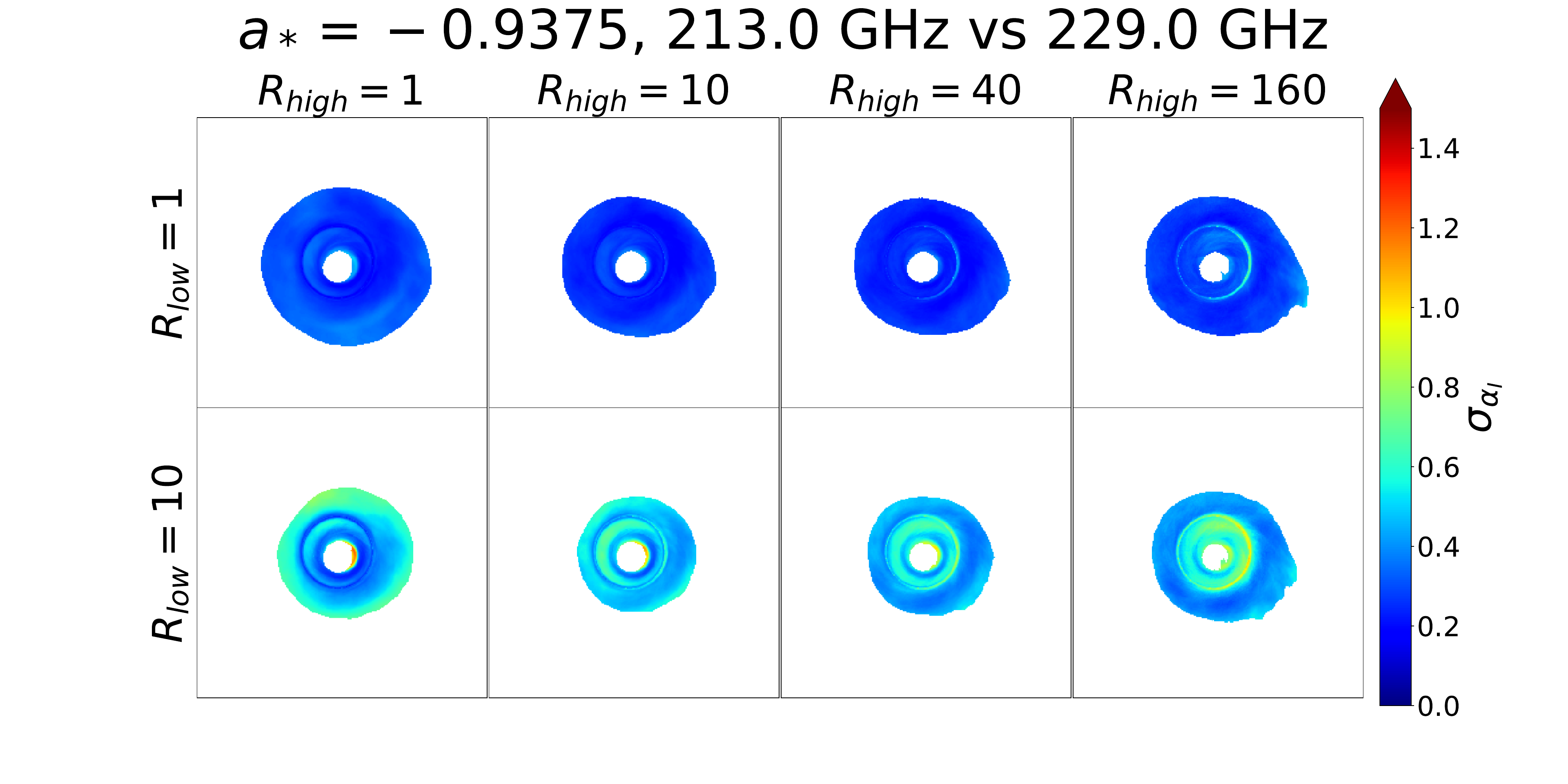}\\  
  \includegraphics[trim={5cm 2.5cm 2cm 0cm},clip,width=0.39\textwidth]{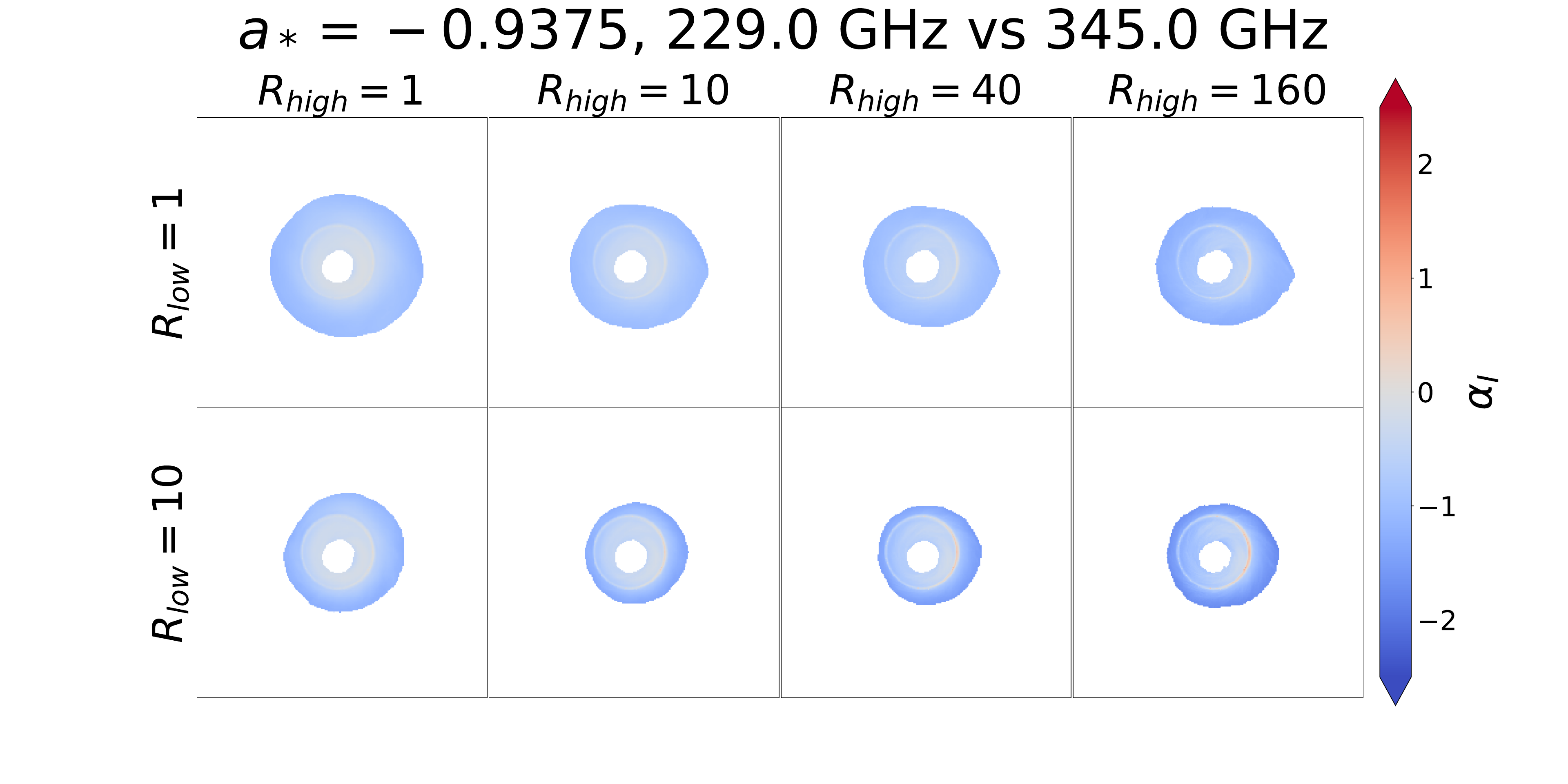}
  \includegraphics[trim={5cm 2.5cm 2cm 0cm},clip,width=0.39\textwidth]{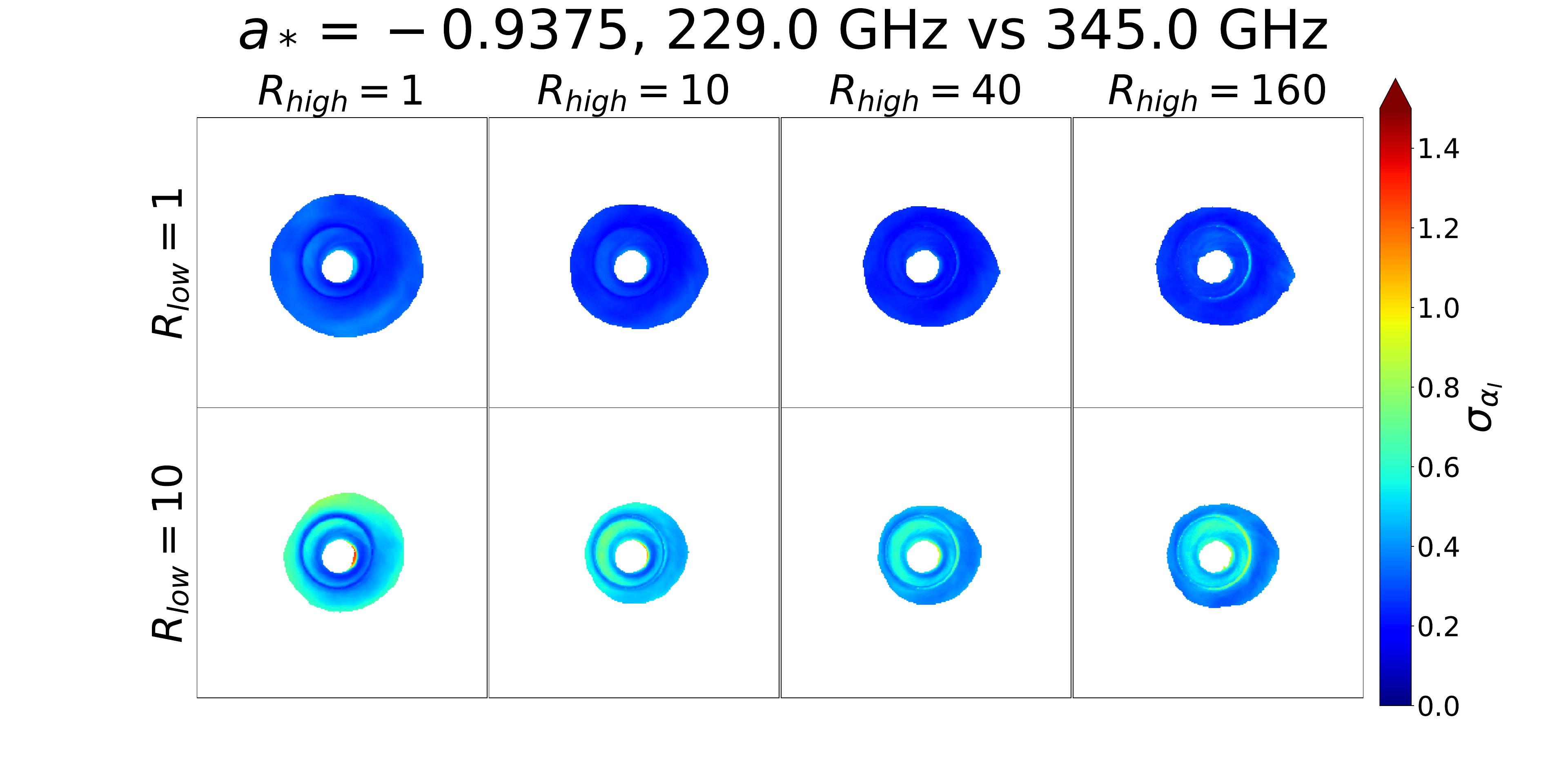}
\caption{Spectral indices of thermal models. Left panels show unblurred, time-averaged spectral index maps for all combinations of parameters $\rlow \in (1, 10)$ and $\rhigh \in (1, 10, 40, 160)$ and frequencies $86-345 \unit{GHz}$ (from top to bottom) for a single black hole spin $a_*=-0.9375$. Right panels show the corresponding standard deviation which indicates the variability of a model. Field of view of each image is $40\times40$M with resolution $256\times256$ pixels. Only the regions where $I > 0.01 I_{\rm max}$ are shown.}
\label{fig:rhigh_rlow_time_averaged_a-0.94}
\end{figure*}

\begin{figure*}
  \centering
  \includegraphics[trim={5cm 2.5cm 2cm 0cm},clip,width=0.39\textwidth]{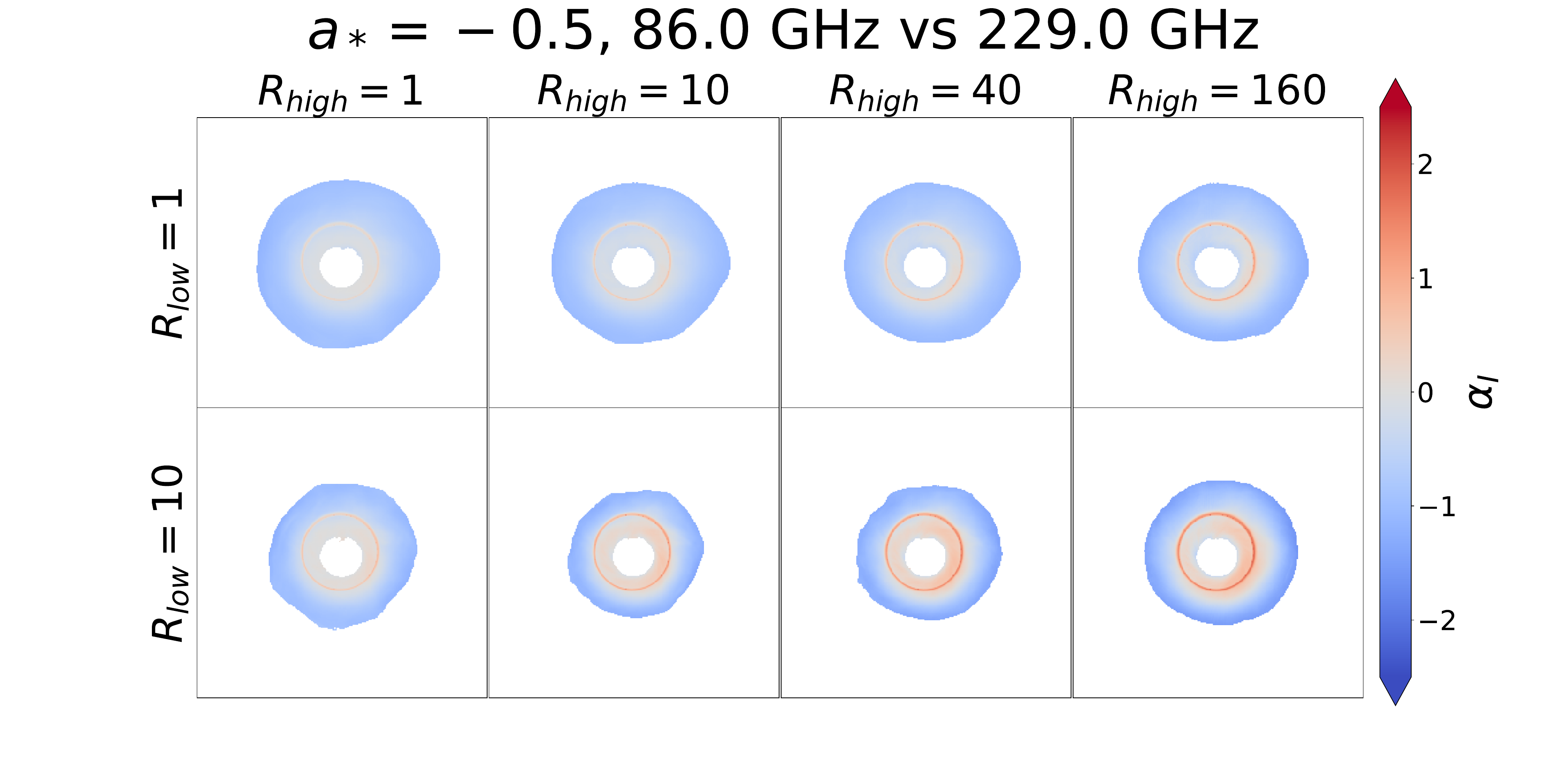}
  \includegraphics[trim={5cm 2.5cm 2cm 0cm},clip,width=0.39\textwidth]{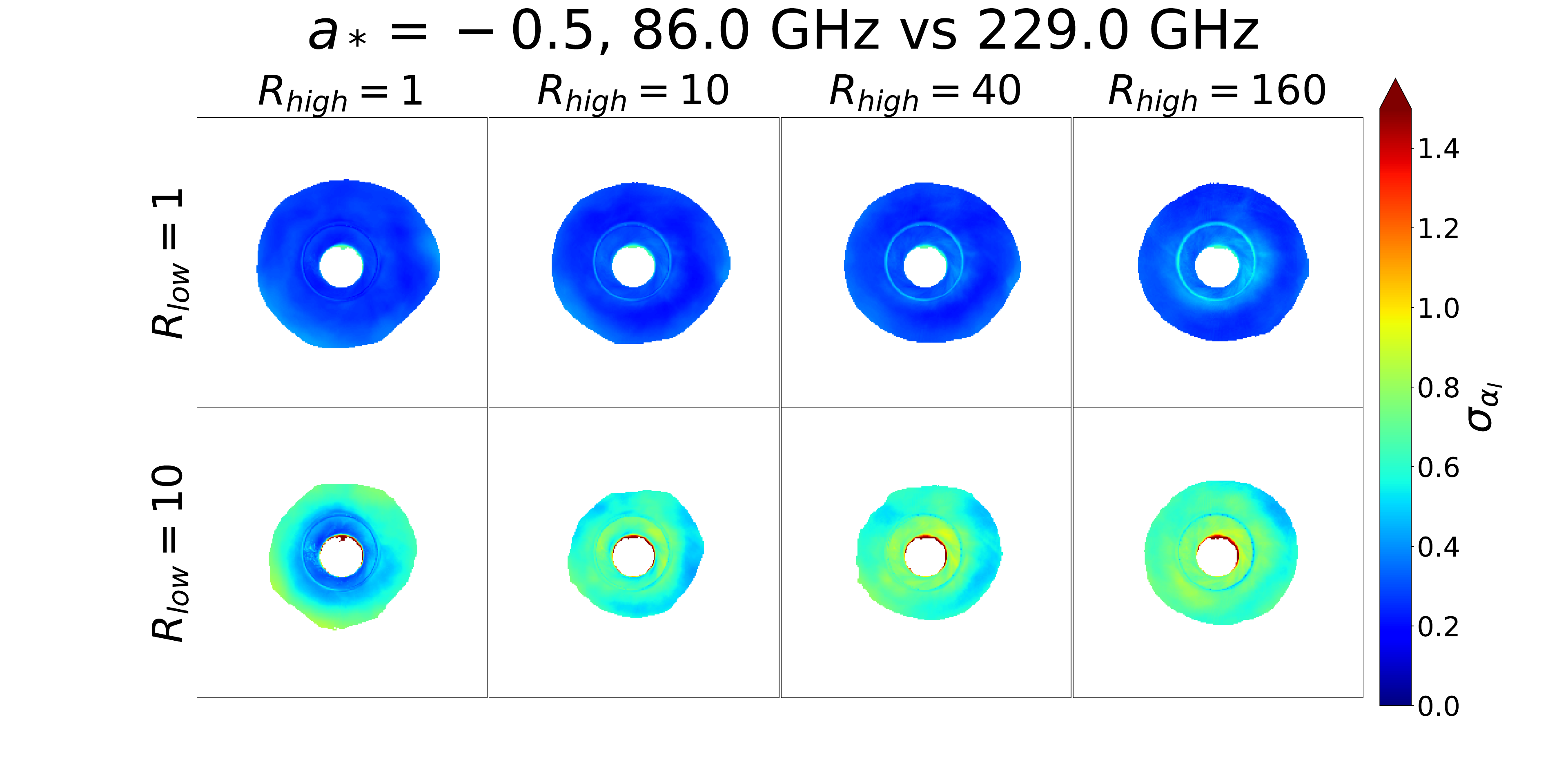}\\
 \includegraphics[trim={5cm 2.5cm 2cm 0cm},clip,width=0.39\textwidth]{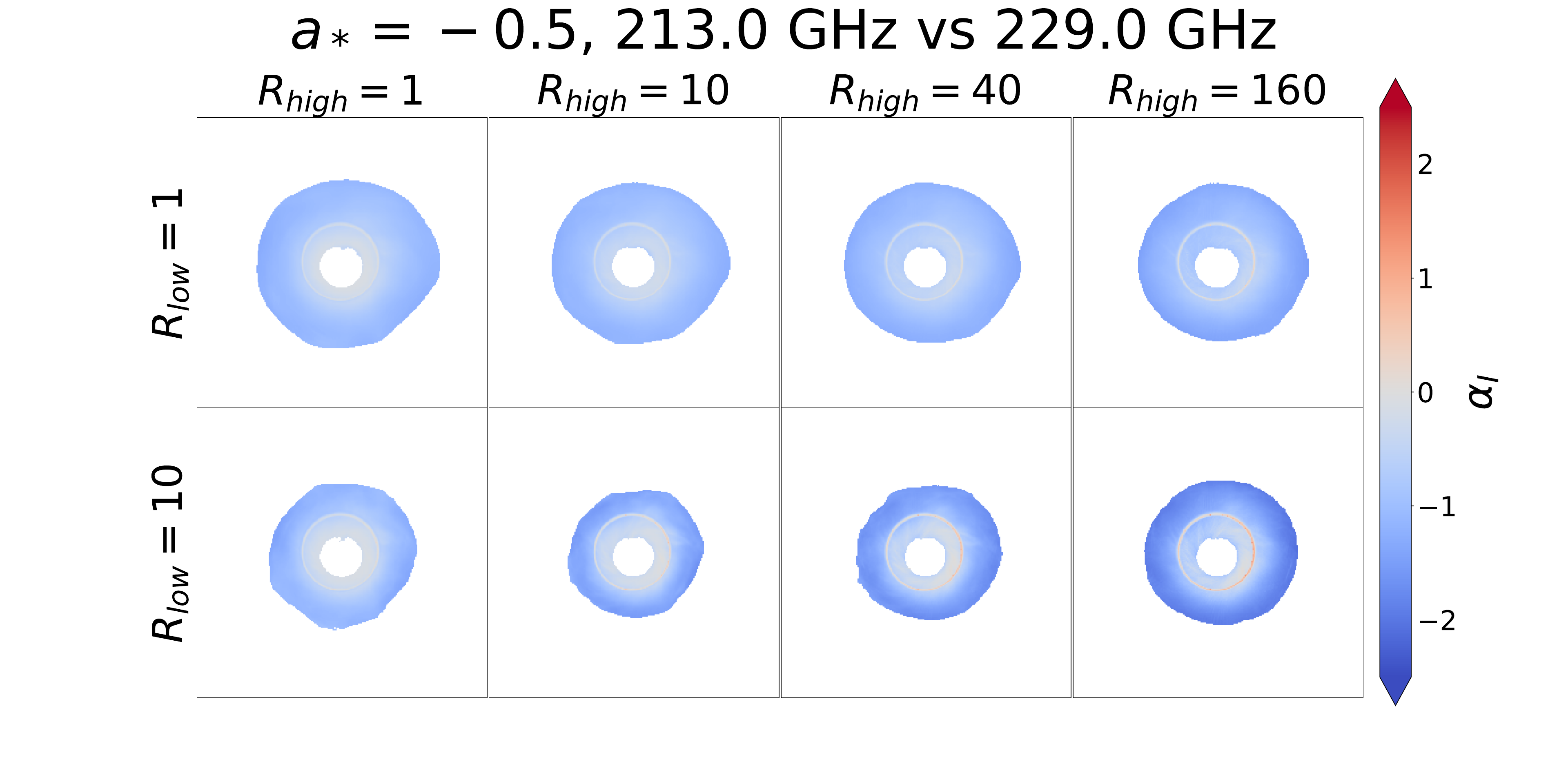}
  \includegraphics[trim={5cm 2.5cm 2cm 0cm},clip,width=0.39\textwidth]{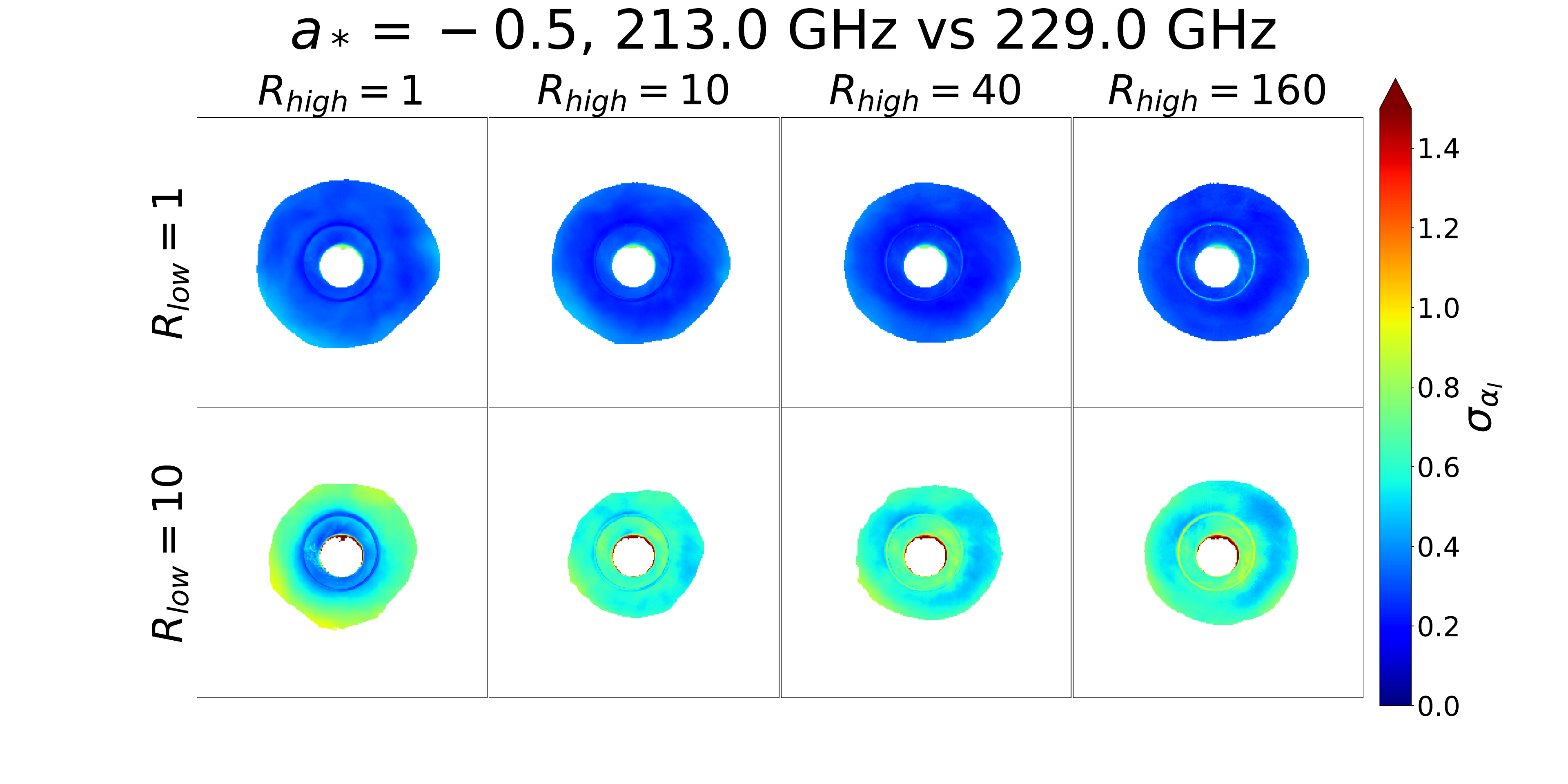}\\  
  \includegraphics[trim={5cm 2.5cm 2cm 0cm},clip,width=0.39\textwidth]{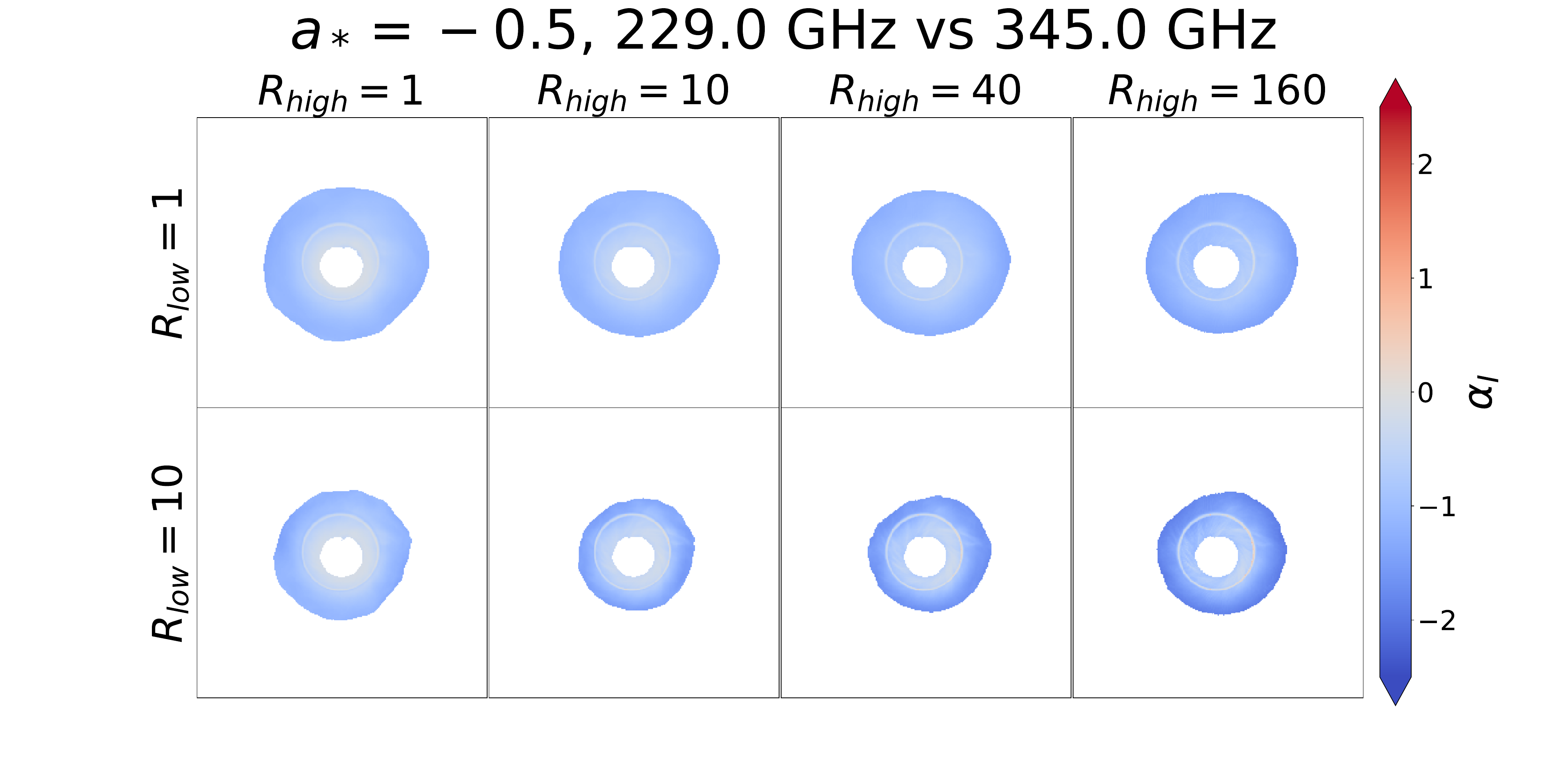}
  \includegraphics[trim={5cm 2.5cm 2cm 0cm},clip,width=0.39\textwidth]{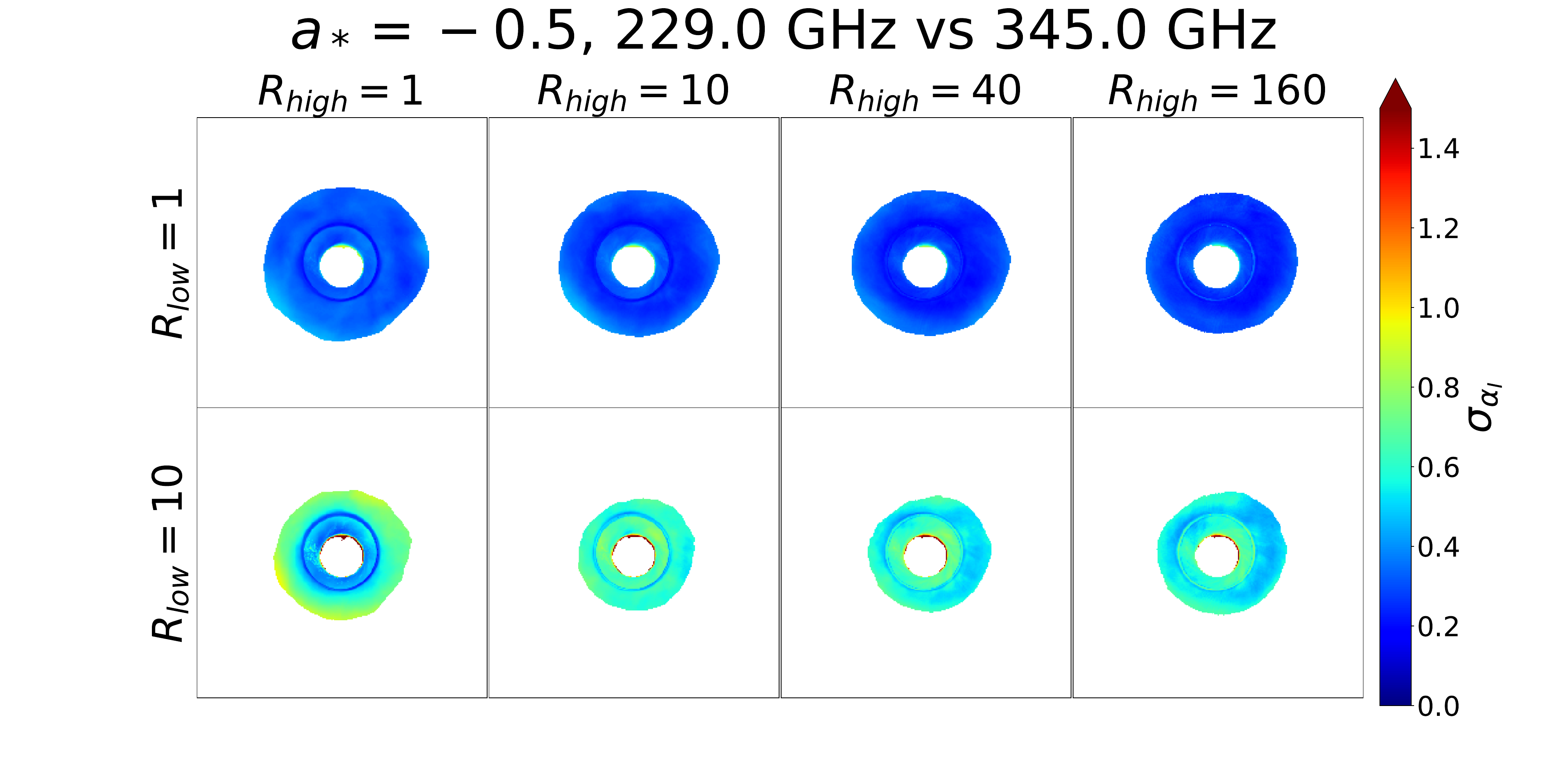}
\caption{Same as in Figure~\ref{fig:rhigh_rlow_time_averaged_a-0.94} but for $a_*=-0.5$.}
\label{fig:rhigh_rlow_time_averaged_a-0.5}
\end{figure*}

\begin{figure*}
  \centering
  \includegraphics[trim={5cm 2.5cm 2cm 0cm},clip,width=0.39\textwidth]{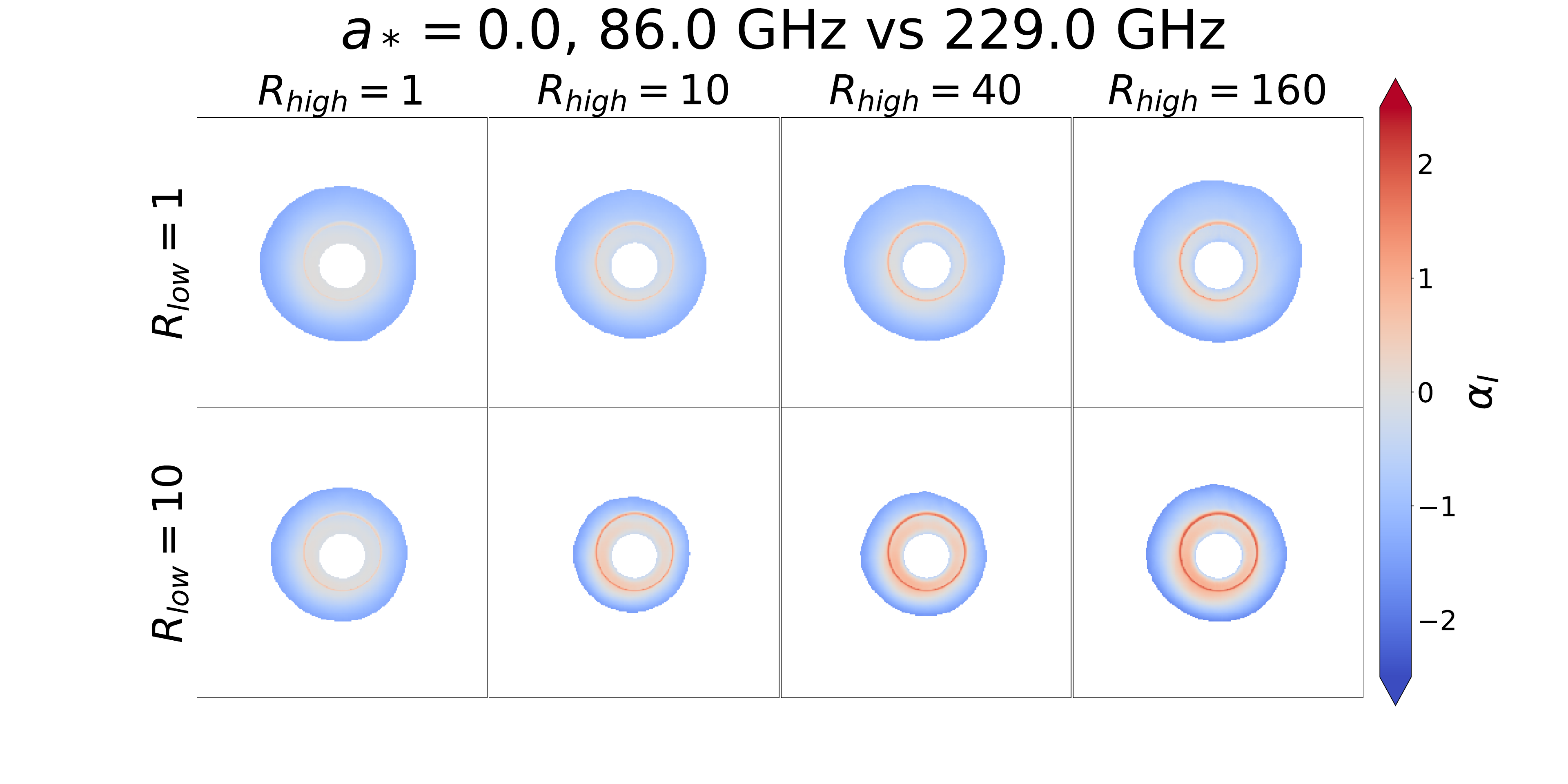}
  \includegraphics[trim={5cm 2.5cm 2cm 0cm},clip,width=0.39\textwidth]{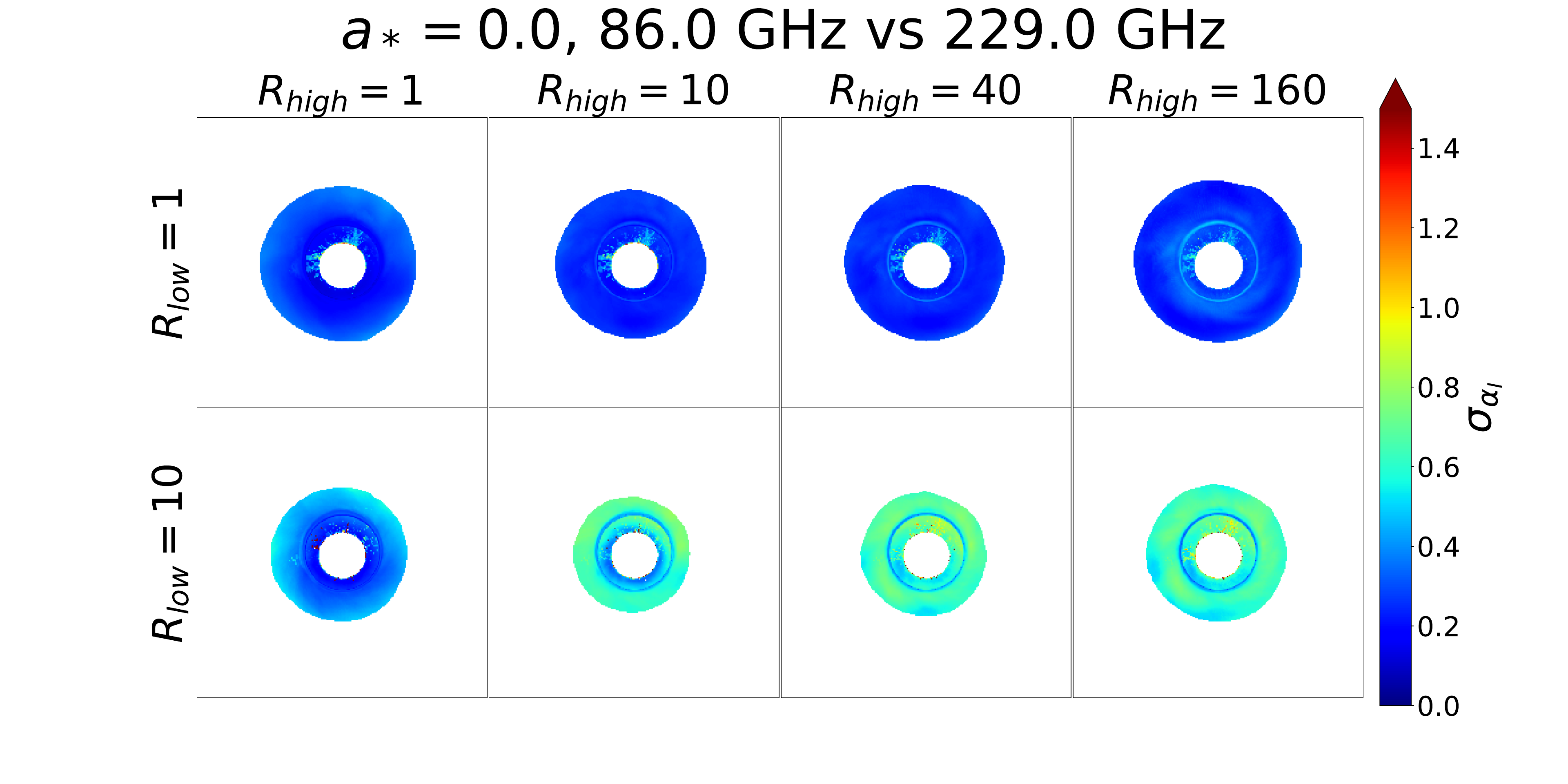}\\
 \includegraphics[trim={5cm 2.5cm 2cm 0cm},clip,width=0.39\textwidth]{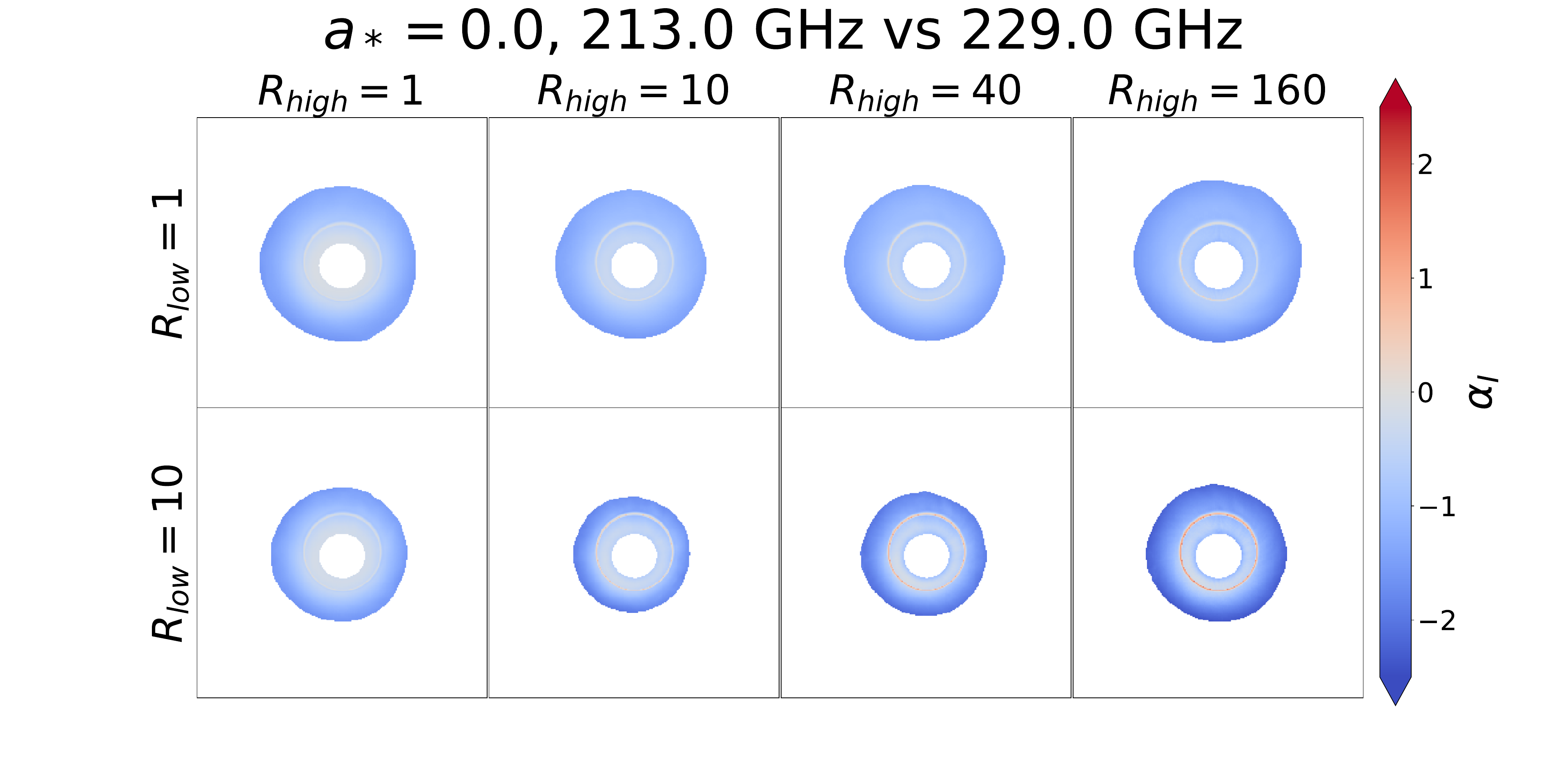}
  \includegraphics[trim={5cm 2.5cm 2cm 0cm},clip,width=0.39\textwidth]{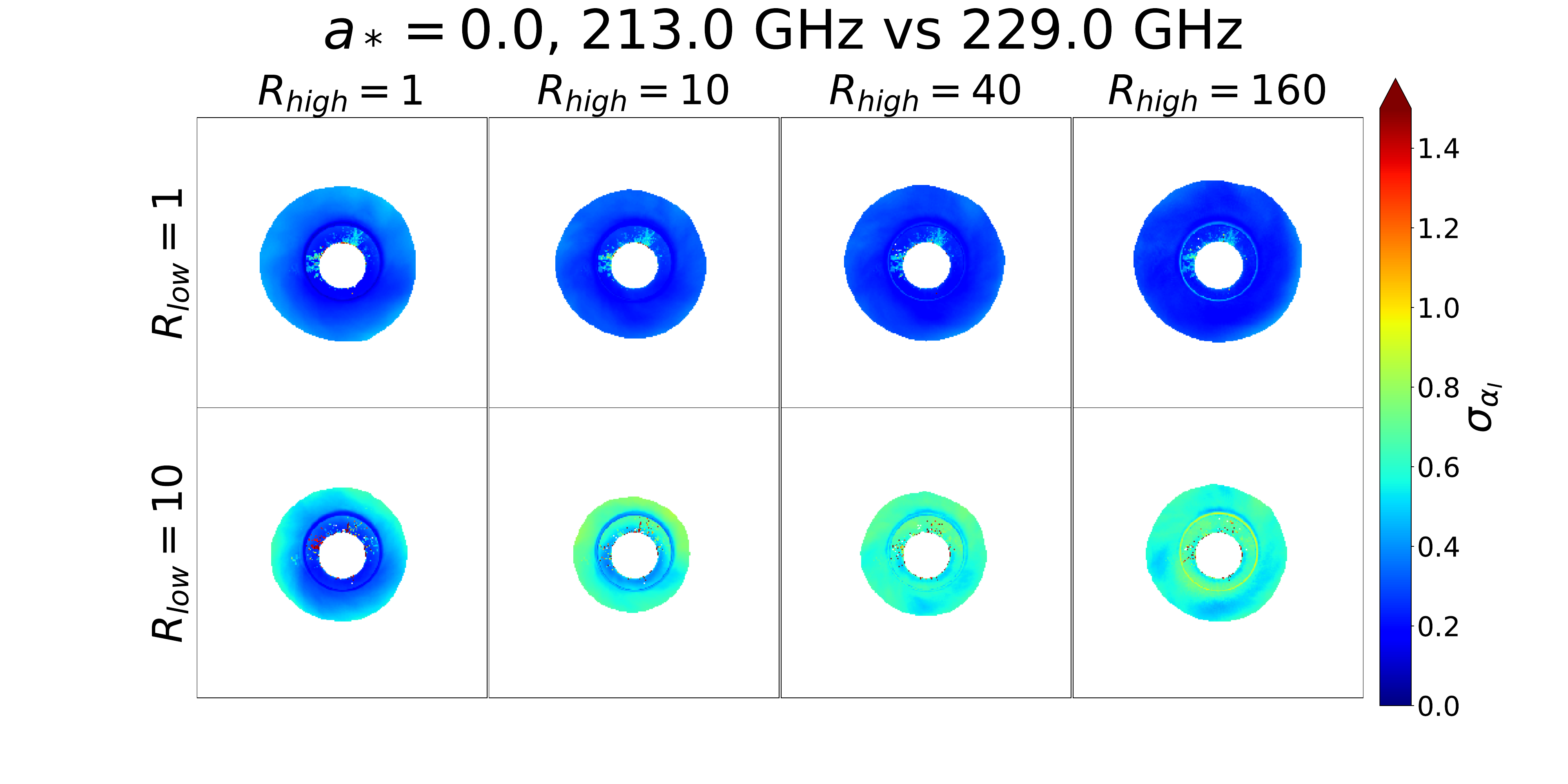}\\  
  \includegraphics[trim={5cm 2.5cm 2cm 0cm},clip,width=0.39\textwidth]{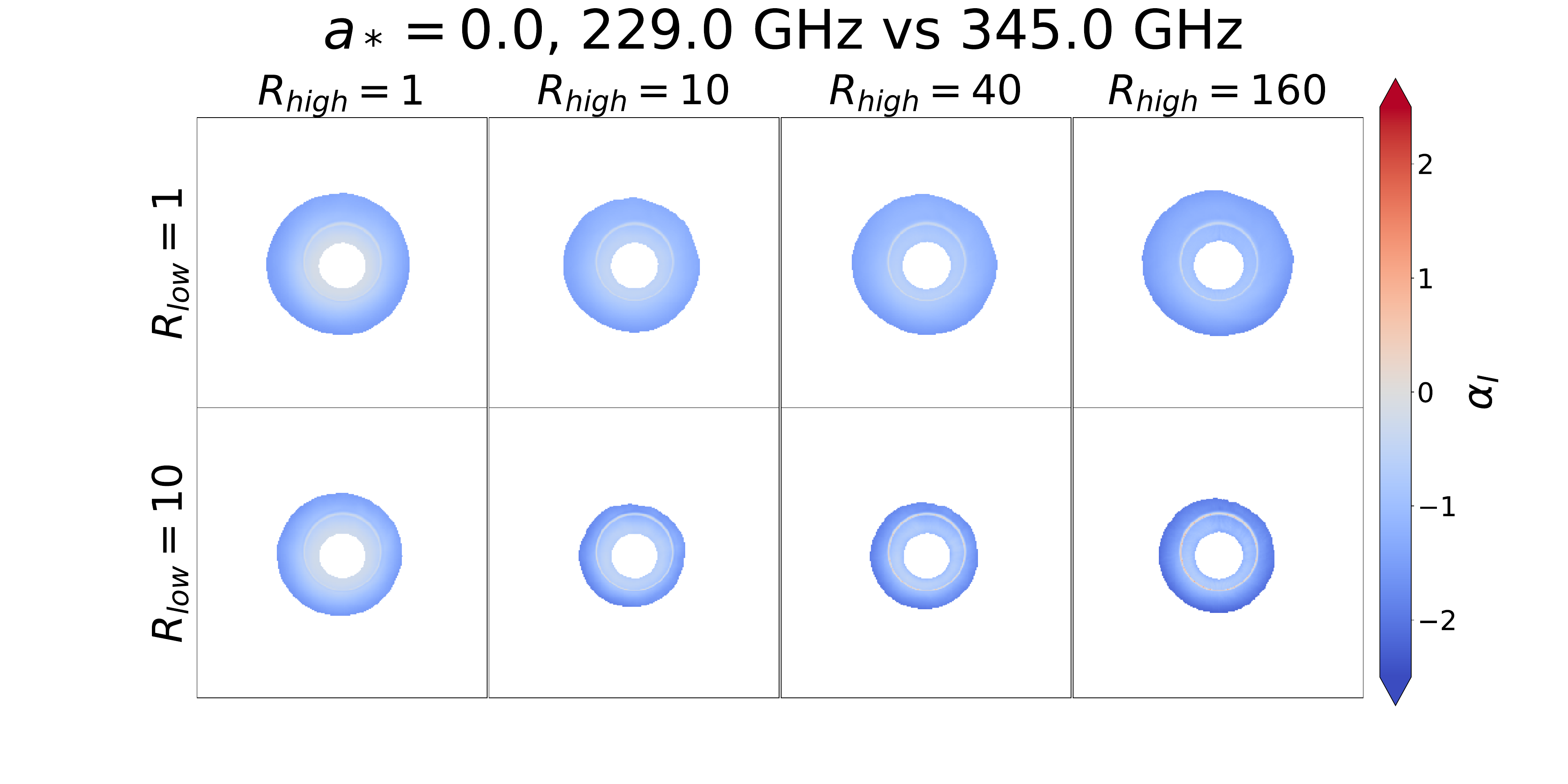}
  \includegraphics[trim={5cm 2.5cm 2cm 0cm},clip,width=0.39\textwidth]{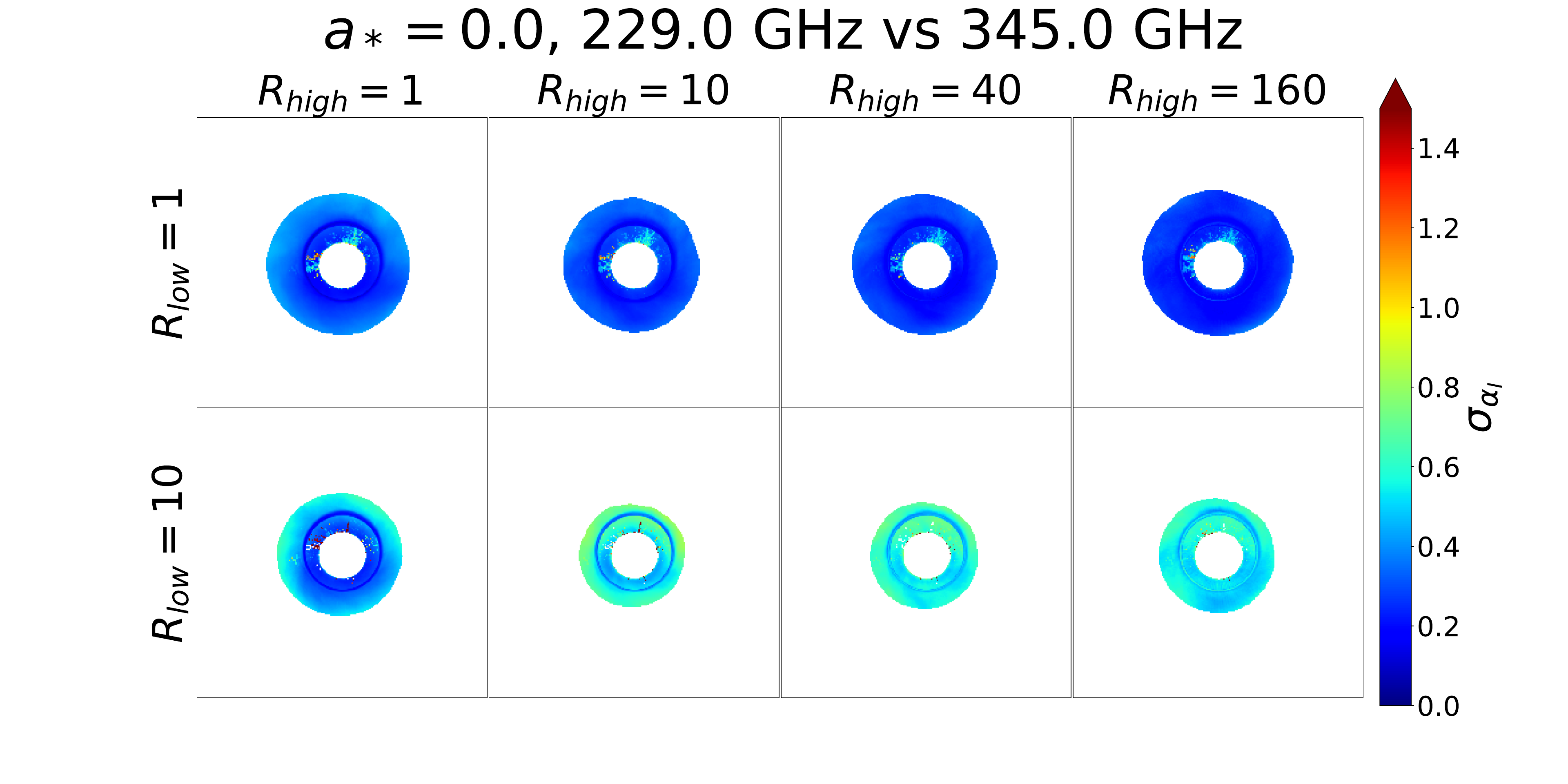}
\caption{Same as in Figure~\ref{fig:rhigh_rlow_time_averaged_a-0.94} but for $a_*=0$.}
\label{fig:rhigh_rlow_time_averaged_a0}
\end{figure*}

\begin{figure*}
  \centering
  \includegraphics[trim={5cm 2.5cm 2cm 0cm},clip,width=0.39\textwidth]{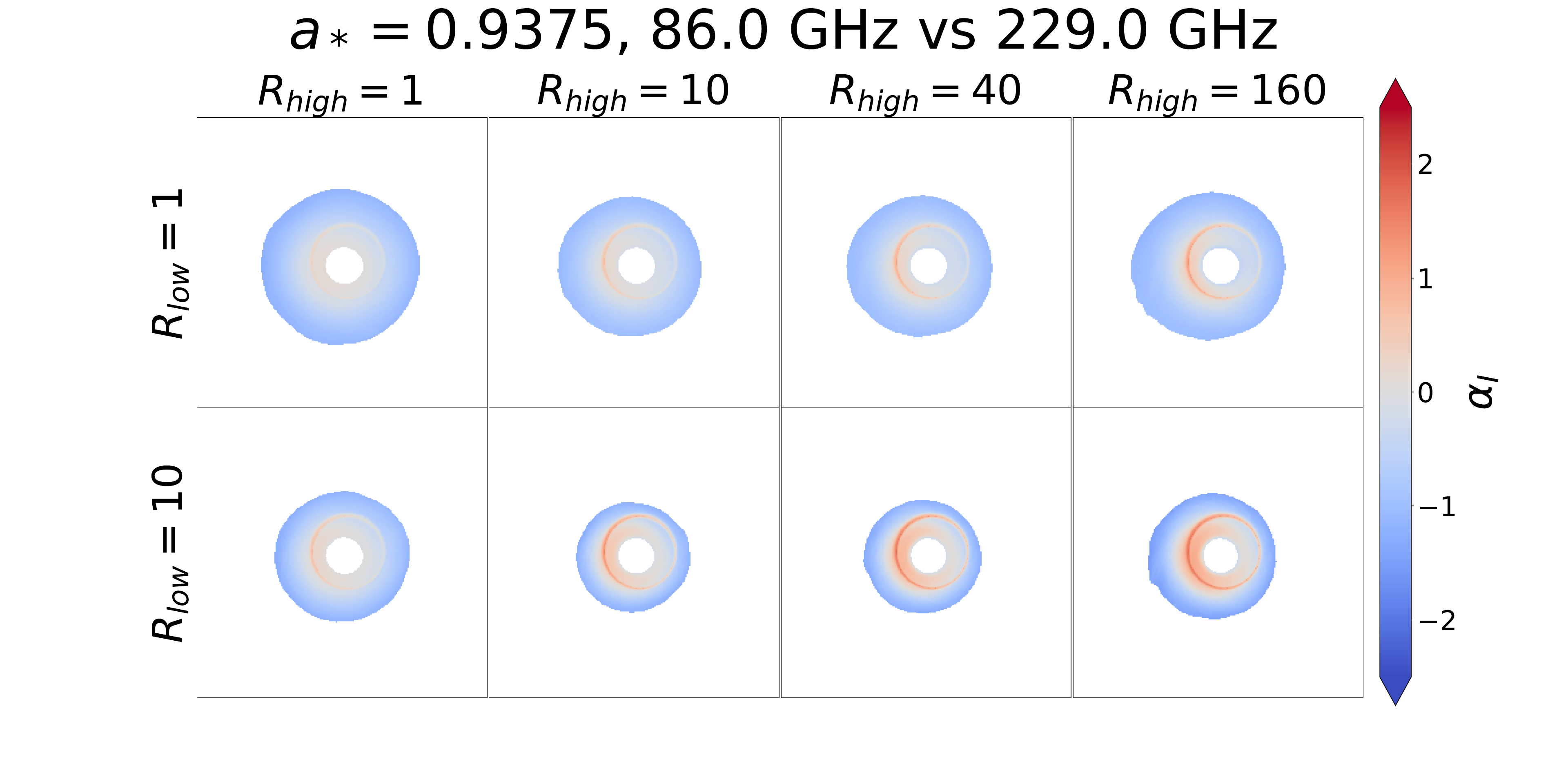}
  \includegraphics[trim={5cm 2.5cm 2cm 0cm},clip,width=0.39\textwidth]{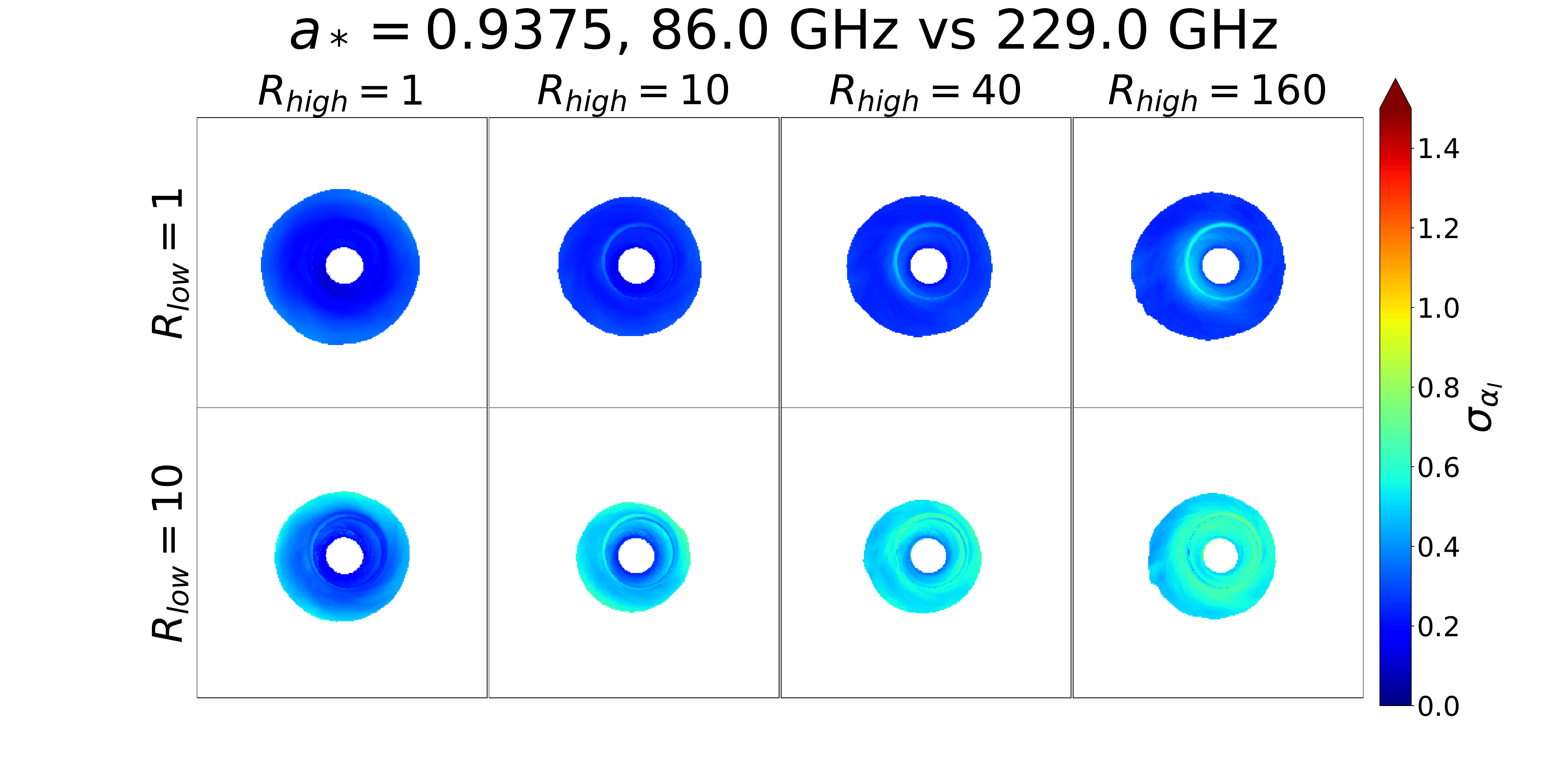}\\
 \includegraphics[trim={5cm 2.5cm 2cm 0cm},clip,width=0.39\textwidth]{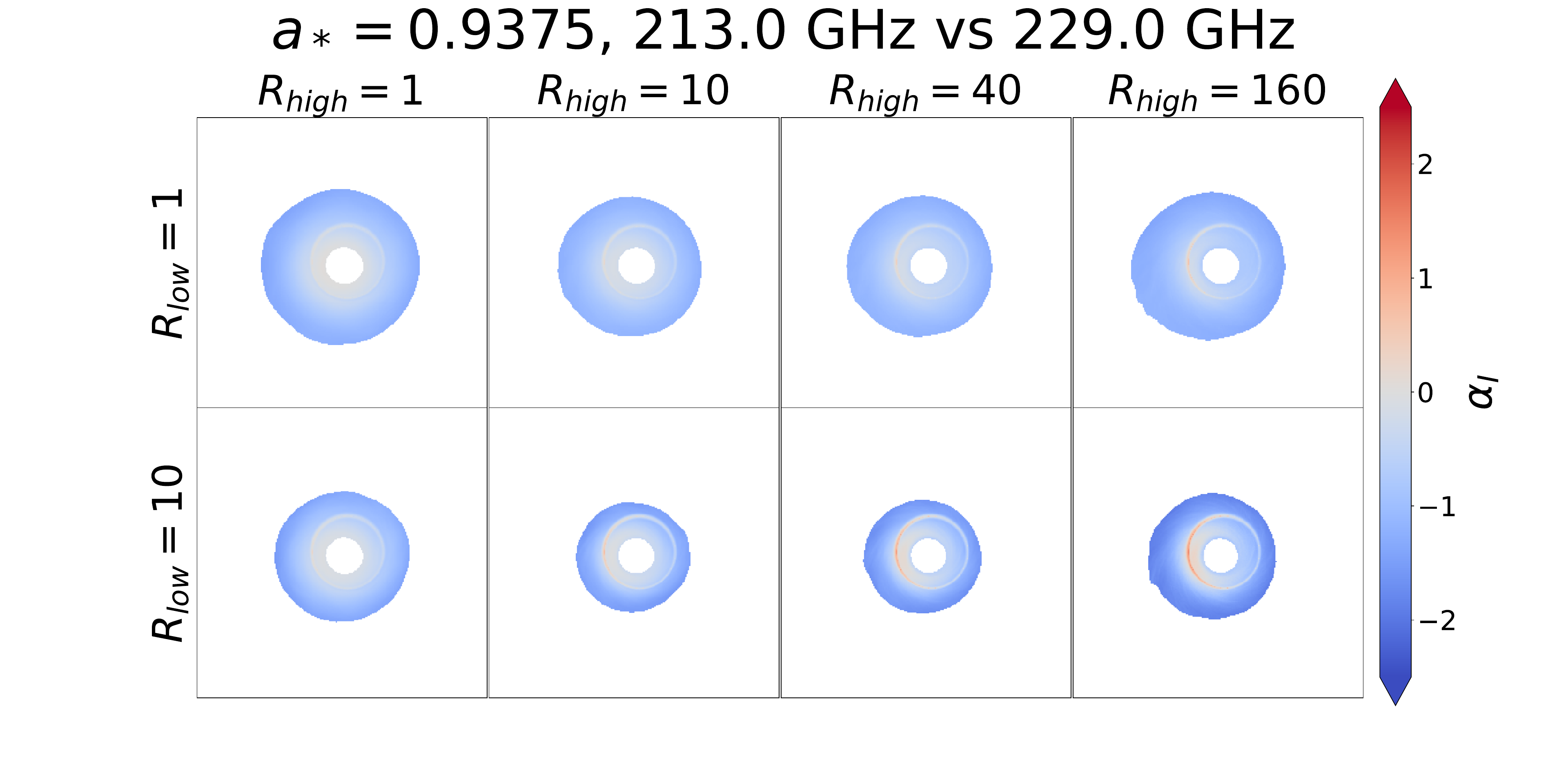}
  \includegraphics[trim={5cm 2.5cm 2cm 0cm},clip,width=0.39\textwidth]{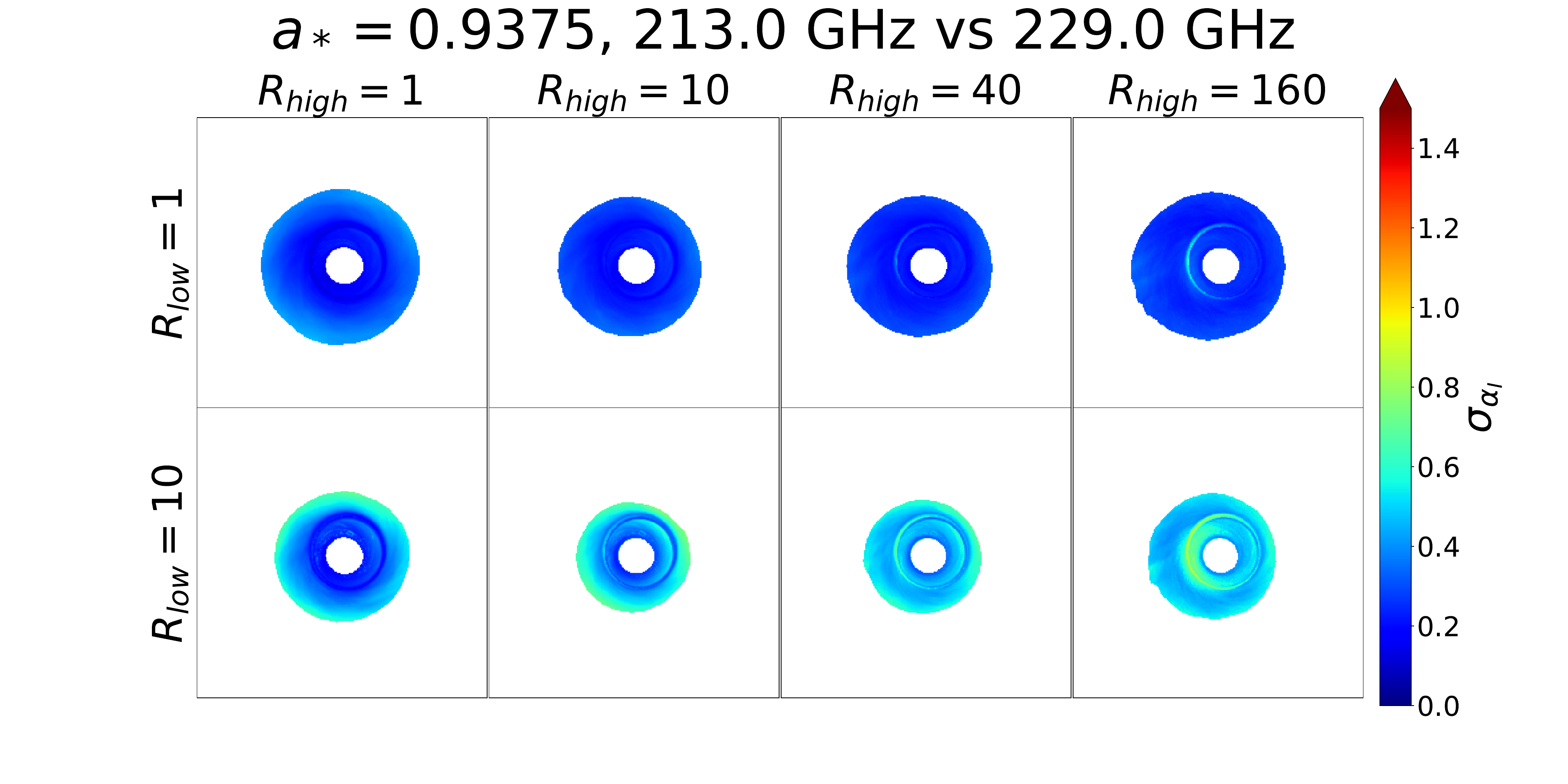}\\  
  \includegraphics[trim={5cm 2.5cm 2cm 0cm},clip,width=0.39\textwidth]{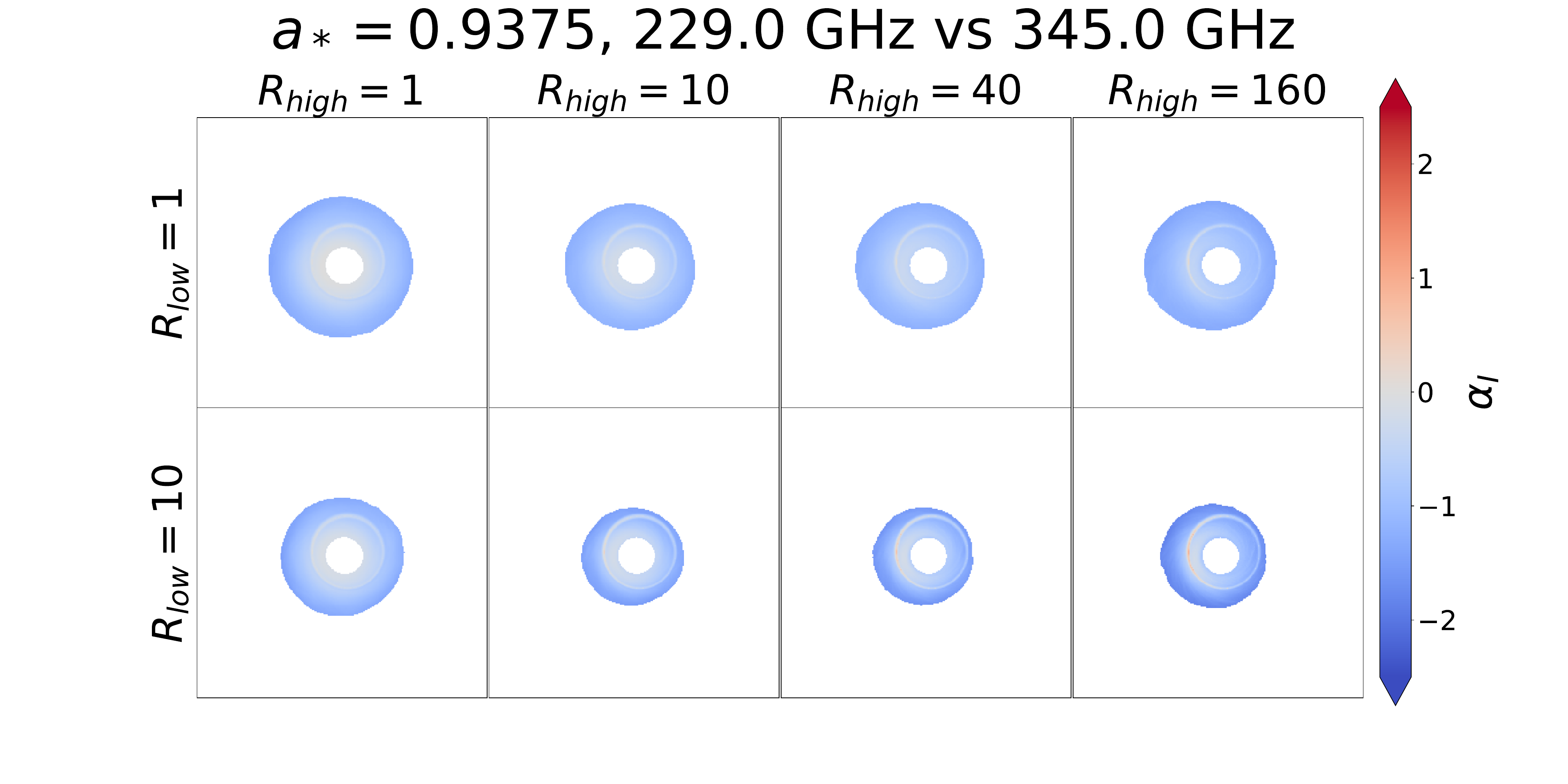}
  \includegraphics[trim={5cm 2.5cm 2cm 0cm},clip,width=0.39\textwidth]{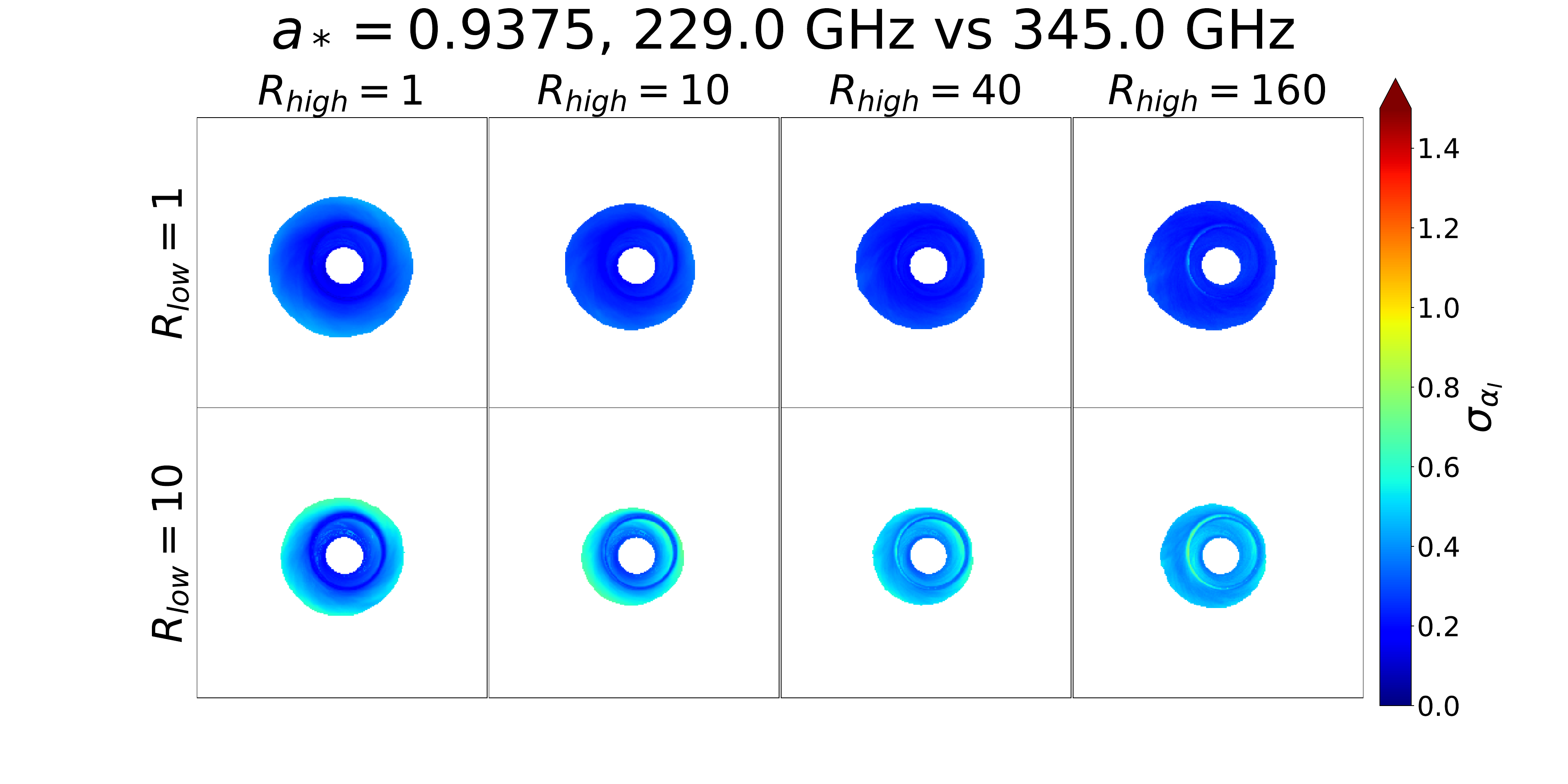}
\caption{Same as in Figure~\ref{fig:rhigh_rlow_time_averaged_a-0.94} but for $a_*=0.9375$.}
\label{fig:rhigh_rlow_time_averaged_a0.94}
\end{figure*}



\bsp	
\label{lastpage}
\end{document}